\documentclass[twocolumn]{aastex62}
\usepackage{float,graphicx,amsmath,multirow,mathtools}
\usepackage{color}
\usepackage[version=3]{mhchem}
\newcommand{\mjb}{mJy~beam$^{-1}$}
\newcommand{\kms}{km~s$^{-1}$}

\begin{document}

\title{Interstellar Glycolaldehyde, Methyl Formate, and Acetic Acid I: A Bi-modal Abundance Pattern in Star Forming Regions}
\author{Samer J. El-Abd}
\altaffiliation{Samer El-Abd is a student of the National Radio\\ Astronomy Observatory}
\affiliation{Department of Astronomy, University of Virginia, Charlottesville, VA 22904, USA}
\author{Crystal L. Brogan}
\affiliation{National Radio Astronomy Observatory, Charlottesville, VA 22903, USA}
\author{Todd R. Hunter}
\affiliation{National Radio Astronomy Observatory, Charlottesville, VA 22903, USA}
\author{Eric R. Willis}
\affiliation{Department of Chemistry, University of Virginia, Charlottesville, VA 22904, USA}
\author{Robin T. Garrod}
\affiliation{Department of Chemistry, University of Virginia, Charlottesville, VA 22904, USA}
\affiliation{Department of Astronomy, University of Virginia, Charlottesville, VA 22904, USA}
\author{Brett A. McGuire}
\altaffiliation{B.A.M. is a Hubble Fellow of the National Radio\\ Astronomy Observatory}
\affiliation{National Radio Astronomy Observatory, Charlottesville, VA 22903, USA}
\affiliation{Harvard-Smithsonian Center for Astrophysics, Cambridge, MA 02138, USA}
\correspondingauthor{Brett A. McGuire}
\email{bmcguire@nrao.edu}

\begin{abstract}

\noindent The relative column densities of the structural isomers methyl formate, glycolaldehyde, and acetic acid are derived for a dozen positions towards the massive star-forming regions MM1 and MM2 in the NGC 6334I complex, which are separated by $\sim$4000 AU. Relative column densities of these molecules are also gathered from the literature for 13 other star-forming regions. In this combined dataset, a clear bi-modal distribution is observed in the relative column densities of glycolaldehyde and methyl formate. No such distribution is evident with acetic acid. The two trends are comprised of star-forming regions with a variety of masses, suggesting that there must be some other common parameter that is heavily impacting the formation of glycolaldehyde. This is indicative of some demonstrable differentiation in these cores; studying the abundances of these isomers may provide a clue as to the integral chemical processes ongoing in a variety of protostellar environments.

\end{abstract}
\keywords{Astrochemistry -- ISM: molecules}

\section{Introduction}
\label{intro}

The formation of complex organic molecules (COMs) - those species with 6 or more atoms \citep{Herbst:2009go} - is a phenomenon known to occur during the early stages of star formation. Early theories on the formation of these COMs favored their production through gas-phase ion-molecule chemistry \citep{Herbst:1977ee}. However this was shown to be too inefficient to replicate the abundances of particular molecules observed in the interstellar medium (ISM) \citep{Horn:2004nr}; this led to grain-surface chemistry being the favored method of formation for many molecules in recent years  \citep[e.g.,][]{Garrod:2008tk,Garrod:2013id,Linnartz:2015ec}. A number of recent laboratory and chemical modeling studies, however, suggest that gas-phase reactions may indeed be non-trivial formation pathways for COMs, particularly in cold environments \citep{Laas:2011yd, Balucani:2015gj,Skouteris:2018de}. 

Many of the radicals that drive the production of these COMs - methyl (\ce{CH3}), hydroxymethyl (\ce{CH2OH}), and methoxy (\ce{CH3O}) - are produced during the photodissociation of methanol (\ce{CH3OH}; \citealt{Laas:2011yd}). For molecules such as the \ce{C2H4O2} isomers methyl formate (\ce{CH3OCHO}), acetic acid (\ce{CH3COOH}), and glycolaldehyde (\ce{CH(O)CH2OH}), the rate at which these radicals are produced, and the relative branching fractions between their production pathways, will directly influence the relative abundance of each isomer, as well as other species containing these functional groups. Understanding the formation of the \ce{C2H4O2} isomers is an important step to understanding the formation of yet more complex molecules that are necessary for life.  

One issue that continues to plague efforts to model the production of the \ce{C2H4O2} isomers is the overabundance of glycolaldehyde in simulations, especially relative to methyl formate which seems to be more accurately reproduced \citep{Laas:2011yd, Garrod:2013id}.  With respect to the reactions forming these species from \ce{CH3OH} photodissociation products, \citet{Garrod:2013id} suggested the primary routes are:
\begin{equation}
    \ce{HCO + CH3O -> HCOOCH3}
    \label{mf_eq}
\end{equation}
and
\begin{equation}
    \ce{HCO + CH2OH -> CH(O)CH2OH}
    \label{ga_eq}
\end{equation}
and
\begin{equation}
    \ce{CH3 + COOH -> CH3COOH}
\end{equation}
for methyl formate, glycolaldehyde, and acetic acid, respectively. All of these reactions are exothermic. Important to note is that while acetic acid can be produced from the photodissociation products of methanol, this is not believed to be its primary formation pathway, which is instead hydrogenation of \ce{CH2COOH} on grain surfaces.

This likely implies some or all of the following possibilities:
\begin{enumerate}
    \item the formation of glycolaldehyde via the above reaction is far less efficient than currently believed,
    \item the abundances of the precursor species (primarily HCO and \ce{CH2OH}) are substantially different than those predicted by the models (perhaps due to incorrect branching ratios from \ce{CH3OH} photodissociation),
    \item there are unknown competing formation pathways that serve as sinks for \ce{CH2OH}, reducing the availability of this radical for forming glycolaldehyde through Reaction~\ref{ga_eq}, or
    \item there is an underlying, unaccounted for physical process that is affecting glycolaldehyde abundances that needs to be better constrained.
\end{enumerate}

Adding further complexity, recent modeling of these species has suggested a number of viable gas-phase formation pathways, in addition to these grain-surface radical-radical recombination reactions \citep{Balucani:2015gj, Skouteris:2018de}. \citet{Rivilla:2019br} study and model the HCO emission from IRAS 16293 and are able to accurately reproduce the abundance of HCO; they also posit that the dominant formation route for glycolaldehyde is the one presented here.

In this paper we perform an in-depth analysis on the abundance of glycolaldehyde, methyl formate, and acetic acid in the massive protocluster NGC 6334I \citep{Brogan:2016cy} by extracting spectra from a dozen positions in the cloud. We also present a systematic study of these molecules across 13 additional interstellar sources using literature data. The new data toward NGC~6334I and their analysis are described in \S\ref{observations}, efforts to standardize the literature sources are presented in \S\ref{lit}, and the results are presented and discussed in \S\ref{results}.

\section{Observations \& Data Reduction}
\label{observations}

\begin{figure*}[!ht]
\centering
\includegraphics[height=0.5\textheight]{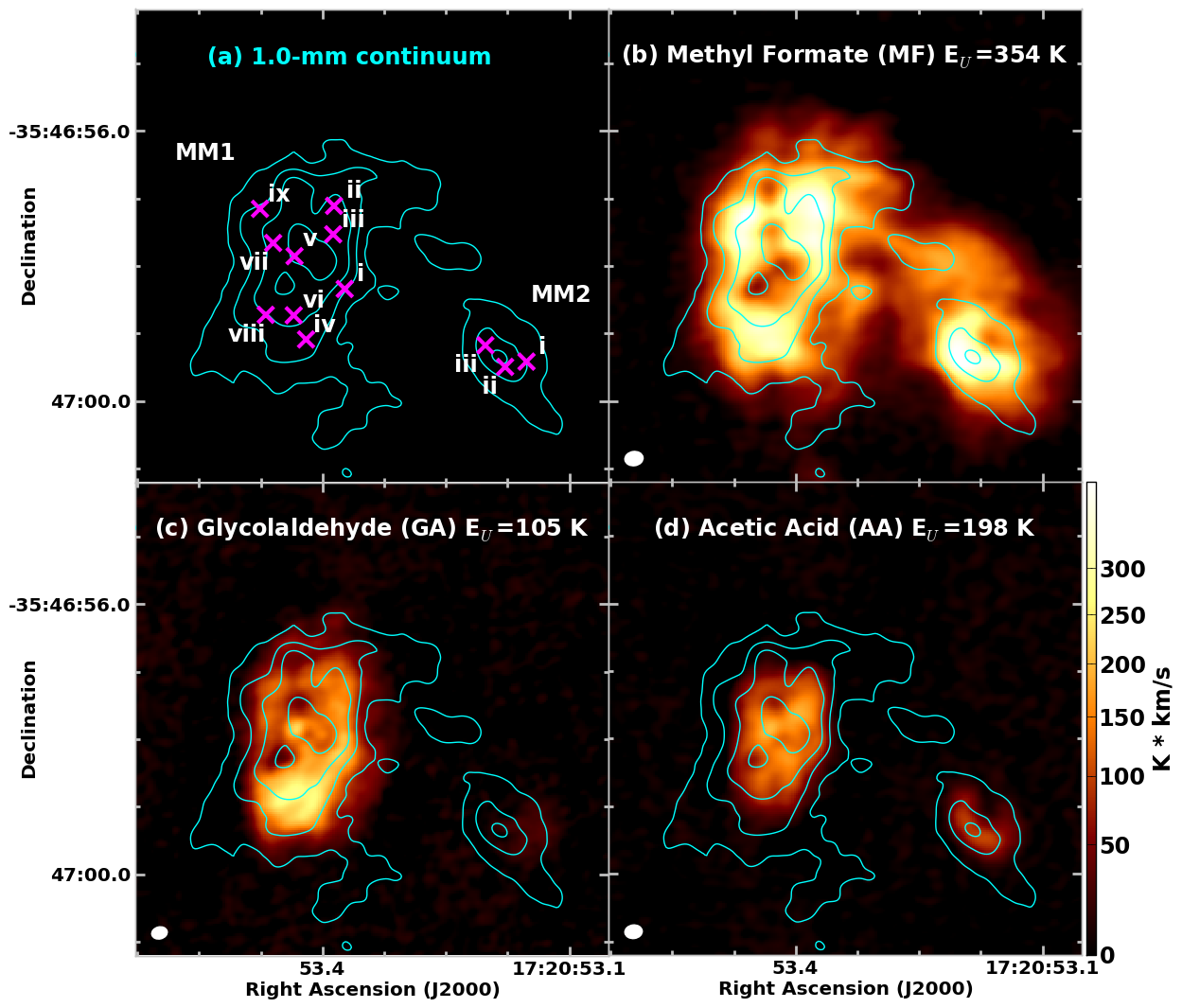}
\caption{Panel (a) shows the locations where spectra where extracted from NGC\,6334\,I MM1 and MM2 (magenta $\times$ symbols and white labels) overlaid on contours of the 1~mm dust continuum (contour levels: 17.5, 52.5, 140.0, and 367.5 \mjb). See Table \ref{params} for the physical conditions in each position. Panels b, c, and d show integrated intensity (moment 0) maps showing the spatial distribution of methyl formate (b), glycolaldehyde (c), and acetic acid (d) integrated from -10.4 to -2.7 \kms; the continuum contours from (a) are also overlaid. See Table \ref{params} for a list of the position coordinates.}
\label{4panel}
\end{figure*}

\subsection{Observations}
\label{obs}
The ALMA data toward NGC~6334I were observed during Cycle 3 in 2016, project code 2015.A.00022.T. The data were calibrated using the ALMA Cycle 4 pipeline (CASA 4.7.2). The observations were centered at $\alpha$(J2000) = 17:20:53.36, $\delta$(J2000) = -35:47:00.0 and had a nominal resolution of $0.24''\times 0.17''$ ($-83^\circ$), a full width half-power (FWHP) of the primary beam of $20''$, a spectral resolution of 1.1 \kms\/, and a rms per channel of 2.0 \mjb\/ (0.62 K). The observations consisted of two tunings, each with four spectral windows, with a bandwidth of 1.87 GHz per window. The first set of spectral windows were centered at 280.1, 282.0, 292.1, and 294.0 GHz. The second set of spectral windows were centered at 337.1, 339.0, 349.1, and 351.0 GHz. Primary beam corrections were applied to the images and the cubes were smoothed to a resolution of $0.26''\times 0.26''$ before analysis. Further details are given in \citet{McGuire:2017gy}, \citet{Hunter:2017th}, and \citet{Brogan:2018wb}.

\subsection{Analysis of NGC 6334I}
\label{analysis}

\subsubsection{Methods}

NGC~6334I contains two prodigious hot core spectral line sources MM1 and MM2 \citep{Bogelund18,Zernickel:2012hx,Beuther07}. As can be seen in Figure \ref{4panel}, the morphology of the three \ce{C2H4O2} isomers varies across each source. To account for this and get an accurate sampling of the chemistry in the cloud we extracted spectra from 9 positions around MM1 and 3 positions around MM2. We began analyzing these spectra by building a simulated model of the observed molecular line emission. We assumed a single excitation temperature and that the molecules were in local thermodynamic equilibrium (LTE). Measurement and subtraction of the background continuum from the observations were done following the methods of \citet{Brogan:2018wb}.


To verify the accuracy of the assumed $T_{ex}$, we use the formalism of \citet{Turner:1991um} for molecules described by a single value of $T_{ex}$.  The pertinent equations are

\begin{equation}
    \Delta T_B = [J_{\nu}(T_{ex}) - J_\nu(T_{bg})](1 - e^{-\tau_0}),
    \label{T_B}
\end{equation}

where

\begin{equation}
    J_\nu(T) \equiv (h\nu/k)[e^{h\nu/kT}- 1]^{-1}.
    \label{eq3}
\end{equation}

Here, $\Delta T_B$ is the observed brightness temperature of the line above the background continuum, $T_{ex}$ is the excitation temperature of that molecule, $T_{bg}$ is the background continuum temperature at the frequency of the transition, and $\tau_0$ is the optical depth of the line.  Given a $T_{ex}$ and $T_{bg}$, the limit of $\Delta T_B$ as lines become optically thick ($\tau_0 \gg 1$) can be calculated from Equation~\ref{T_B}.  Alternatively, and in the case of the observations described here, if $\Delta T_B$ can be measured from optically thick lines, and $T_{bg}$ is known, $T_{ex}$ can be inferred. 
This analysis assumes that all molecules within the $0.26''\times 0.26''$ beam share the same value of $T_{ex}$ (see Appendix~\ref{spectra} for a more complete discussion of this assumption).

Assuming the molecules can be described by a single excitation temperature, the column density (cm$^{-2}$) can be determined using the formalisms outlined in \citet{Hollis:2004uh} for a molecule described by a single $T_{ex}$ in the presence of a background continuum $T_{bg}$, and given in Equation~\ref{N_T}.

\begin{equation}
    N_T = \frac{1}{2} \frac{3k}{8\pi^3} \sqrt{\frac{\pi}{\rm{ln}2}} \frac{Qe^{E_u/T_{ex}}\Delta T_B \Delta V}{ \nu S \mu^2 \eta_B} \frac{1}{1-\frac{e^{h \nu / kT_{ex}}-1}{e^{h \nu / kT_{bg}}-1}}
    \label{N_T}
\end{equation}
Here, $Q$ is the partition function, $E_u$ is the upper state energy (K), $\Delta T_B$ is the  brightness temperature of the line above the continuum (K), converted from the measured intensity in Jy beam$^{-1}$ using the Planck scale, $\Delta V$ is the linewidth (km~s$^{-1}$), $S$ is the intrinsic line strength, $\mu^2$ is the transition dipole moment (Debye$^2$), $\eta_B$ is the beam efficiency, and $T_{bg}$ is the background continuum temperature (K). The partition function for each molecule was calculated through a direct summation of states as described in \citet{Gordy:1984uy}, and includes contributions from excited vibrational states with non-trivial populations.  Given a column density, this formalism can also be used to obtain a simulated, predicted spectrum of a molecule ($\Delta T_B$ for each transition with an applied Gaussian lineshape function of width $\Delta V$).

We sought to identify those transitions of methyl formate, glycolaldehyde, and acetic acid that were least blended with other molecular features.  To do this, we performed a zeroth-order, by-eye fit to the data for a number of different interstellar molecules by generating simulated spectra at fixed values of $T_{ex}$, $\Delta V$, and $v_{lsr}$. In total, at least 50 molecular species, including vibrational states and isotopologues, were identified in each MM1 and MM2 extraction position. For each species, the value of $N_T$ was then varied until the best visual match between a detected molecule (typically for many transitions) and the observations was found.  The spectral line properties for all of the simulated molecules were obtained from the CDMS \citep{Muller:2005ii} and JPL \citep{Pickett:1998cp} databases, accessible at \url{www.splatalogue.net}. While the exact column densities for these molecules were not derived, the simulated spectra served to identify the transitions of our target species that were the most useful for analysis (i.e. least contaminated by spectral blending with other lines). A number of spectra, including the modeled molecular spectra, are shown in Appendix~\ref{spectra}. 

Next, a least-squares optimization of the column densities of methyl formate, glycolaldehyde, and acetic acid was performed using only those lines that were identified as likely unblended, and which were seen to be optically thin (see Table \ref{freqs}). The molecular line properties of methyl formate, acetic acid, and glycolaldehyde were based on the laboratory works of \citet{Ilyushin:2009kg}, \citet{Ilyushin:2013gh}, and \citet{Carroll:2010gt}, respectively, and previous work referenced therein.

\subsubsection{Exploring the Parameter Space of NGC 6334I}

The molecular emission models that we constructed for each extraction position across MM1 and MM2 serve as valuable tools with which to explore the variation in physical conditions across the two sources. None of the parameters that were used to create the emission model - $T_{ex}$, $\Delta V$, and $v_{lsr}$ - were consistent across the entire region which is indicative of the prodigious effect of protostars on their surroundings. Excitation temperatures in MM1, for instance, were found to vary between 135-285 K while the excitation temperatures in MM2 varied from 152-200 K. The background temperature (i.e. continuum brightness temperature) was also found to vary significantly across NGC 6334I; increases in excitation temperature largely tracked changes in the background continuum which varied anywhere from 26.9-192.7 K in MM1 and 21.8-58.6 K in MM2. Fit $v_{lsr}$ values spanned a range of 3.8 \kms~in MM1 and 1.2 \kms~in MM2 which demonstrates the complex bulk motions of the gas in the source. The linewidths in MM2 were able to be consistently fit at 2.80 \kms~while in MM1 they varied from 3.00-4.50~\kms. For a full breakdown of the measured physical conditions in each region, refer to Table \ref{params}.

\begin{table*}
    \centering
    \caption{Physical Parameters for Spectra Extraction Locations in NGC 6334I}
    \begin{tabular}{cccccccc}
    \hline\hline
    Region  &   RA  &   Dec &   $T_{ex}$    & $T_{bg, Low}$  &   $T_{bg, High}$    &   $v_{LSR}$   &   $\Delta V$ \\
        &   hh:mm:ss    &   dd.mm.ss   &    (K)   &   (K)   &   (K)   &   (\kms)    &   (\kms)    \\ 
    \hline
    MM1-i &  17:20:53.373   &   -35.46.58.341   &   135 &   26.9    &   31.3    &   -7.0  &   3.25 \\
    MM1-ii &  17:20:53.386   &   -35.46.57.112   &  175  &  79.7    &   94.6    &  -5.0   &   3.25 \\
    MM1-iii &  17:20:53.387   &   -35.46.57.533   &   225 & 108.2   &   129.7    &   -5.2    &   3.00 \\
    MM1-iv &  17:20:53.420   &   -35.46.59.088   &   150 &  38.4    &   44.8    &   -8.2   &   4.50 \\
    MM1-v &  17:20:53.434   &   -35.46.57.856   &   285 &   159.3   &   192.7    &   -4.4   &   3.25 \\
    MM1-vi &  17:20:53.435   &   -35.46.58.731   &   190 &  77.6    &   88.6    &   -7.0   &   3.25 \\
    MM1-vii &  17:20:53.460   &   -35.46.57.661   &   185 & 96.5    &   112.4    &   -4.8    &   3.00 \\
    MM1-viii &  17:20:53.469   &   -35.46.58.724   &   150 & 57.1   &   64.7    &   -6.8 &   3.00 \\
    MM1-ix &  17:20:53.476   &   -35.46.57.156   &   150 &  36.8    &   45.2    &   -5.0   &   2.50 \\
    MM2-i &  17:20:53.152   &   -35.46.59.416   &   150 &   21.8    &   27.8    &   -9.0   &   2.80 \\
    MM2-ii &  17:20:53.178   &   -35.46.59.494   &   200 &   44.4   &   58.6    &   -9.0  &   2.80 \\
    MM2-iii &  17:20:53.202   &   -35.46.59.175   &   180 &  44.2   &   57.7    &   -7.8   &   2.80 \\
    \hline
    \multicolumn{8}{l}{The background temperature of the observations varied between the lower and}\\
    \multicolumn{8}{l}{upper sidebands of the ALMA data.}
    \end{tabular}
    \label{params}
\end{table*}

\begin{table*}
    \centering
    \footnotesize
    \caption{Transitions Used to Calculate Isomer Column Densities in NGC 6334I}
    \begin{tabular}{ccccccccccccc}
    \hline\hline
    Source & Molecule & $J'$ & $K_a'$ & $K_c'$ & $J''$ & $K_a''$ & $K_c''$ & A/E & Frequency & $I$ & $S_{ij} \mu^2$ & $E_{upper}$\\
    & & & & & & & & & (MHz) & (Jy beam$^{-1}$) & (D$^2$) & (K)\\
    \hline
    &\vspace{-0.75em}\\
    MM1-i & Methyl Formate  & 24 & 3 & 22 & 23 & 3 & 21 & A & 279294.919 & 0.086 & 7.29 & 178.53 \\
                &    & 23 & 17 & * & 22 & 17 & * & A & 282510.849 & 0.202 & 55.80 & 354.42 \\
                &    & 24 & 20 & * & 23 & 20 & * & E & 294678.389 & 0.097 & 19.66 & 442.06 \\
                &    & 24 & 18 & * & 23 & 18 & * & A & 294769.955 & 0.184 & 56.17 & 391.75 \\
    & Acetic Acid     & 23 & * & 20 & 22 & * & 19 & E & 279775.7366 & 0.080 & 111.46 & 177.86 \\
                &    & 26 & * & 26 & 25 & * & 25 & E & 281891.4488 & 0.089 & 48.83 & 186.17 \\
                &    & 26 & * & 25 & 25 & * & 24 & E & 291814.6653 & 0.102 & 141.3 & 198.33 \\
    & Glycolaldehyde  & 25 & 3 & 22 & 24 & 4 & 21 & ... & 279230.1847 & 0.184 & 74.73 & 189.12 \\
                &    & 25 & 4 & 22 & 24 & 3 & 21 & ... & 282760.5619 & 0.237 & 74.86 & 189.19 \\
                &    & 15 & 4 & 11 & 14 & 3 & 12 & ... & 292924.2749 & 0.052 & 11.55 & 77.67 \\
                &    & 16 & 7 & 10 & 15 & 6 & 9 & ... & 348314.054 & 0.380 & 37.24 & 105.43 \\
    \hline
    \multicolumn{13}{l}{The transitions that were used in the least squares analysis (The full table is available online in a machine-readable format).}
    \end{tabular}
    \label{freqs}
\end{table*}


\section{Literature Sources}
\label{lit}

For our comparison of the \ce{C2H4O2} isomers, we compiled a list of star-forming regions for which the column density of methyl formate was known and at least an upper limit was available for the column density of glycolaldehyde. Values for the column densities of methyl formate, acetic acid, and glycolaldehyde in the literature are presented here unchanged from the original analysis with few exceptions, detailed in Table \ref{cds_table}. For the acetic acid column densities provided by \citet{Remijan:2002tv} and \citet{Remijan:2003wf}, a correction was applied to the partition function used in those works to make it consistent with the partition function used in other sources, and by our analysis. A correction for the beam size was also made to the glycolaldehyde abundance reported in \citet{Fuente:2014je} for NGC 7129 FIRS 2 in order to more accurately compare with their methyl formate value. 
For the source G31.41+0.31, measurements of the methyl formate and glycolaldehyde column densities were available from \citet{Calcutt:2014ug}, however \citet{Rivilla:2017vg} provide a more recent measurement with a multi-line analysis, so their value was adopted for this work. Even so, the ratio of methyl formate to glycolaldehyde abundances between the two works only differed by $\sim$20\%. Errors on the column densities were generally provided in the literature. When they were not, a 30\% uncertainty was assumed. 

\section{Results \& Discussion}
\label{results}


Presented in Table \ref{cds_table} are the compiled column densities of methyl formate, acetic acid, and glycolaldehyde for each source that was used in this analysis. The molecules span several orders of magnitude in column density, reflective primarily of the span of absolute gas number densities in these sources.  Critically, however, the ratios of the molecular column densities in each source vary significantly.  Rather than introduce additional uncertainty in the source-to-source comparisons by attempting to derive abundances relative to \ce{H2} from yet other disparate literature studies for each source, we choose to focus our analysis on the relative behavior of the three molecules to each other.  This brings with it the assumption that all three molecules are co-spatial in each observation.  A more accurate approach would require a detailed, self-consistent, interferometric survey of these species from a single study.

To test for any relationship between the isomers among the column densities provided in the literature, we plotted the column density of methyl formate as a function of the acetic acid and glycolaldehyde column densities. In the plot of methyl formate vs acetic acid (Figure 2a), the column densities increase at a similar rate.  In the plot of methyl formate vs glycolaldehyde (Figure 2b), however, there is a clear bi-modal distribution in the glycolaldehyde column densities. The two trends are characterized by a number of sources with a ``small" methyl formate to glycolaldehyde (MF/GA) ratio, and a number of sources with a ``large" MF/GA ratio. The ``small" MF/GA sources are a mixture of low and high-mass star-forming regions, while the ``large" MF/GA sources are exclusively high-mass star-forming regions.

\begin{figure*}
\centering
\includegraphics[width=\textwidth]{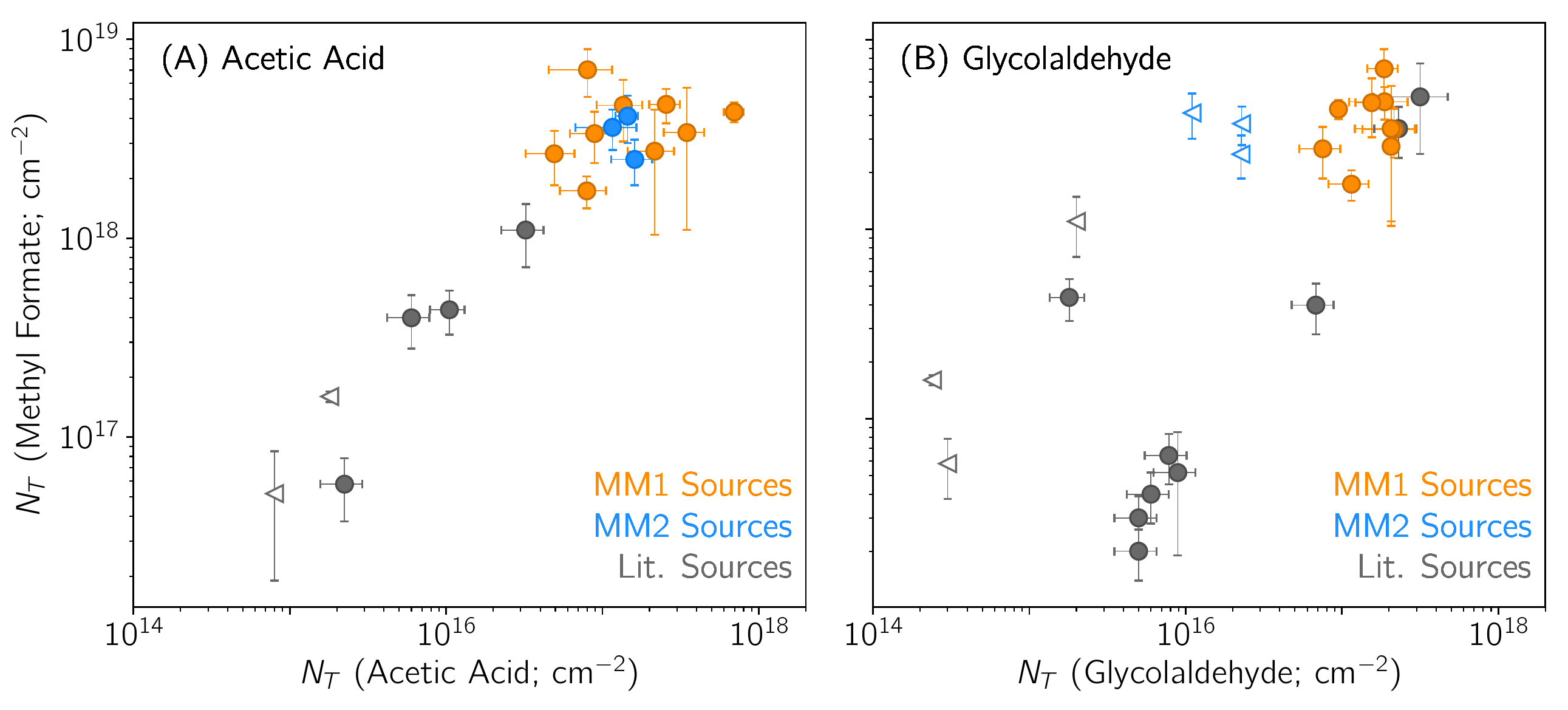}
\caption{Scatter plots of methyl formate  column density versus \textbf{(A)} acetic acid and \textbf{(B)} glycolaldehyde column densities in our sample of sources.  Filled symbols are detections, open symbols are upper limits for the x-axis (see Table~\ref{cds_table}).  Data points from MM1 are colored in orange, data points from MM2 are colored in blue, and data points from sources in the literature are in gray. There is a clear bifurcation in the GA data that does not exist in the AA data.}
\label{fits}
\end{figure*}

The trends shown in Figure \ref{fits} raise several questions about the nature of the chemistry occurring in the compiled list of star-forming regions. In the case of methyl formate versus acetic acid, the single trend is likely a simple statement that the column densities of both species are tied to the bulk amount of gas in the cloud: i.e., ``more is more."

Far more interesting is the clear demarcation in the column density of methyl formate versus glycolaldehyde in the star-forming regions.  We can discern no single parameter which explains the observed differentiation. While the regions with a large MF/GA ratio are exclusively high-mass star-forming regions, there are also several high-mass star-forming regions in the group of sources with a small MF/GA ratio. This would seem to indicate that while mass is certainly an important factor in the ongoing processes of these protostellar regions, there must be some additional factor(s) that are impacting the chemistry. 

That the two trends are not comprised only of sources with a specific mass means that there must be some other underlying factor through which the sources are chemically related. There was no significant trend in the excitation temperature of the sources, and there was not enough information on their ages to draw any meaningful conclusion. The inhomogeneity of the current dataset makes it an impossible task to reliably compare further properties among all of the sources. Particularly interesting is the fact that NGC~6334I-MM1 and -MM2, despite their proximity ($\sim 4000$~AU), do not share the same trend in their MF/GA ratios (being small and large, respectively). Significantly, this was consistent across all of the positions from which we extracted spectra (Figure \ref{4panel}). In a forthcoming paper (Willis et al. 2019, in prep), we examine the differentiation observed between MM1 and MM2 with a detailed physical and chemical model of NGC~6334I. This work will present a substantially updated chemical network and modelling study on all three isomers, incorporating a range of recent laboratory and theorized studies on these reactions (e.g. \citet{Shannon:2013gr,Skouteris:2018de}).

The likely implication of the bi-modal distribution is that the protostellar environment in which these molecules are forming is heavily influencing the production of glycolaldehyde, more so than the other \ce{C2H4O2} isomers. While it cannot be ruled out completely based on current data, the other possibility is that the protostellar environment is instead affecting the production of methyl formate and acetic acid in extremely similar ways. If there were evidence for a single parent molecule that was responsible for the formation of a significant fraction of the interstellar methyl formate and acetic acid, this would be a much more attractive possibility. However, to our knowledge no such precursor is currently proposed in the literature. 

Recent work by \citet{Skouteris:2018de} suggests ethanol (\ce{C2H5OH}) is a precursor molecule for both glycolaldehyde and acetic acid. A preliminary examination regarding the abundances of ethanol, acetic acid, and glycolaldehyde in these sources did not suggest a strong correlation, but the sample size was too small to draw any concrete conclusions as only six sources had measured abundances for all three molecules. 

Dimethyl ether (\ce{CH3OCH3}) has been proposed as a precursor molecule to methyl formate \citep{Balucani:2015gj}. The link between the two molecules is supported by observations of their relative abundance across a number of sources \citep{Jaber:2014ty}, as well as their spatial distribution \citep{Brouillet:2013bh}. The spatial distribution of the molecules is tightly correlated in NGC~6334I, and the relative abundances of dimethyl ether and methyl formate would also appear to be related in the sources used for our analysis. While there is no known process from which to form acetic acid directly from dimethyl ether, the radicals which go on to form dimethyl ether, methyl (\ce{CH3}) and methoxy (\ce{CH3O}), are direct photodissociation products of \ce{CH3OH}, and are thus in competition with formation routes for glycolaldehyde, methyl formate, and acetic acid.

The difficulty of accurately reproducing glycolaldehyde abundances in chemical models may be due partly to factors that are displayed in this data. If the production of glycolaldehyde is significantly altered depending on some physical factor that was previously unaccounted for, then reproducing the proper abundances becomes a much more difficult task. Although many models are now incorporating time-dependent temperature alterations, density gradients, and even cosmic-ray-induced secondary electron processes (e.g., \citealt{Shingledecker:2018fm}), constructing detailed physical 

\noindent models for each individual source is time-consuming and still involves many assumptions and simplifications.  If the root cause for the dichotomy in glycolaldehyde abundances observed here is determined to be physical in nature as opposed to being a result of the atomic inventory of the regions, it is likely influencing many other molecular pathways as well.  Observational determination of the cause, or at least a constraint on the possible causes, would significantly narrow the phase-space needed to be explored by models.

It must be noted that while we believe we have located a majority of the reliable literature reports of sources with at least two of these species, we are still working with a relatively small number of sources. Additionally, because the measurements of these column densities were compiled from a list of 11 different papers spanning over a decade, using many disparate facilities, and with different linear resolutions on the sky, the techniques used to measure the column densities are not self-consistent.  Especially in the case of single-dish observations, there is a real possibility that the results are confused by underlying, unresolved substructure in the chemical distribution. 

Nevertheless, the clear discrepancy between the glycolaldehyde and methyl formate column densities, especially when compared to the tightly-correlated acetic acid and methyl formate column densities, strongly suggests that there is a real chemical and/or physical cause underlying these results. Most previous studies of the molecular composition of star-forming regions have either observed sources with a beam unable to resolve the source or extracted spectra from a single location within a resolved source. The best path forward is clearly a dedicated, self-consistent study of these three isomers in a much larger sample set of sources, using ALMA to account for underlying source substructure, and to ensure the observations are conducted on consistent linear spatial scales.

\section{Conclusions}

We have conducted a literature search and presented new observations comparing the relative column densities of the three \ce{C2H4O2} isomers methyl formate, glycolaldehyde, and acetic acid in a total of fifteen various protostellar environments.  The column densities of methyl formate and acetic acid were well-correlated, and are likely simply tracking the relative total gas mass in each source.  Methyl formate and glycolaldehyde, however, display a stark dichotomy in their relative column densities. One group of sources, for which the MF/GA ratio was small, was comprised of star-forming regions with a variety of masses. The other group, for which the MF/GA ratio was much larger, was comprised entirely of high-mass star-forming regions. That the trends could not be entirely ascribed to the mass of the star-forming region suggests the existence of another parameter by which these regions can be linked. This is an excellent indicator of the stellar environment impacting the column densities of at least one of the aforementioned molecules. A dedicated, self-consistent follow-up observational study of the \ce{C2H4O2} isomeric family, combined with chemical simulations, has the potential to constrain formation pathways for a variety of interstellar molecules.

\acknowledgements

This paper makes use of the following ALMA data: \#2015.A.00022.T. ALMA is a partnership of ESO (representing its member states), NSF (USA) and NINS (Japan), together with NRC (Canada) and NSC and ASIAA (Taiwan) and KASI (Republic of Korea), in cooperation with the Republic of Chile. The Joint ALMA Observatory is operated by ESO, AUI/NRAO and NAOJ.   The National Radio Astronomy Observatory is a facility of the National Science Foundation operated under cooperative agreement by Associated Universities, Inc.  Support for B.A.M. was provided by NASA through Hubble Fellowship grant \#HST-HF2-51396 awarded by the Space Telescope Science Institute, which is operated by the Association of Universities for Research in Astronomy, Inc., for NASA, under contract NAS5-26555. This research made use of NASA’s Astrophysics Data System  Bibliographic  Services,  Astropy,  a community-developed core Python package for Astronomy \citep{astropy}, and APLpy, an open-source plotting package for Python hosted at http://aplpy.github.com.

\clearpage

\begin{table*}
\centering
\footnotesize
\caption{\ce{C2H4O2} Isomer Column Densities in Each Source}
\label{cds_table}
\begin{tabular*}{\textwidth}{cccccccccc}
\hline\hline
Source & MF & GA & AA & MF/GA & MF/GA & Region Mass & Obs. & Beam & Ref. \\
    &   \multicolumn{3}{c}{Column Density [cm$^{-2}$]} &    &   Class.   &   &  Type & Corrected  &   \\
\hline

NGC 6334I MM1-i & $1.7 (3) \times 10^{18}$ & $1.2 (3) \times 10^{17}$ & $ 8 (3) \times 10^{16}$ & 15 (5) & Small & High & Int. & M & ... \\
NGC 6334I MM1-ii & $3 (1) \times 10^{18}$ & $2.2 (8) \times 10^{17}$ & $ 9 (3) \times 10^{16}$ & 16 (7) & Small & High & Int. & M & ... \\
NGC 6334I MM1-iii & $3 (2) \times 10^{18}$ & $2.1 (8) \times 10^{17}$ & $ 3 (1) \times 10^{17}$ & 17 (13) & Small & High & Int. & M & ... \\
NGC 6334I MM1-iv & $7 (2) \times 10^{18}$ & $1.9 (4) \times 10^{17}$ & $ 8 (3) \times 10^{16}$ & 38 (13) & Small & High & Int. & M & ... \\
NGC 6334I MM1-v & $4.3 (5) \times 10^{18}$ & $9.5 (8) \times 10^{16}$ & $ 7 (1) \times 10^{17}$ & 45 (6) & Small & High & Int. & M & ... \\
NGC 6334I MM1-vi & $3 (2) \times 10^{18}$ & $2.1 (2) \times 10^{17}$ & $ 2.2 (7) \times 10^{17}$ & 13 (8) & Small & High & Int. & M & ... \\
NGC 6334I MM1-vii & $4.7 (9) \times 10^{18}$ & $1.9 (8) \times 10^{17}$ & $ 2.6 (6) \times 10^{17}$ & 25 (11) & Small & High & Int. & M & ... \\
NGC 6334I MM1-viii & $5 (2) \times 10^{18}$ & $1.6 (3) \times 10^{17}$ & $ 1.4 (4) \times 10^{17}$ & 30 (12) & Small & High & Int. & M & ... \\
NGC 6334I MM1-ix & $2.7 (8) \times 10^{18}$ & $8 (2) \times 10^{16}$ & $ 5 (2) \times 10^{16}$ & 35 (15) & Small & High & Int. & M & ... \\
NGC 6334I MM2-i & $3.6 (8) \times 10^{18}$ & $< 2.3 \times 10^{16}$ & $1.2 (5) \times 10^{17}$ & $\geq$158.3 & Large & High & Int. & M & ... \\
NGC 6334I MM2-ii & $2.5 (6) \times 10^{18}$ & $< 2.3 \times 10^{16}$ & $1.6 (5) \times 10^{17}$ & $\geq$110.2 & Large & High & Int. & M & ... \\
NGC 6334I MM2-iii & $4 (1) \times 10^{18}$ & $< 1.1 \times 10^{16}$ & $1.5 (2) \times 10^{17}$ & $\geq$373.6 & Large & High & Int. & M & ... \\
Orion-KL & $1.6 (1) \times 10^{17}$ & $< 2.4 \times 10^{14}$ & $< 1.8 \times 10^{15}$ & $\geq$667 & Large & High & Int. & M & 1 \\
Sgr-B2 (N) & $4 (1) \times 10^{17}$ & $1.8 (5) \times 10^{15}$ & $1.1 (2) \times 10^{16}$ & 243 (85) & Large & High & S.D. & No & 2 \\
W51/e2 & $1.1 (4) \times 10^{18}$ & $< 2.0 \times 10^{15}$ & $3 (1) \times 10^{16}$ & $\geq$550.0 & Large & High & S.D. & Yes & 3, 4 \\
G34.3+0.2 & $6 (2) \times 10^{16}$ & $< 3.0 \times 10^{14}$ & $2.2 (7) \times 10^{15}$ & $\geq$193.3 & Large & High & S.D. & Yes & 3, 5 \\
G31.41+0.31 & $5 (3) \times 10^{18}$ & $3 (2) \times 10^{17}$ & ... & 16 (10) & Small & High & S.D. & Yes & 6 \\
G29.96-0.02 & $2.0 (6) \times 10^{16}$ & $5 (2) \times 10^{15}$ & ... & 4 (1) & Small & High & Int. & Yes & 7 \\
G24.78+0.08A1 & $3.0 (9) \times 10^{16}$ & $5 (2) \times 10^{15}$ & ... & 6 (3) & Small & High & Int. & Yes & 7 \\
G24.78+0.08A2 & $4 (1) \times 10^{16}$ & $6 (2) \times 10^{15}$ & ... & 7 (3) & Small & High & Int. & Yes & 7 \\
NGC 7129 FIRS 2 & $3 (1) \times 10^{18}$ & $2.3 (7) \times 10^{17}$ & ... & 15 (6) & Small & Intermediate & Int. & $^{\ddag}$ & 8 \\
IRAS 16293 & $4 (1) \times 10^{17}$ & $7 (2) \times 10^{16}$ & $6 (2) \times 10^{15}$ & 6 (2) & Small & Low & Int. & Yes & 9 \\
L1157-B1 & $5.4 (8) \times 10^{13}$ & $3.4 (7) \times 10^{13}$ & $< 2.1 \times 10^{13}$ & 1.6 (0.4) & Small & Low & S.D. & No & 10 \\
NGC 1333 IRAS 4A & $5 (3) \times 10^{16}$ & $9 (3) \times 10^{15}$ & $< 8.0 \times 10^{14}$ & 6 (4) & Small & Low & Int. & M & 11, 5 \\
NGC 1333 IRAS 2A & $6 (2) \times 10^{16}$ & $8 (2) \times 10^{15}$ & ... & 8 (3) & Small & Low & Int. & M & 11 \\
\hline\hline
\multicolumn{10}{l}{Please refer to \S\ref{results} for an explanation of the MF/GA classification.}\\
\multicolumn{10}{l}{The region masses given are taken directly from the respective reference.}\\
\multicolumn{10}{l}{Int. and S.D. - Interferometer and Single-Dish, respectively.}\\
\multicolumn{10}{l}{M - These beams appear well-matched to the source size as observed, and thus no correction was made by the authors.}\\
\multicolumn{10}{l}{$^{\ddag}$A beam correction was applied to glycoladehyde for this work. See Appendix \ref{beam}.}\\
\multicolumn{10}{l}{$^1$\citet{Favre:2011fg}, $^2$\citet{Belloche:2013eba}, $^3$\citet{Lykke:2015ig}, $^4$\citet{Remijan:2002tv}, $^5$\citet{Remijan:2003wf}, $^6$\citet{Rivilla:2017vg},}\\
\multicolumn{10}{l}{$^7$\citet{Calcutt:2014ug}, $^8$\citet{Fuente:2014je}, $^9$\citet{Jorgensen:2016cq}, $^{10}$\citet{Lefloch:2017jd}, $^{11}$\citet{Taquet:2015en}}\\

\end{tabular*}
\end{table*}


\appendix

\renewcommand\thefigure{\thesection\arabic{figure}}   
\renewcommand\thetable{\thesection\arabic{table}}    

\setcounter{figure}{0}    
\setcounter{table}{0} 

\section{Beam Corrections}
\label{beam}

\citet{Fuente:2014je} provide column densities for both methyl formate and glycolaldehyde in NGC 7129 FIRS 2.  For methyl formate, sufficient transitions were observed to allow the source size to vary, and they fit a source size of 0.12$^{\prime\prime}$ (Table~3 of \citealt{Fuente:2014je}).  For glycolaldehyde, sufficient transitions were seen to derive a column density, but not to fit a source size, and \citet{Fuente:2014je} provided instead a beam-averaged column density.  Here, we have made the assumption that the glycolaldehyde and methyl formate will have a similar spatial distribution, and have corrected the reported glycolaldehyde abundance assuming a source size of 0.12$^{\prime\prime}$, matching that of methyl formate.

\section{Full NGC 6334I MM1 Analysis Results}
\label{spectra}

As described in \S\ref{analysis}, we conducted an exhaustive search of the databases to identify all lines belonging to known interstellar species in the spectra from NGC 6334I MM1 and MM2 in an attempt to simultaneously identify those lines of methyl formate, glycolaldehyde, and acetic acid most appropriate (unblended and optically thin) for conducting a column density analysis.  The full results of the model are beyond the scope of this paper, and for all the species except these targets, the analysis was done by-eye: the primary goal was line identification, not quantification.  Nevertheless, the resulting agreement of the model with the observations is excellent, even with the assumption of a uniform $T_{ex}$, $\Delta V$, and $v_{lsr}$ for all species.  

\begin{figure}
    \centering
    \includegraphics[width=0.9\textwidth]{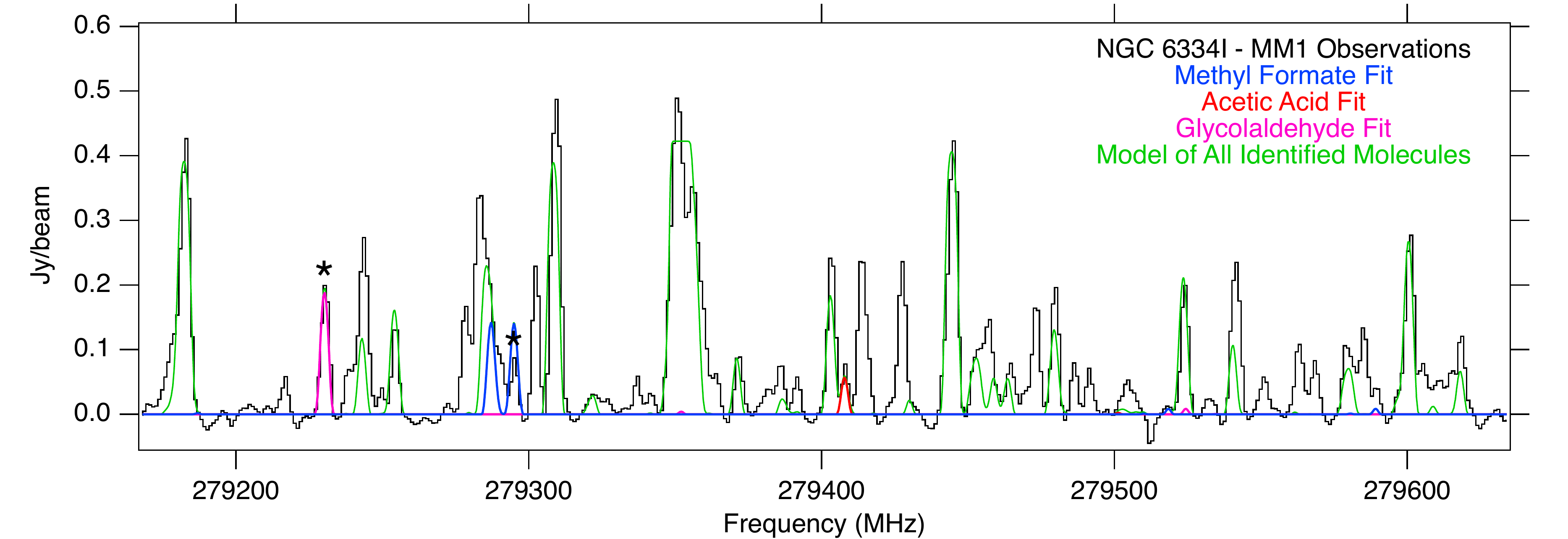}
    \includegraphics[width=0.9\textwidth]{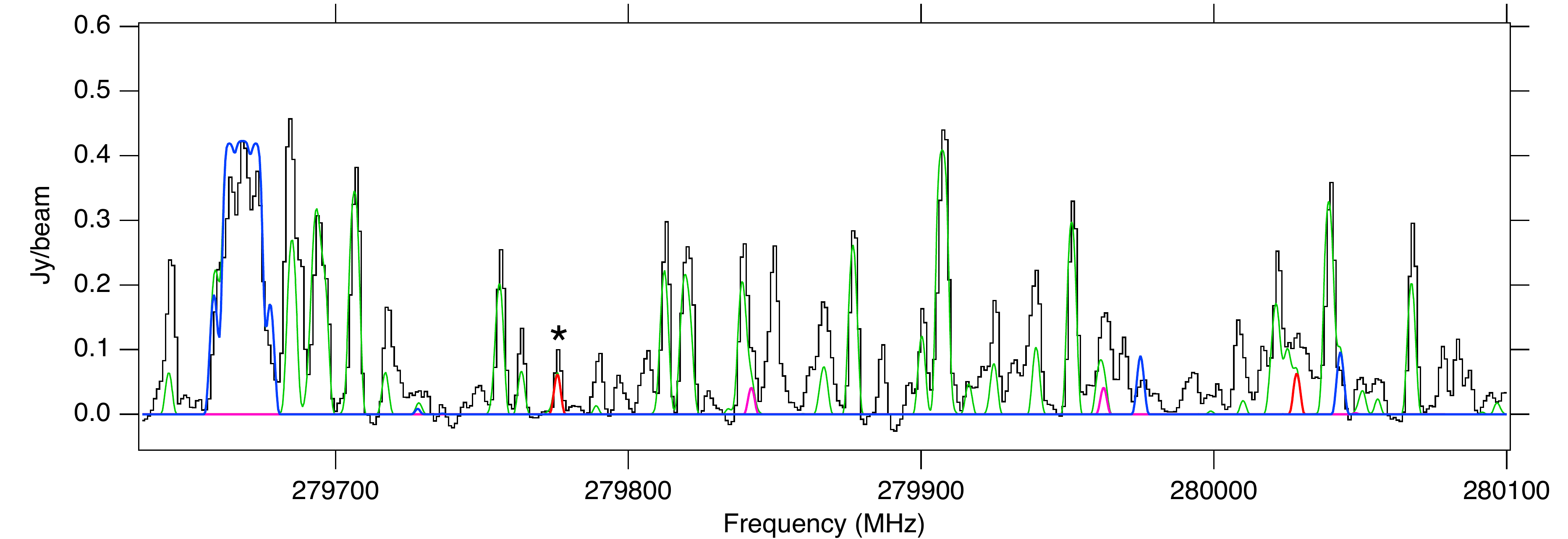}
    \includegraphics[width=0.9\textwidth]{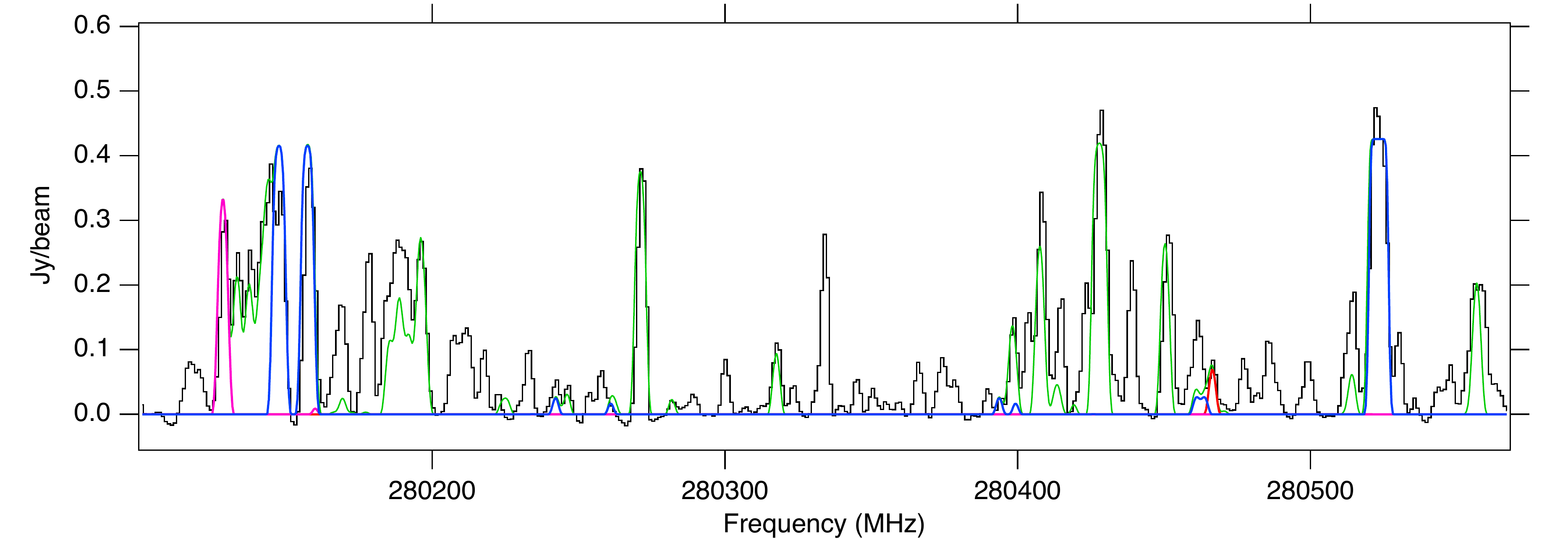}
    \includegraphics[width=0.9\textwidth]{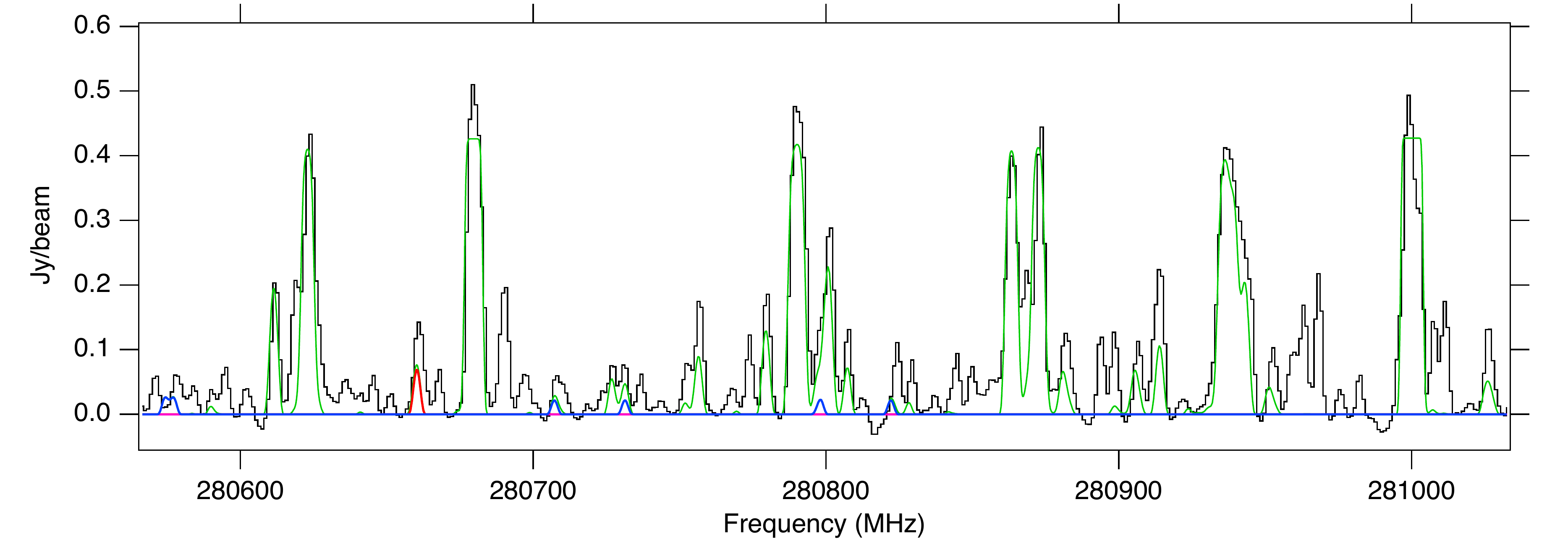}
    \caption{Spectra extracted toward NGC 6334I MM1-i (black).  Overlaid in green is the full model of all assigned molecules in the spectrum (see text), and methyl formate, glycolaldehyde, and acetic acid are shown in color.  Transitions marked with an asterisk were identified as the least blended and optically thin, and were used for the column density analysis (see Table~\ref{freqs}).  Spectra were offset to a $v_{lsr}$~=~-7~km~s$^{-1}$.   }
    \label{mm1_1}
\end{figure}

\begin{figure}
    \centering
    \includegraphics[width=0.9\textwidth]{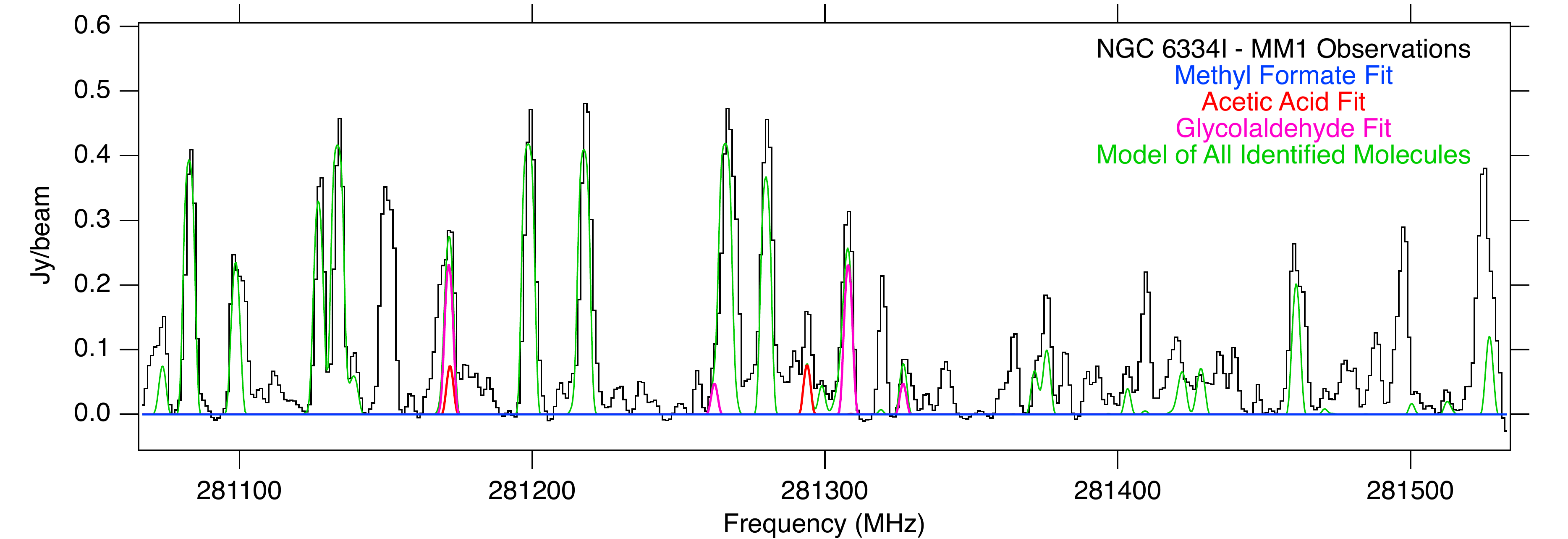}
    \includegraphics[width=0.9\textwidth]{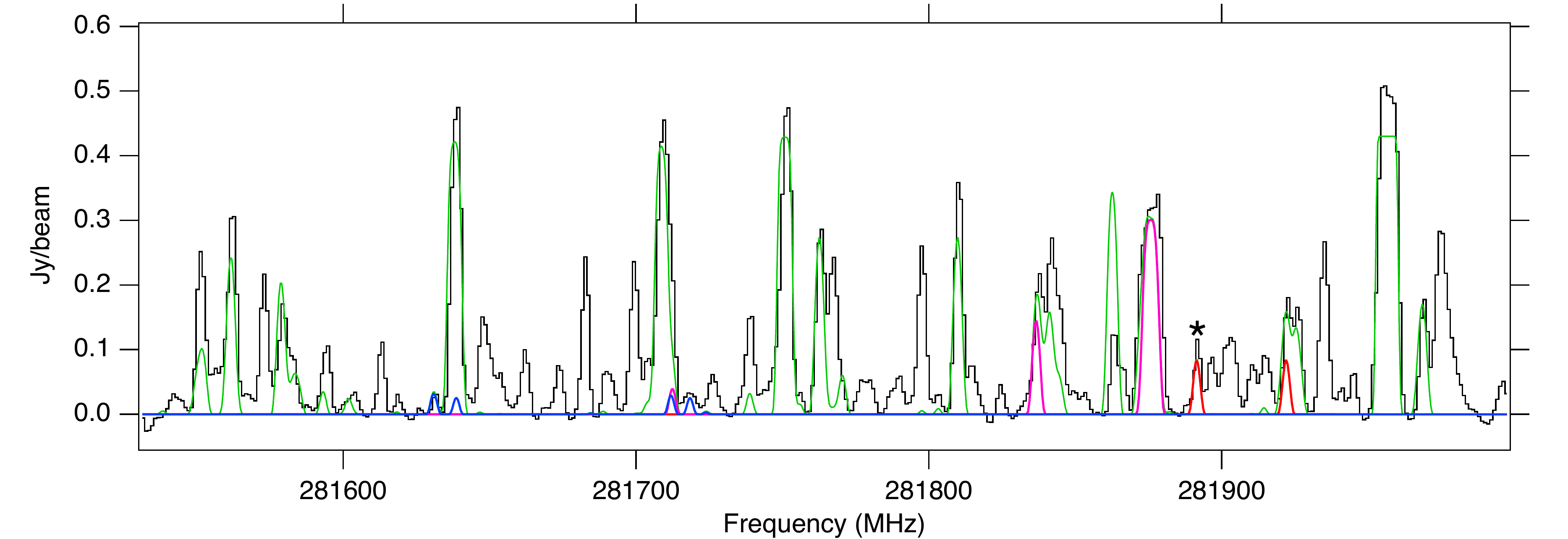}
    \includegraphics[width=0.9\textwidth]{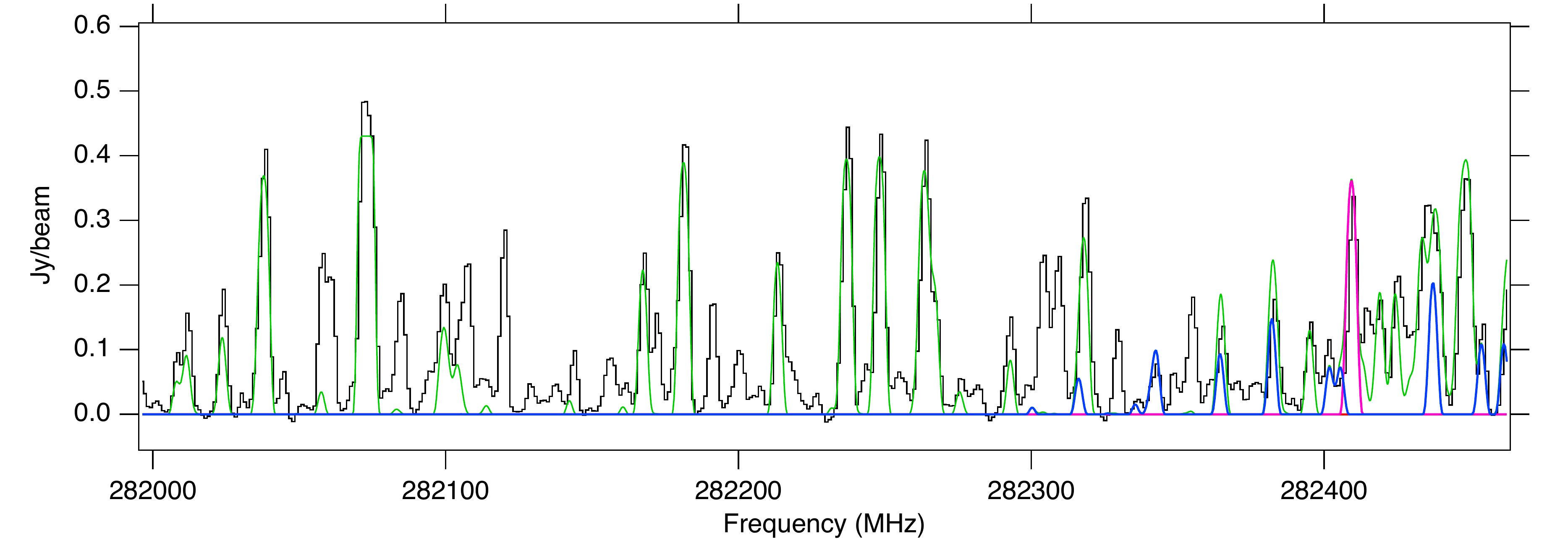}
    \includegraphics[width=0.9\textwidth]{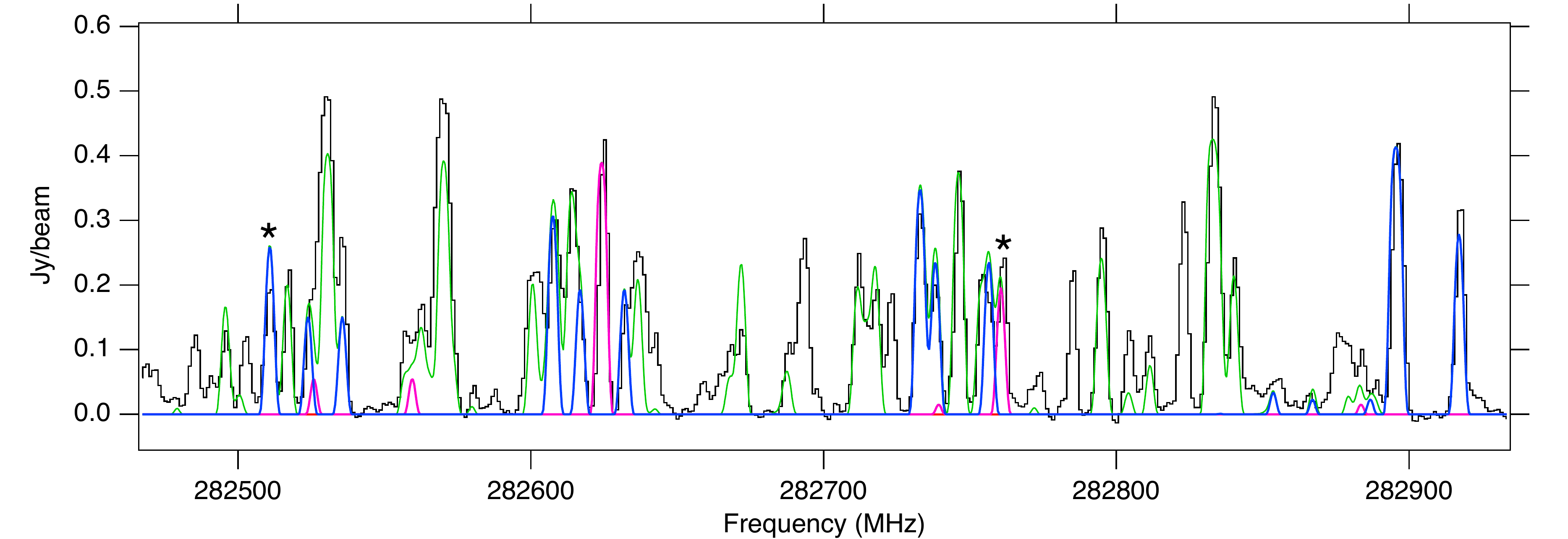}
    \caption{Spectra extracted toward NGC 6334I MM1-i (black).  Overlaid in green is the full model of all assigned molecules in the spectrum (see text), and methyl formate, glycolaldehyde, and acetic acid are shown in color.  Transitions marked with an asterisk were identified as the least blended and optically thin, and were used for the column density analysis (see Table~\ref{freqs}).  Spectra were offset to a $v_{lsr}$~=~-7~km~s$^{-1}$.   }
    \label{mm1_2}
\end{figure}

\begin{figure}
    \centering
    \includegraphics[width=0.9\textwidth]{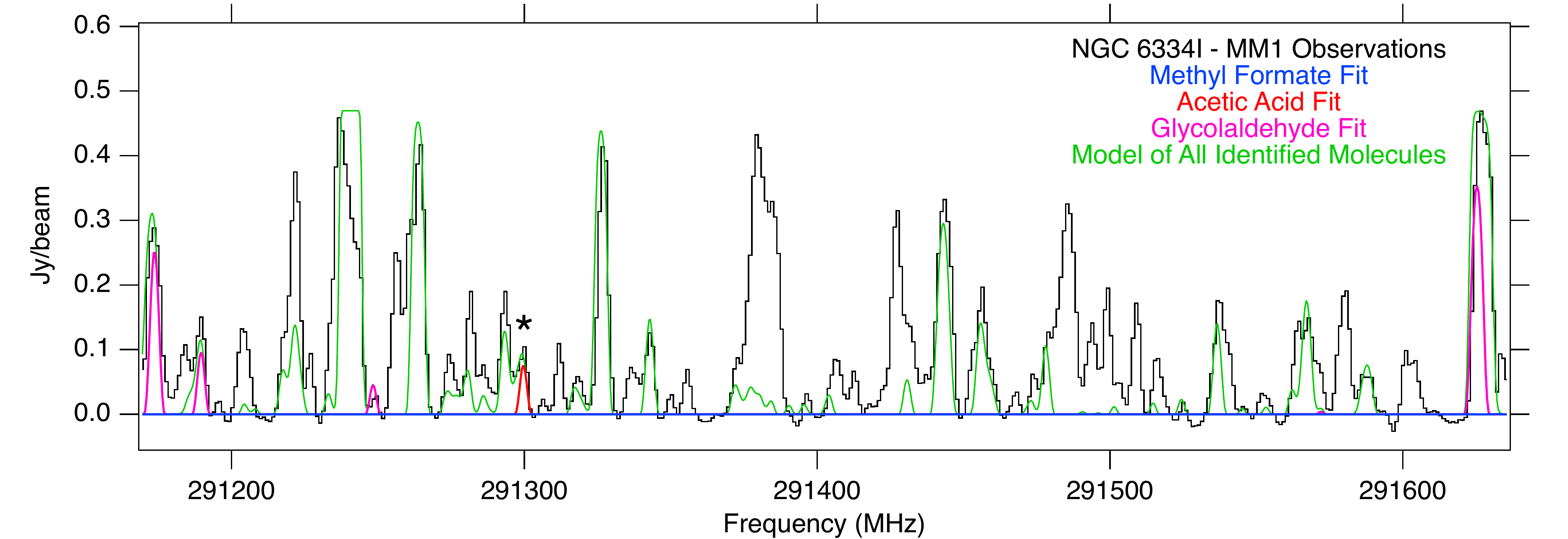}
    \includegraphics[width=0.9\textwidth]{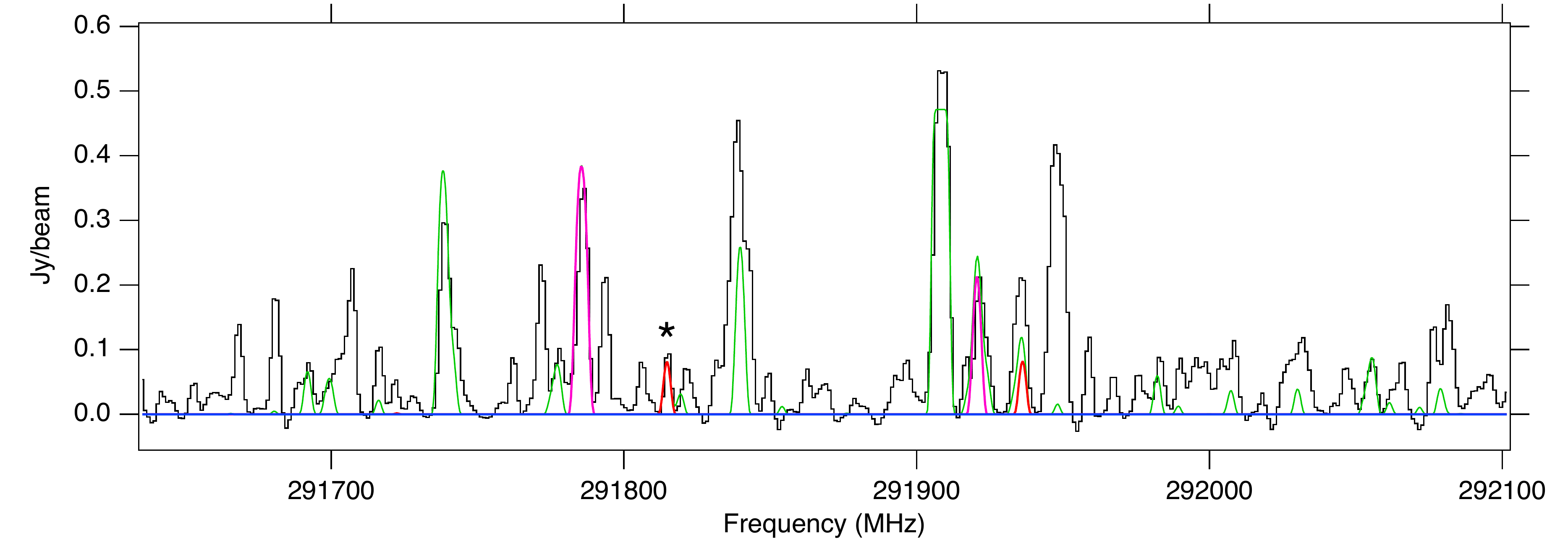}
    \includegraphics[width=0.9\textwidth]{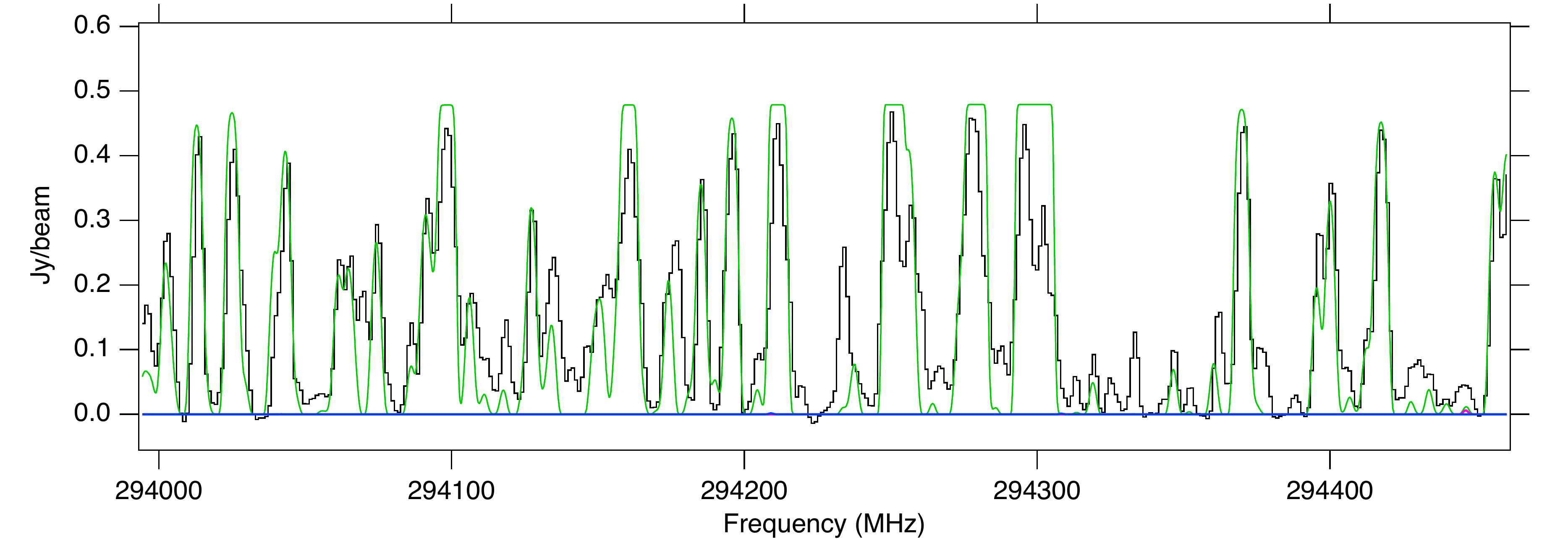}
    \includegraphics[width=0.9\textwidth]{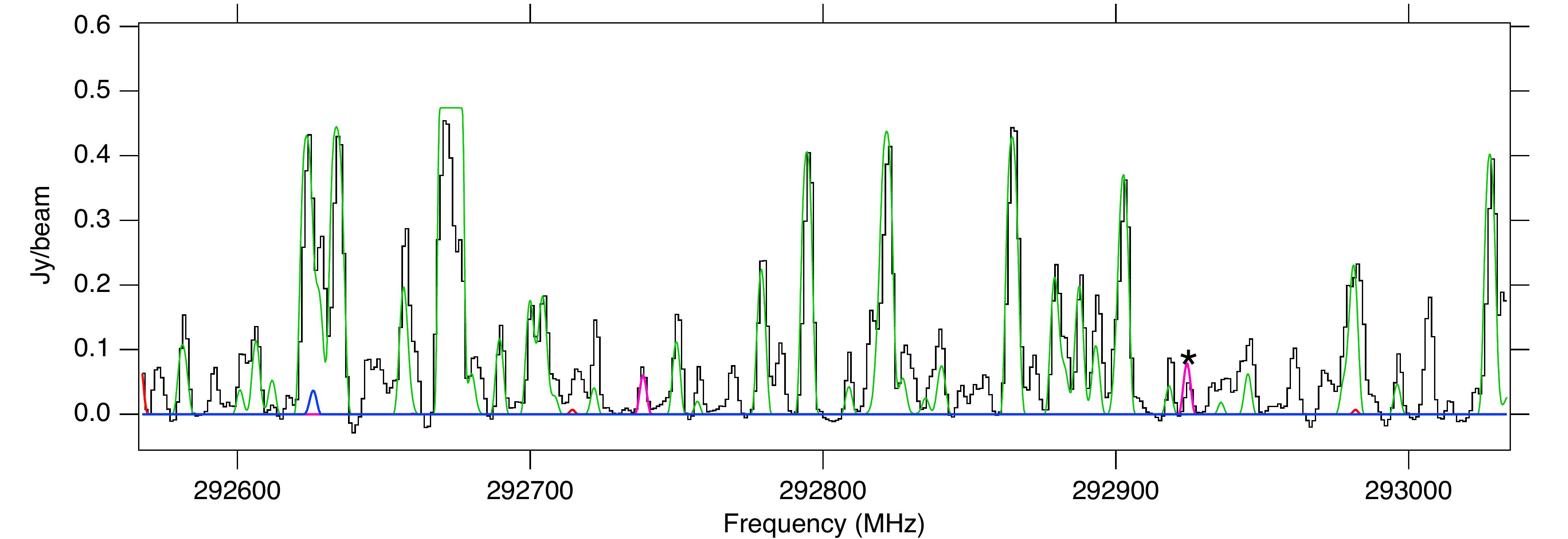}
    \caption{Spectra extracted toward NGC 6334I MM1-i (black).  Overlaid in green is the full model of all assigned molecules in the spectrum (see text), and methyl formate, glycolaldehyde, and acetic acid are shown in color.  Transitions marked with an asterisk were identified as the least blended and optically thin, and were used for the column density analysis (see Table~\ref{freqs}).  Spectra were offset to a $v_{lsr}$~=~-7~km~s$^{-1}$.   }
    \label{mm1_3}
\end{figure}

\begin{figure}
    \centering
    \includegraphics[width=0.9\textwidth]{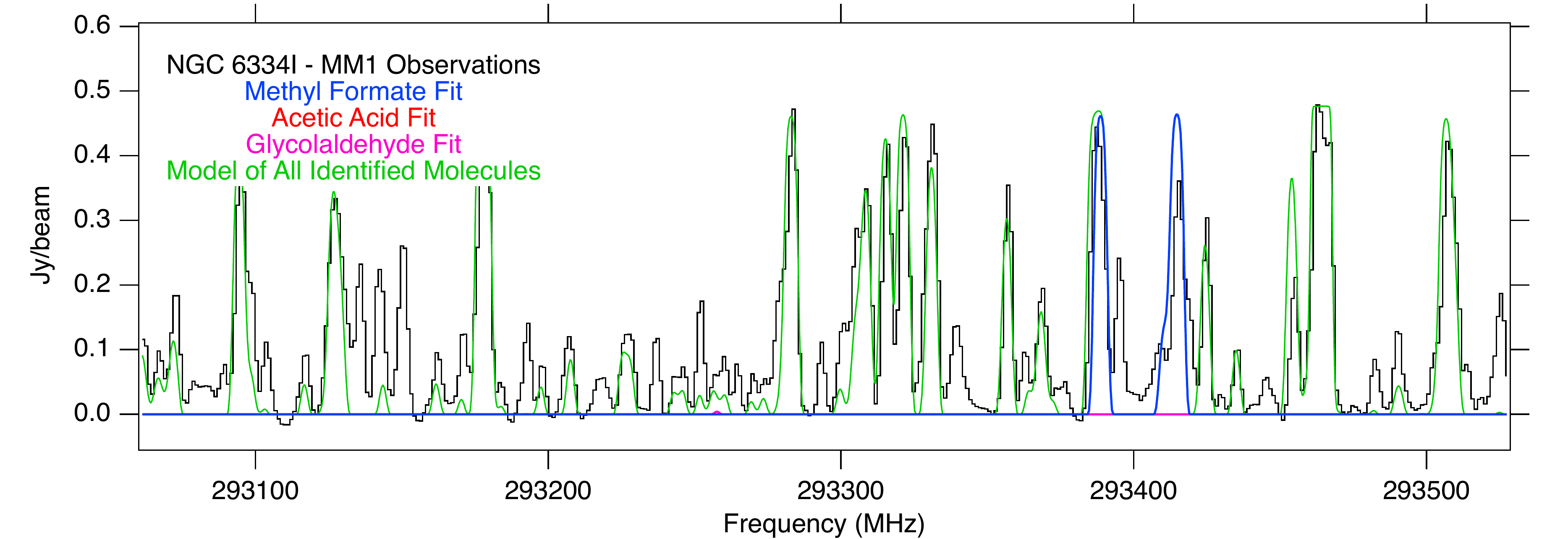}
    \includegraphics[width=0.9\textwidth]{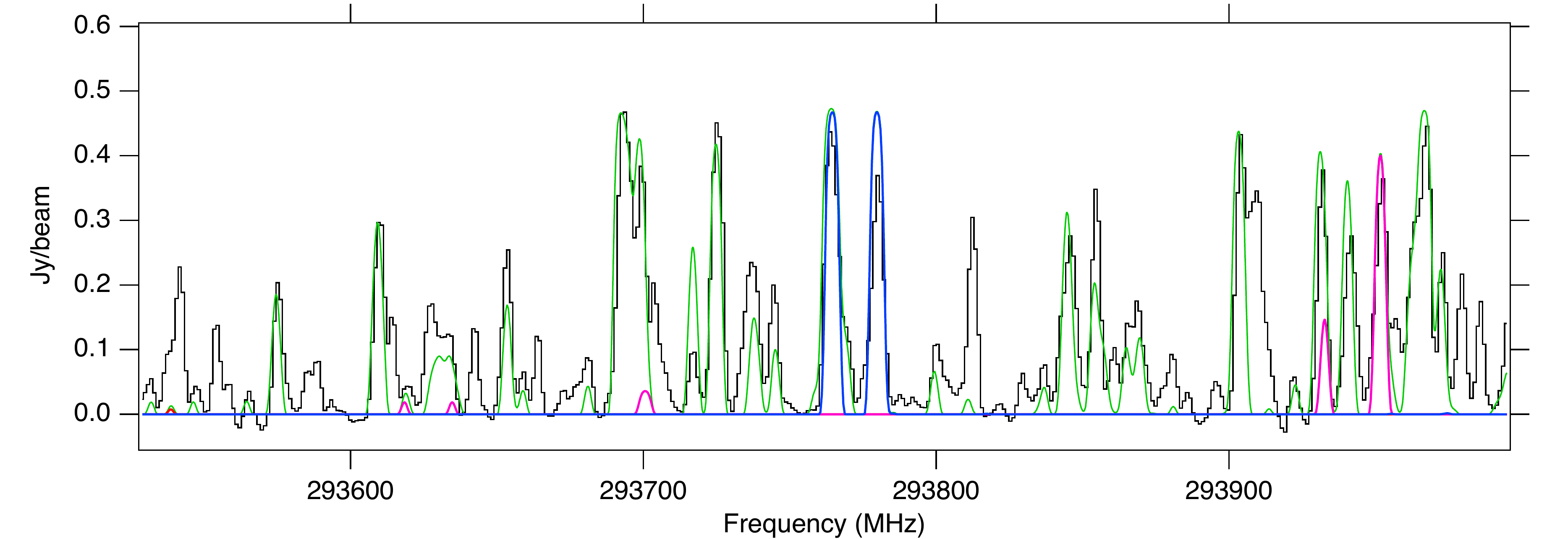}
    \includegraphics[width=0.9\textwidth]{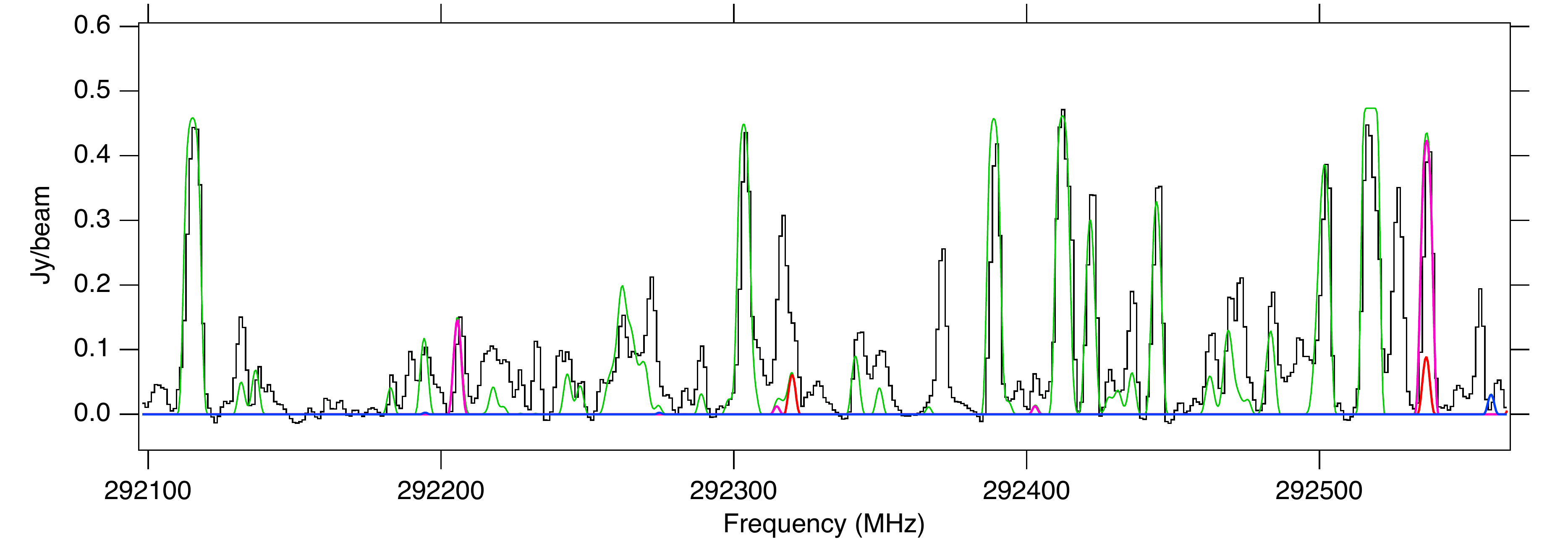}
    \includegraphics[width=0.9\textwidth]{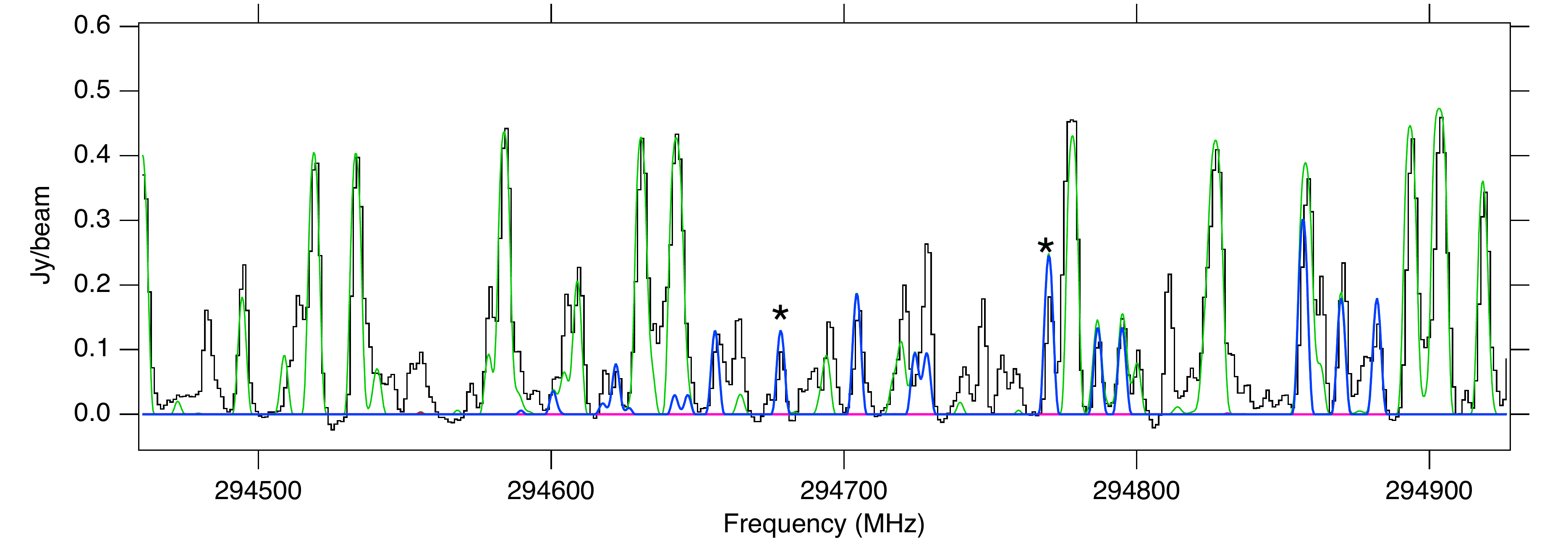}
    \caption{Spectra extracted toward NGC 6334I MM1-i (black).  Overlaid in green is the full model of all assigned molecules in the spectrum (see text), and methyl formate, glycolaldehyde, and acetic acid are shown in color.  Transitions marked with an asterisk were identified as the least blended and optically thin, and were used for the column density analysis (see Table~\ref{freqs}).  Spectra were offset to a $v_{lsr}$~=~-7~km~s$^{-1}$.   }
    \label{mm1_4}
\end{figure}

\begin{figure}
    \centering
    \includegraphics[width=0.9\textwidth]{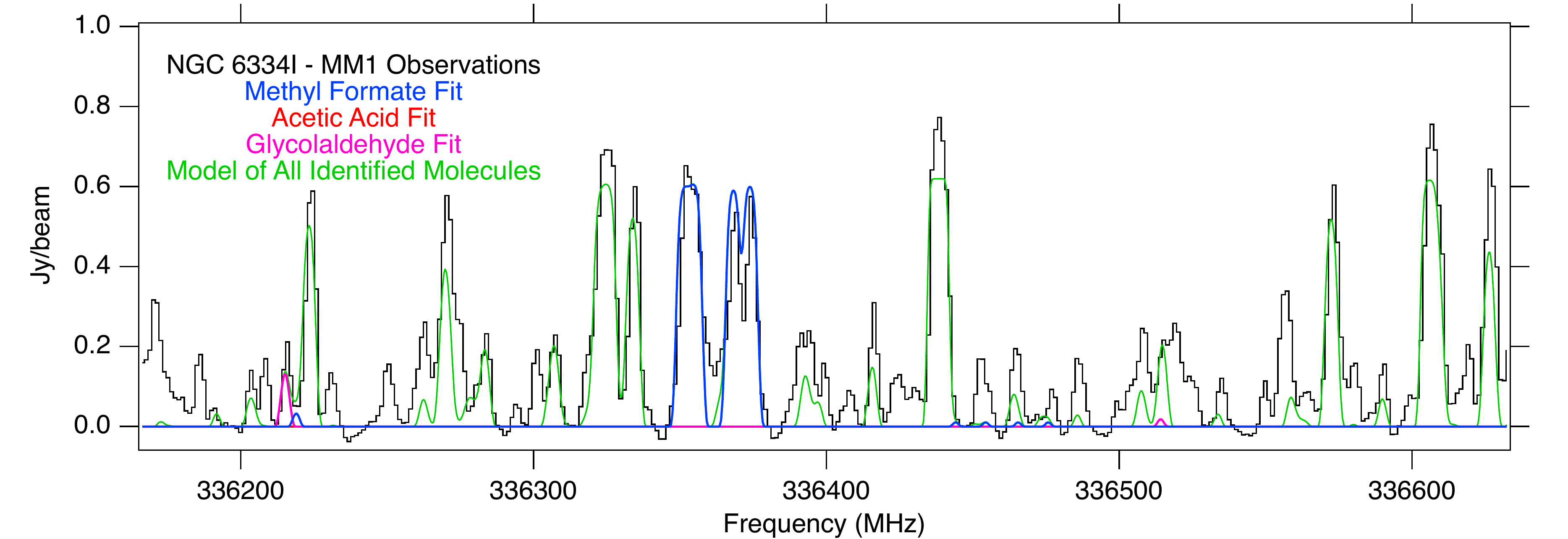}
    \includegraphics[width=0.9\textwidth]{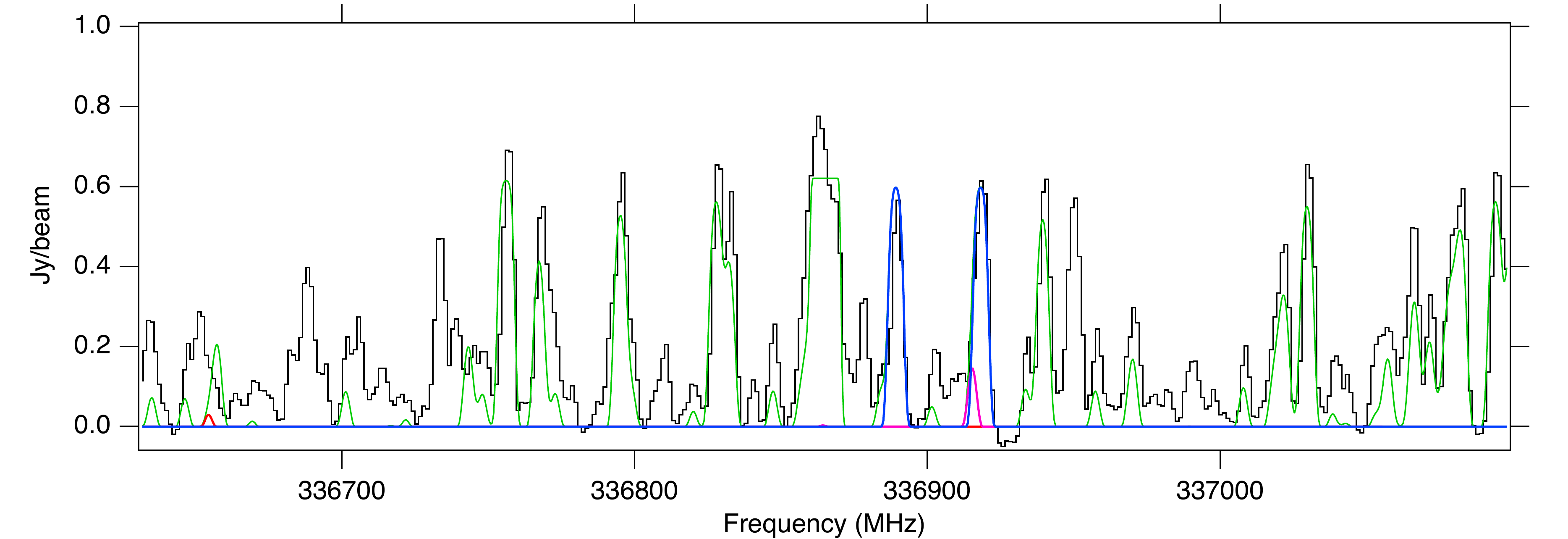}
    \includegraphics[width=0.9\textwidth]{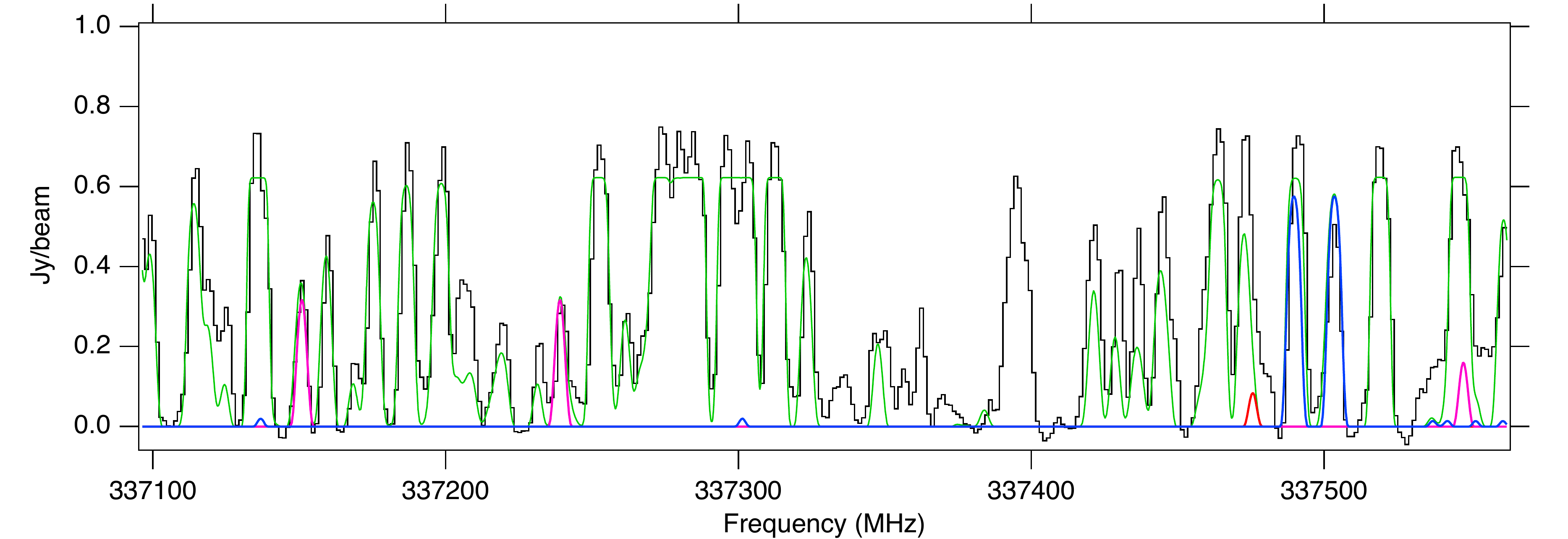}
    \includegraphics[width=0.9\textwidth]{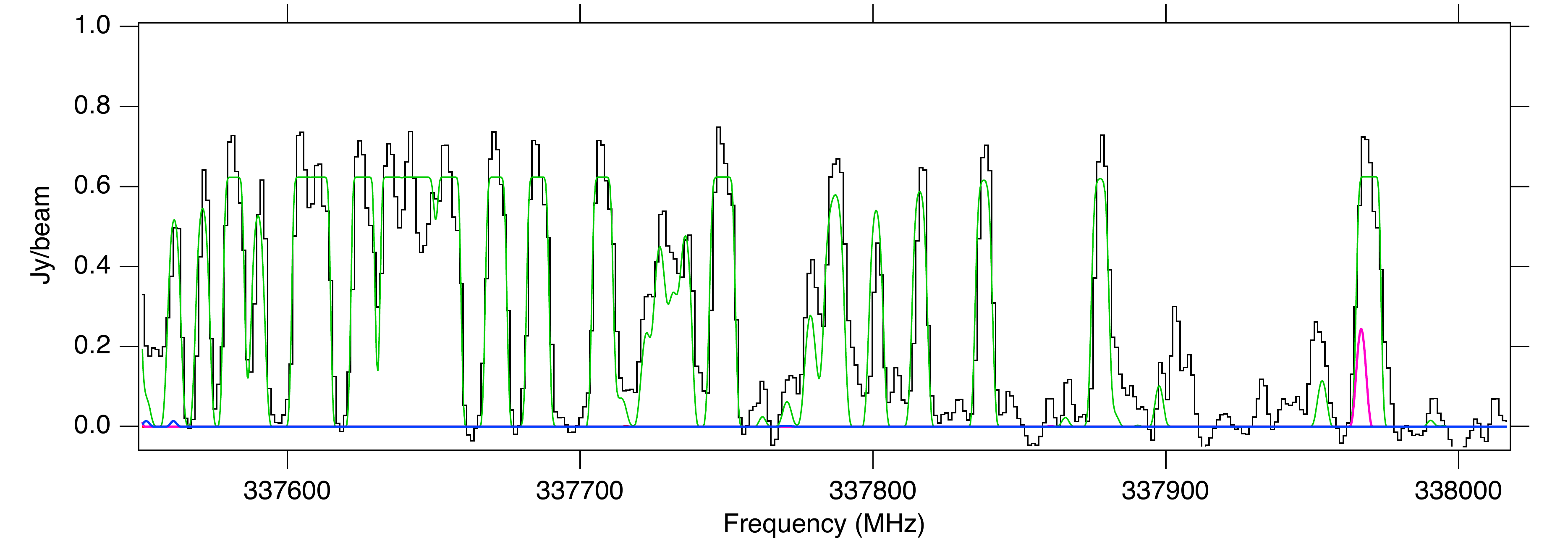}
    \caption{Spectra extracted toward NGC 6334I MM1-i (black).  Overlaid in green is the full model of all assigned molecules in the spectrum (see text), and methyl formate, glycolaldehyde, and acetic acid are shown in color.  Transitions marked with an asterisk were identified as the least blended and optically thin, and were used for the column density analysis (see Table~\ref{freqs}).  Spectra were offset to a $v_{lsr}$~=~-7~km~s$^{-1}$.   }
    \label{mm1_5}
\end{figure}

\begin{figure}
    \centering
    \includegraphics[width=0.9\textwidth]{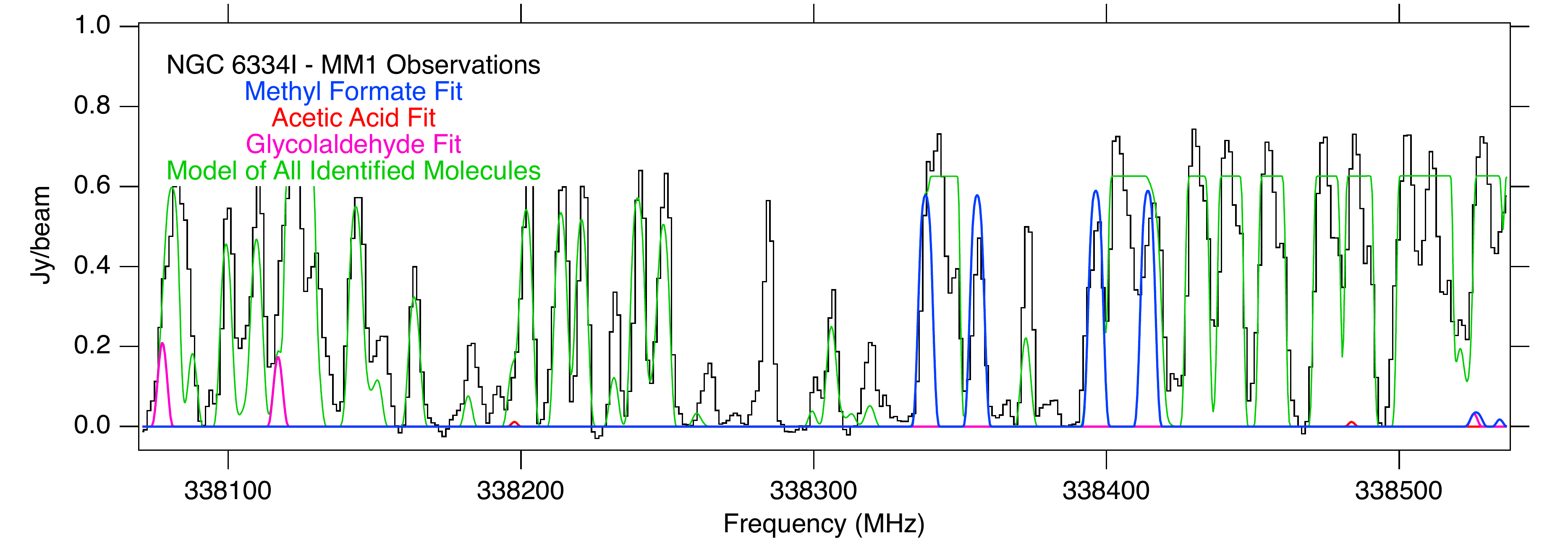}
    \includegraphics[width=0.9\textwidth]{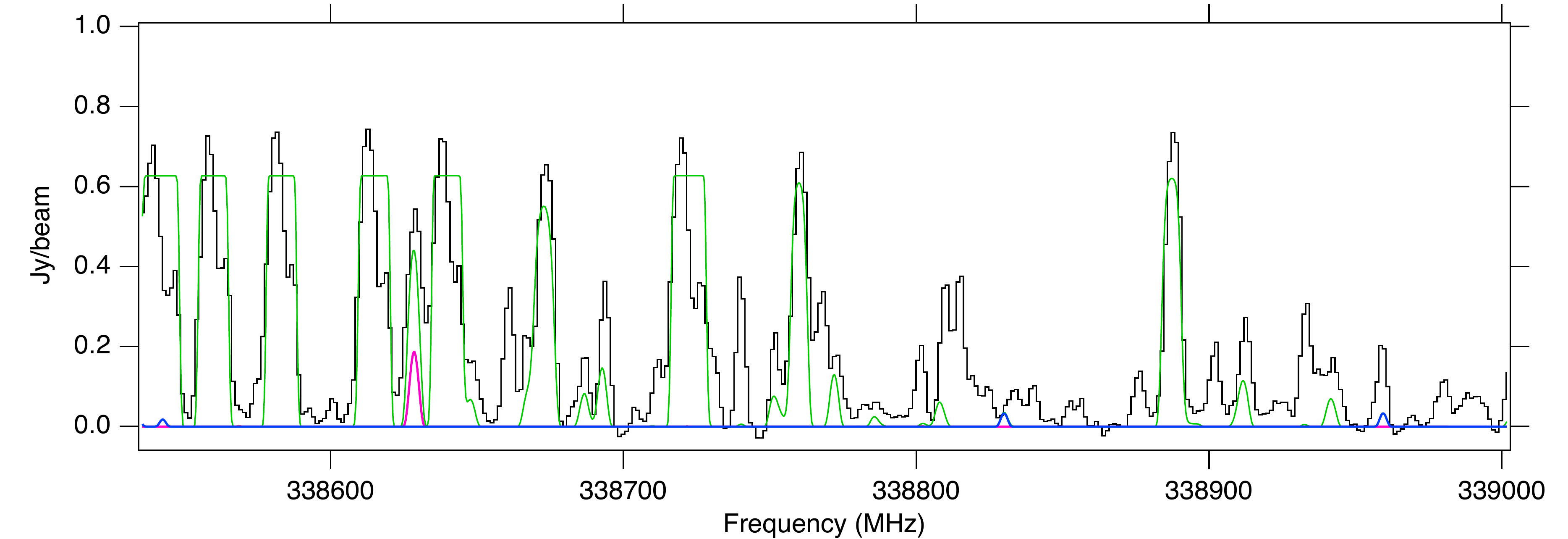}
    \includegraphics[width=0.9\textwidth]{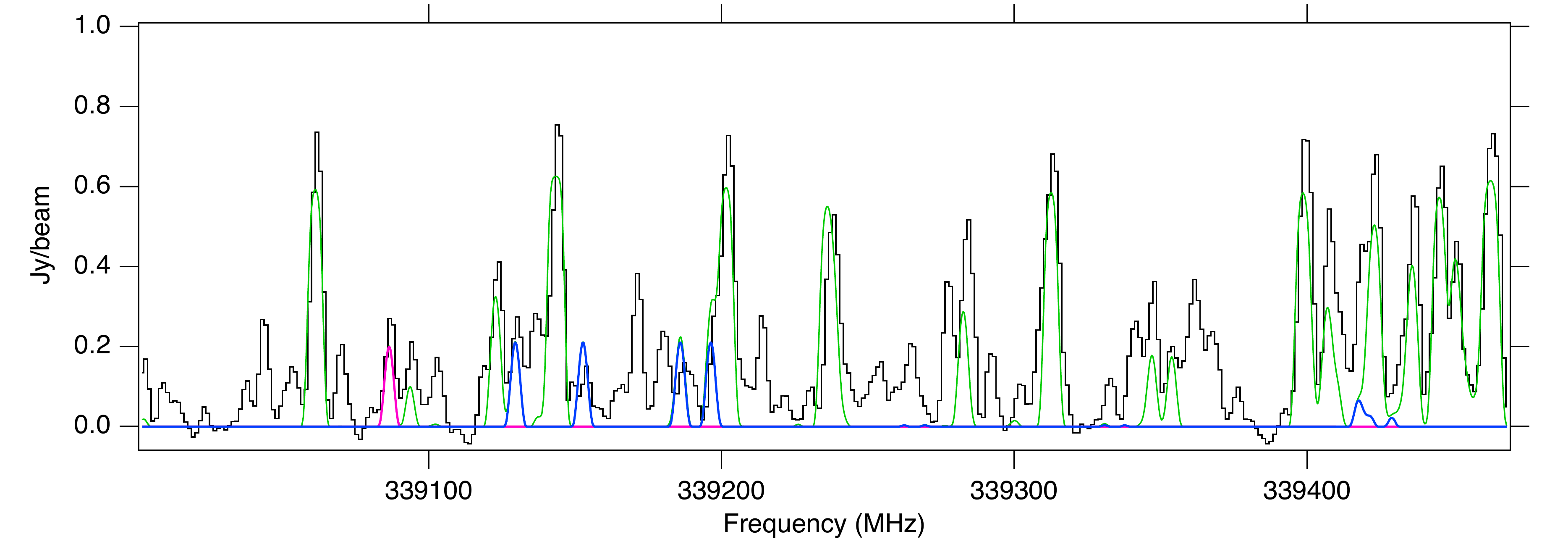}
    \includegraphics[width=0.9\textwidth]{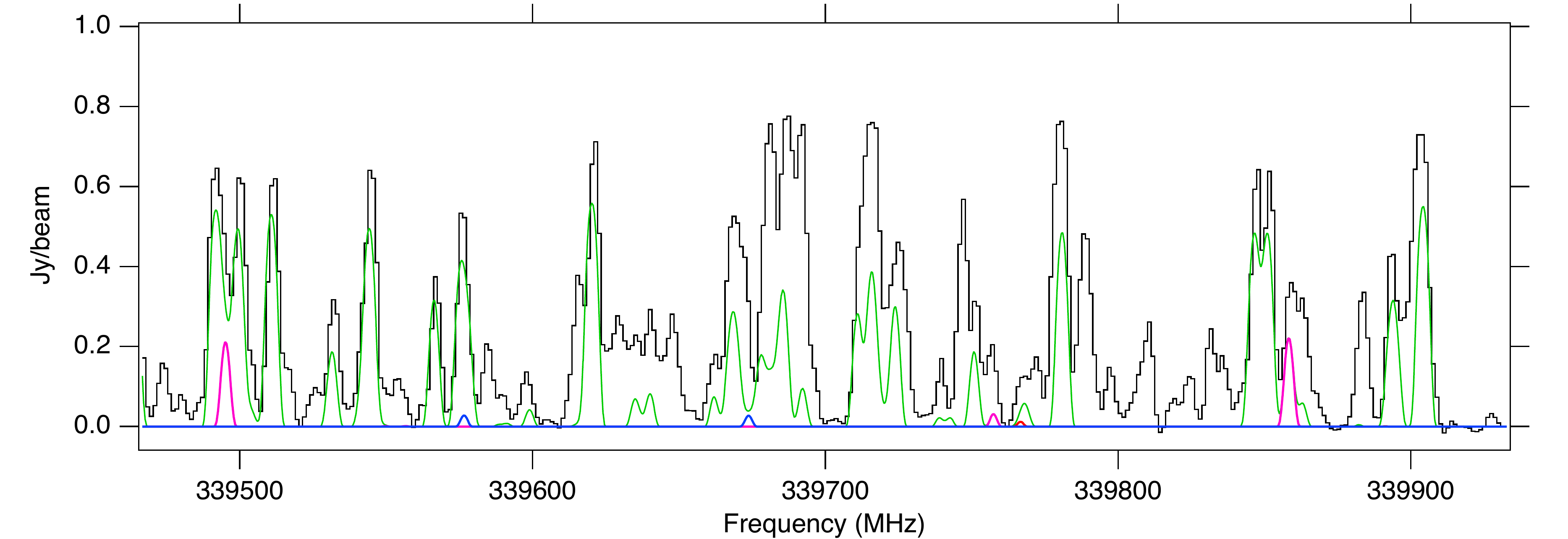}
    \caption{Spectra extracted toward NGC 6334I MM1-i (black).  Overlaid in green is the full model of all assigned molecules in the spectrum (see text), and methyl formate, glycolaldehyde, and acetic acid are shown in color.  Transitions marked with an asterisk were identified as the least blended and optically thin, and were used for the column density analysis (see Table~\ref{freqs}).  Spectra were offset to a $v_{lsr}$~=~-7~km~s$^{-1}$.   }
    \label{mm1_6}
\end{figure}

\begin{figure}
    \centering
    \includegraphics[width=0.9\textwidth]{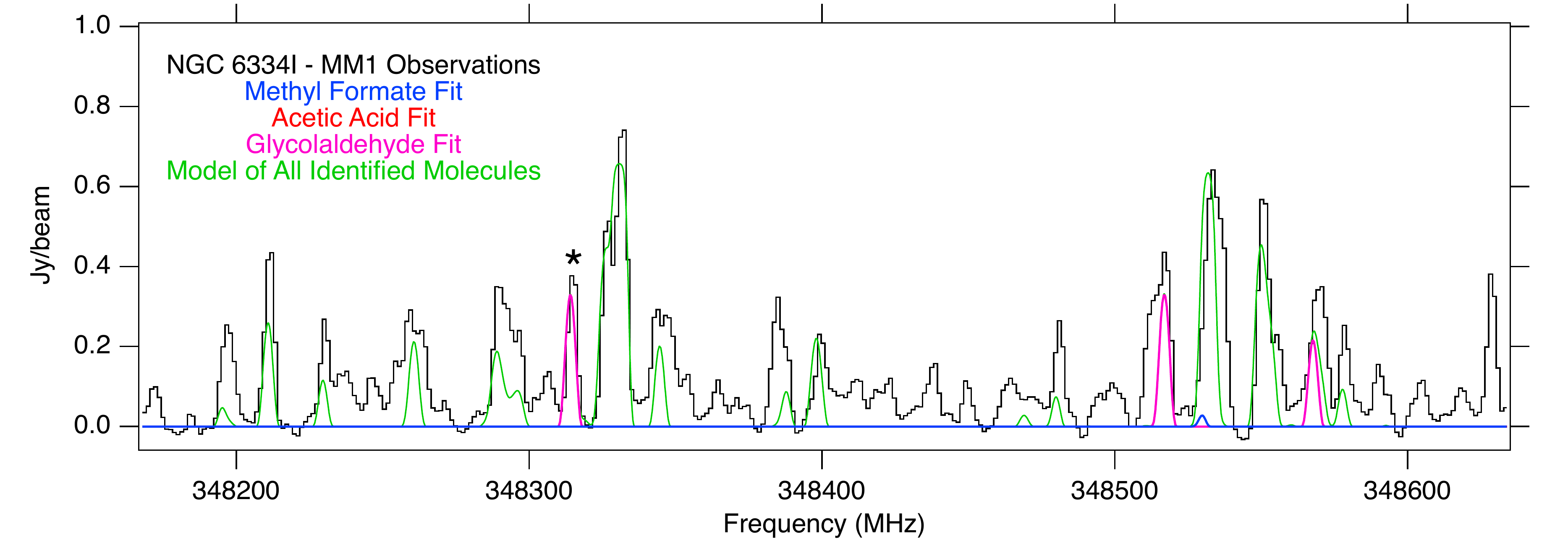}
    \includegraphics[width=0.9\textwidth]{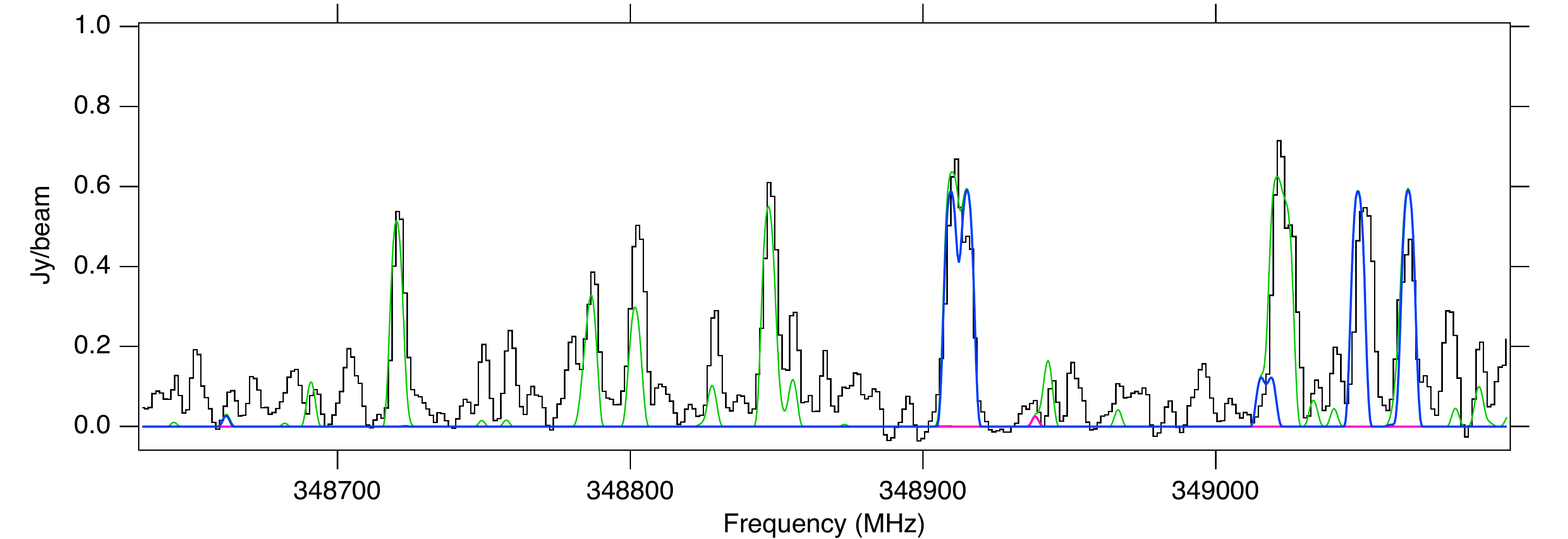}
    \includegraphics[width=0.9\textwidth]{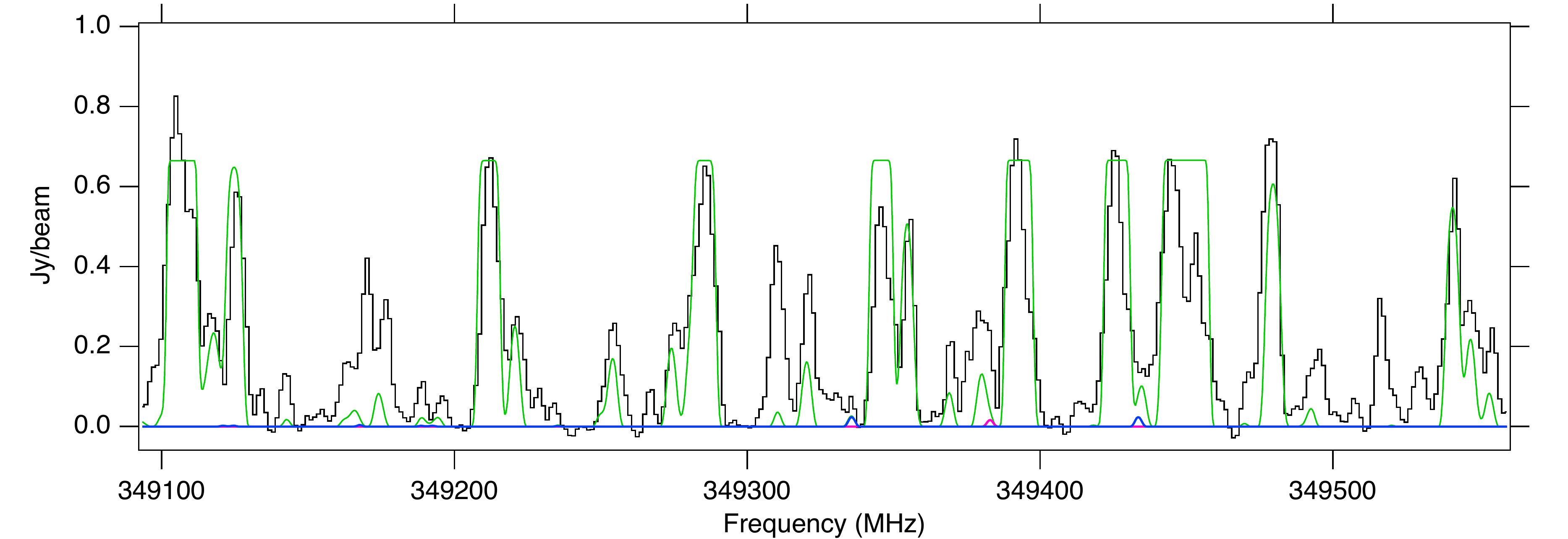}
    \includegraphics[width=0.9\textwidth]{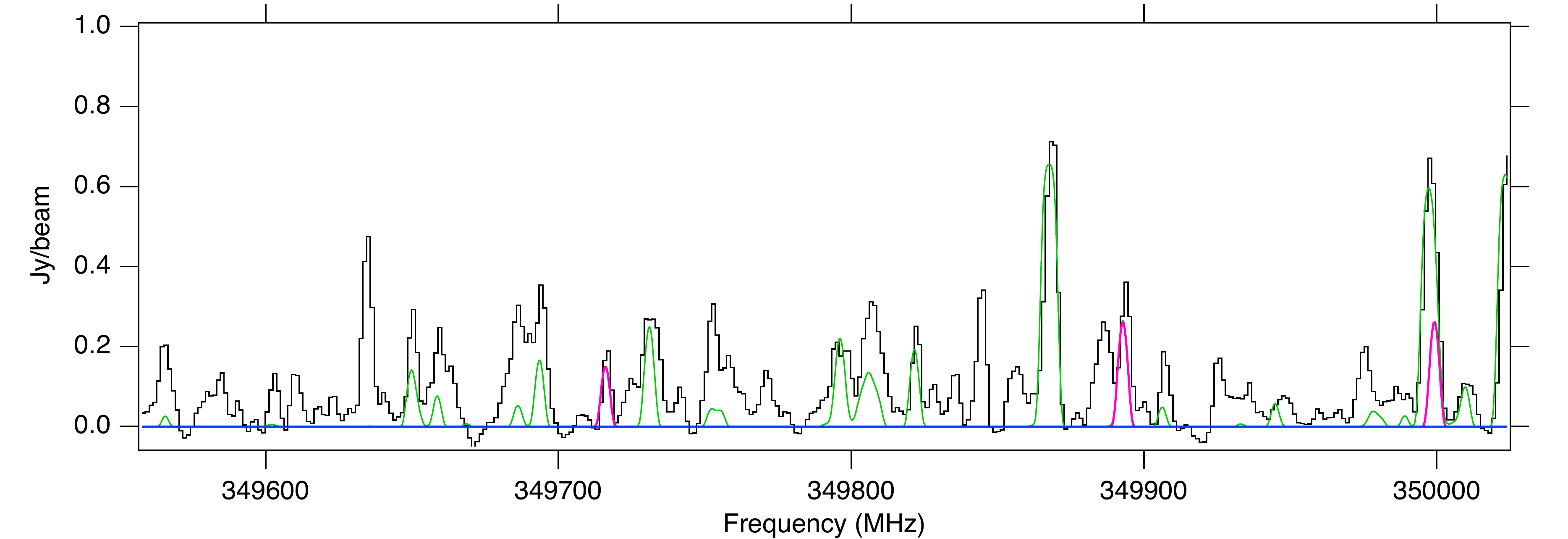}
    \caption{Spectra extracted toward NGC 6334I MM1-i (black).  Overlaid in green is the full model of all assigned molecules in the spectrum (see text), and methyl formate, glycolaldehyde, and acetic acid are shown in color.  Transitions marked with an asterisk were identified as the least blended and optically thin, and were used for the column density analysis (see Table~\ref{freqs}).  Spectra were offset to a $v_{lsr}$~=~-7~km~s$^{-1}$.   }
    \label{mm1_7}
\end{figure}

\begin{figure}
    \centering
    \includegraphics[width=0.9\textwidth]{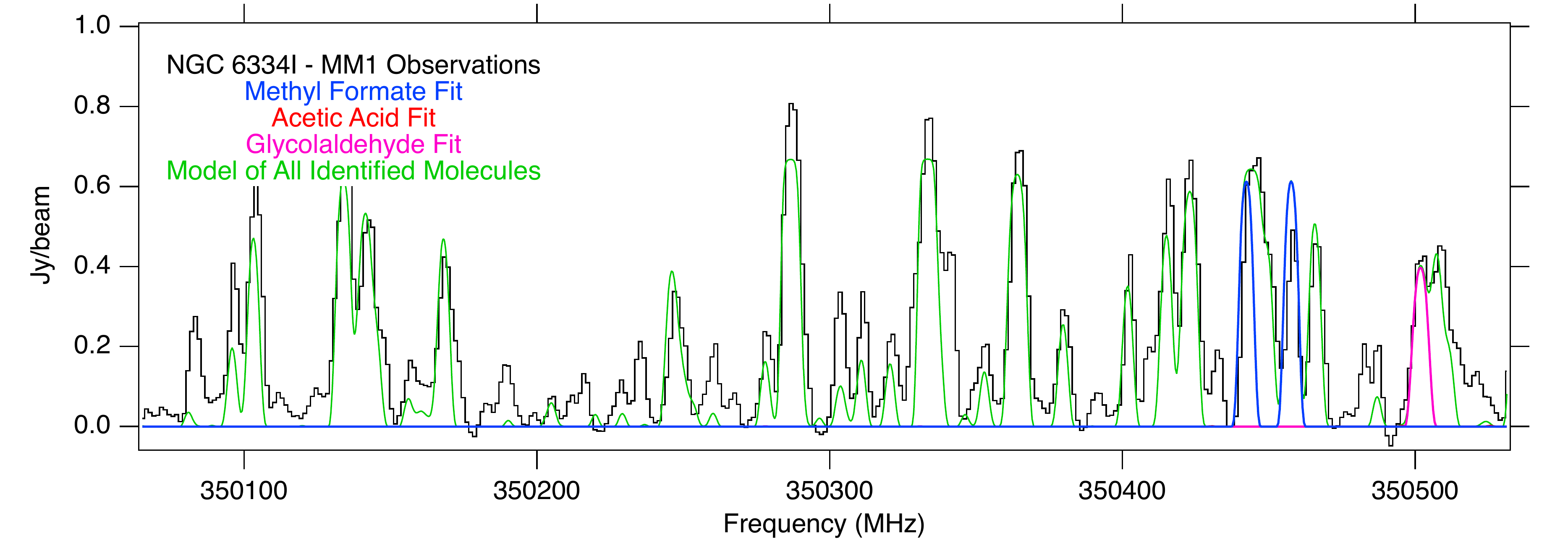}
    \includegraphics[width=0.9\textwidth]{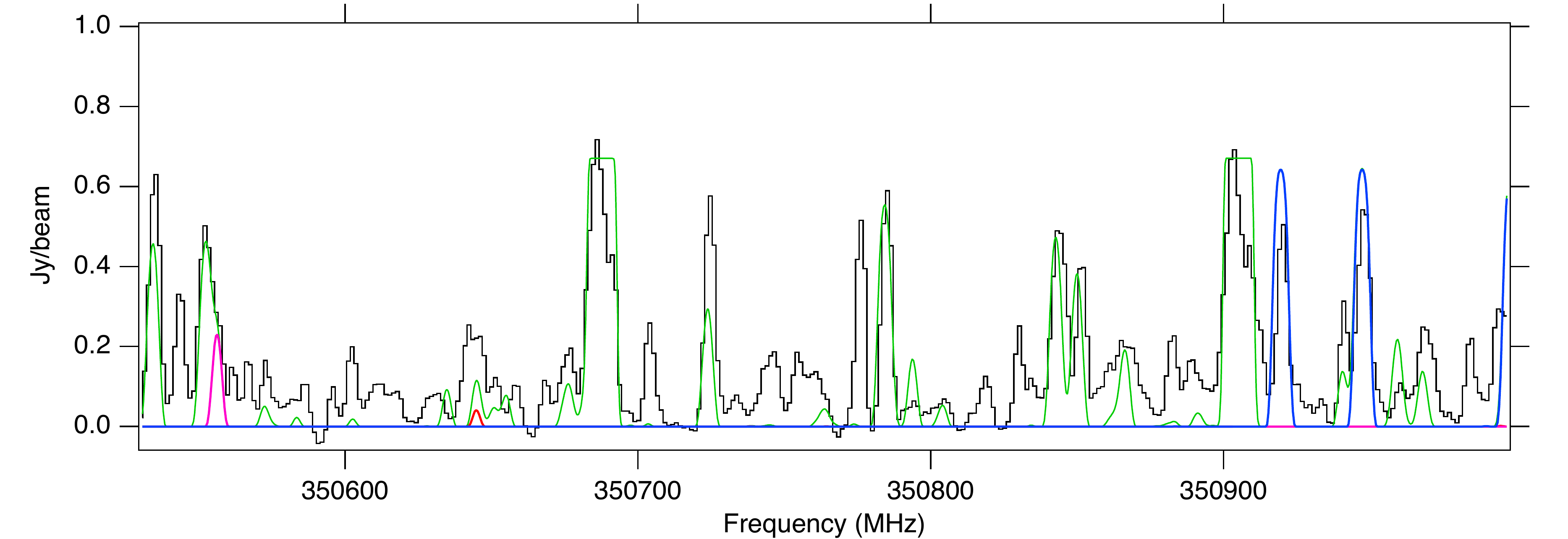}
    \includegraphics[width=0.9\textwidth]{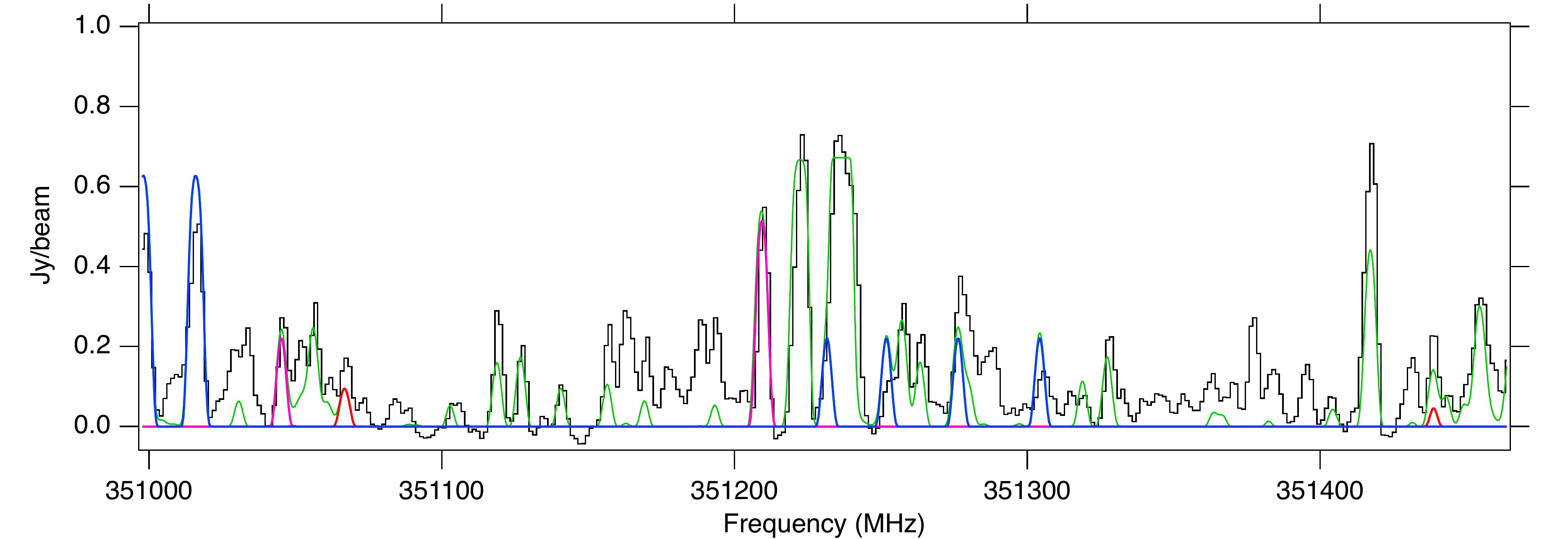}
    \includegraphics[width=0.9\textwidth]{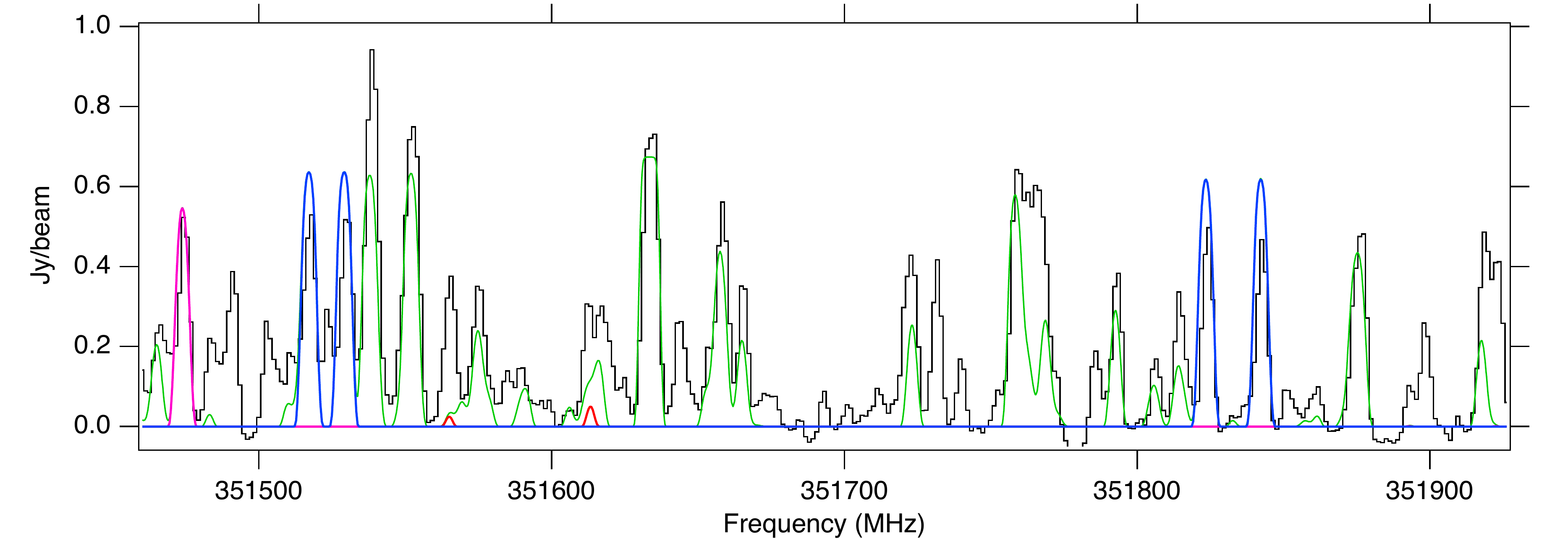}
    \caption{Spectra extracted toward NGC 6334I MM1-i (black).  Overlaid in green is the full model of all assigned molecules in the spectrum (see text), and methyl formate, glycolaldehyde, and acetic acid are shown in color.  Transitions marked with an asterisk were identified as the least blended and optically thin, and were used for the column density analysis (see Table~\ref{freqs}).  Spectra were offset to a $v_{lsr}$~=~-7~km~s$^{-1}$.   }
    \label{mm1_8}
\end{figure}

\begin{figure}
    \centering
    \includegraphics[width=0.9\textwidth]{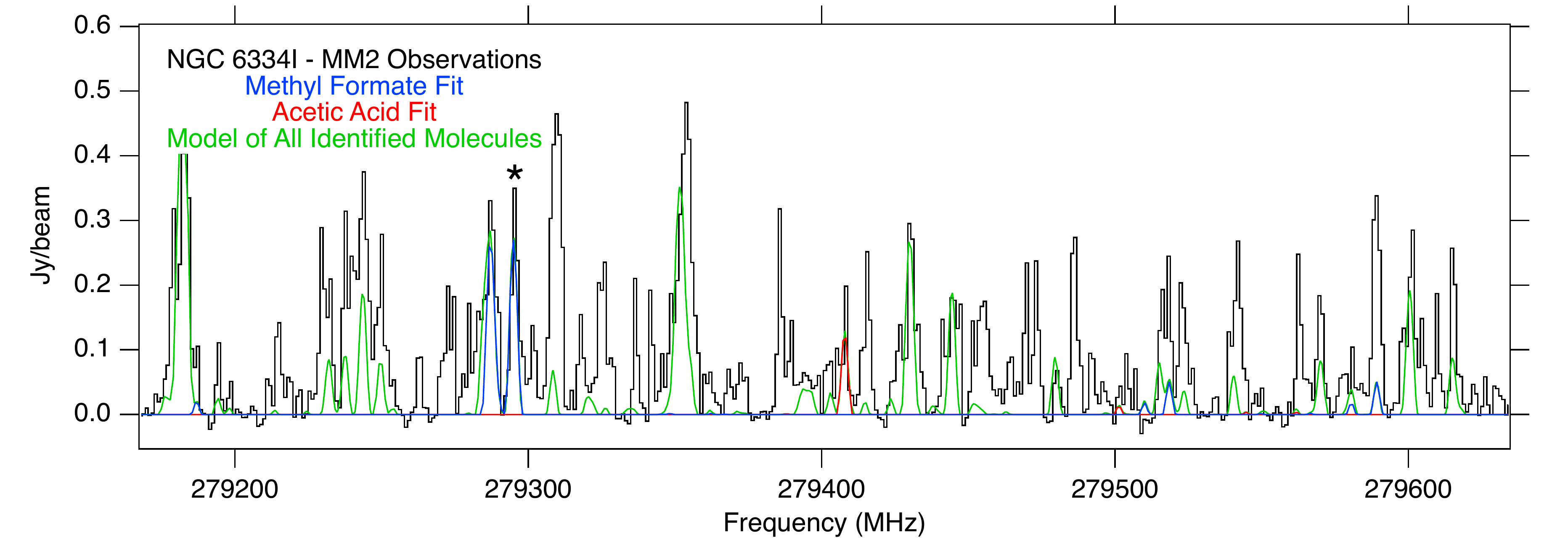}
    \includegraphics[width=0.9\textwidth]{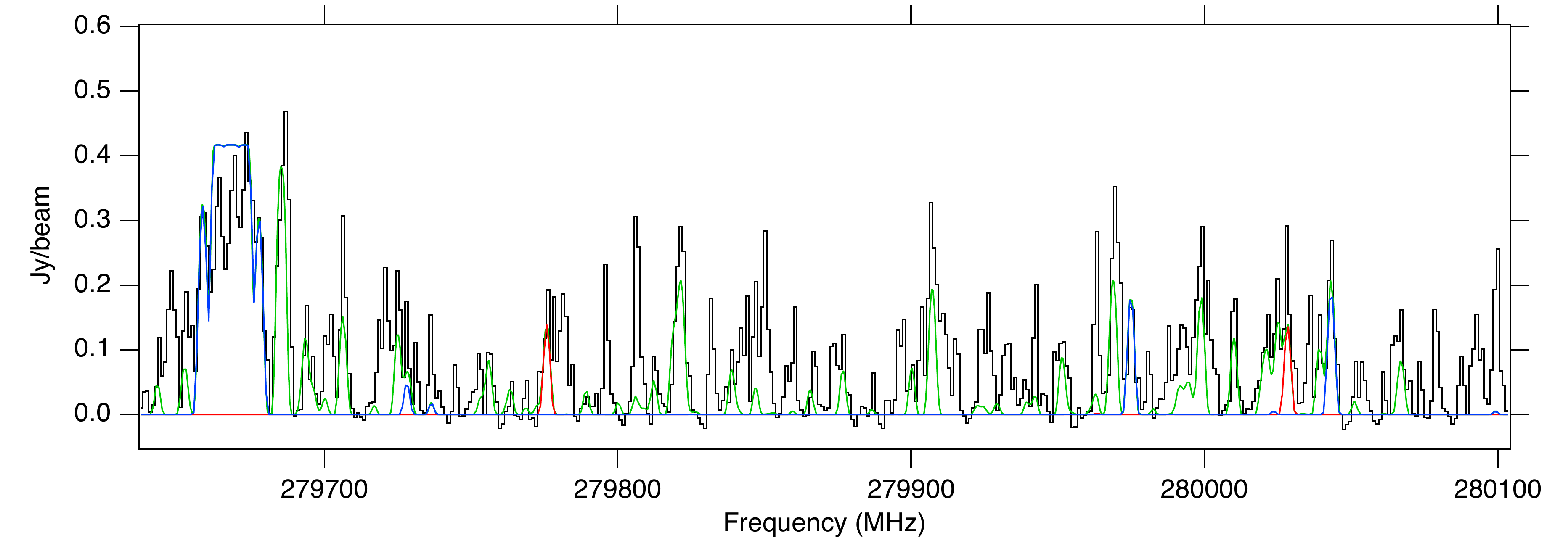}
    \includegraphics[width=0.9\textwidth]{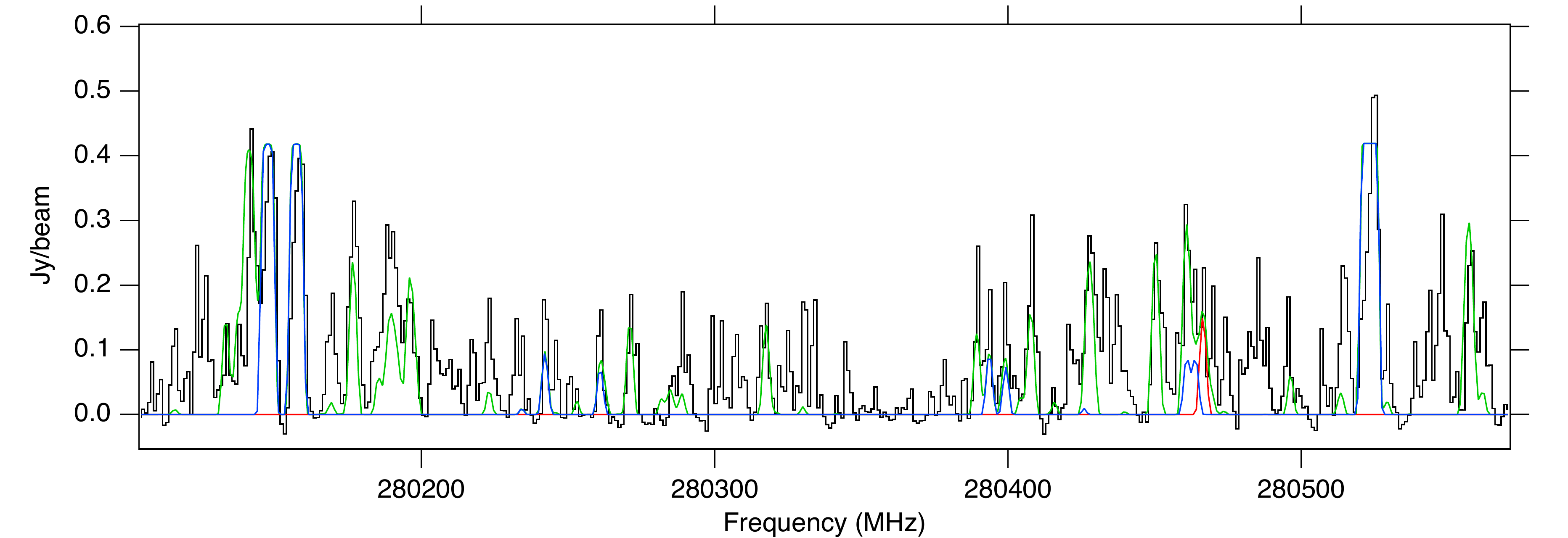}
    \includegraphics[width=0.9\textwidth]{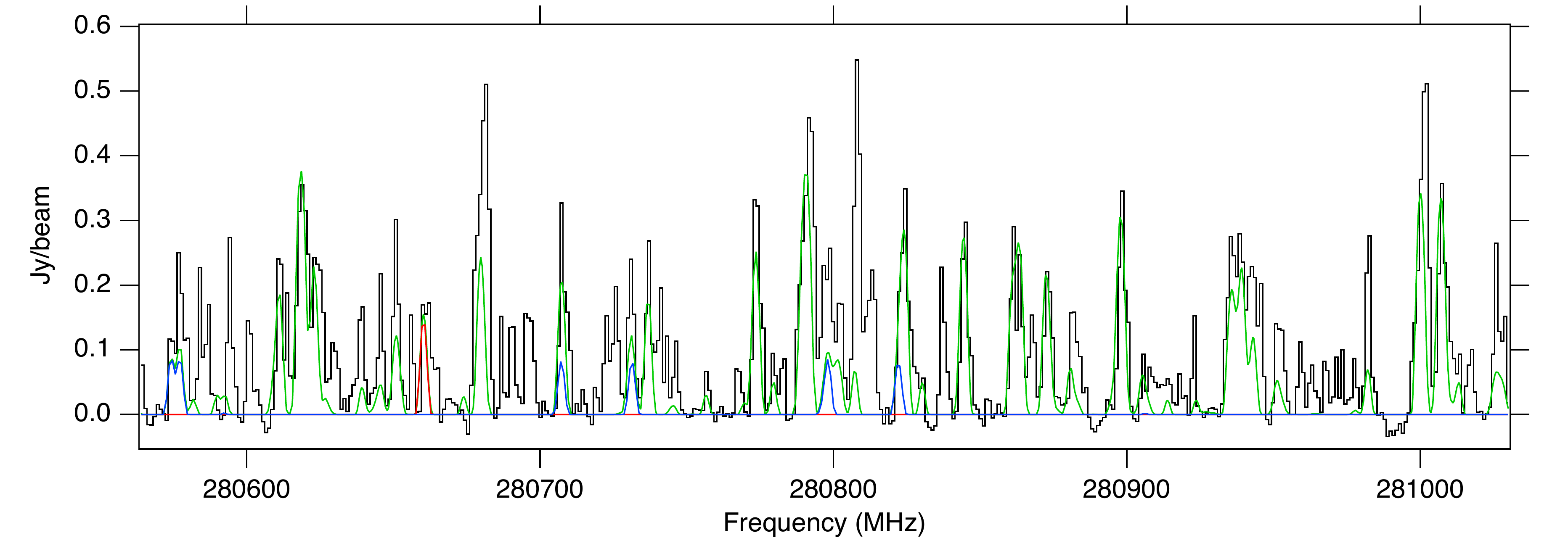}
    \caption{Spectra extracted toward NGC 6334I MM2-ii (black).  Overlaid in green is the full model of all assigned molecules in the spectrum (see text), and methyl formate and acetic acid are shown in color.  Transitions marked with an asterisk were identified as the least blended and optically thin, and were used for the column density analysis (see Table~\ref{freqs}).  Spectra were offset to a $v_{lsr}$~=~-9~km~s$^{-1}$.   }
    \label{mm2_1}
\end{figure}

\begin{figure}
    \centering
    \includegraphics[width=0.9\textwidth]{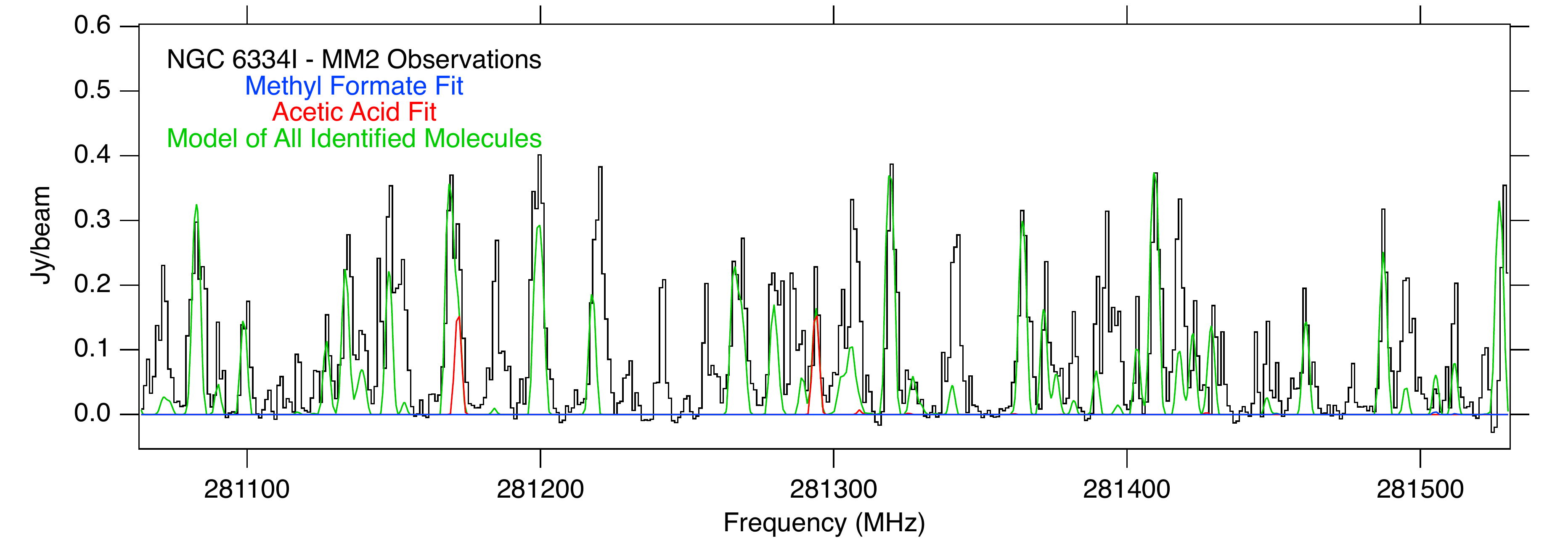}
    \includegraphics[width=0.9\textwidth]{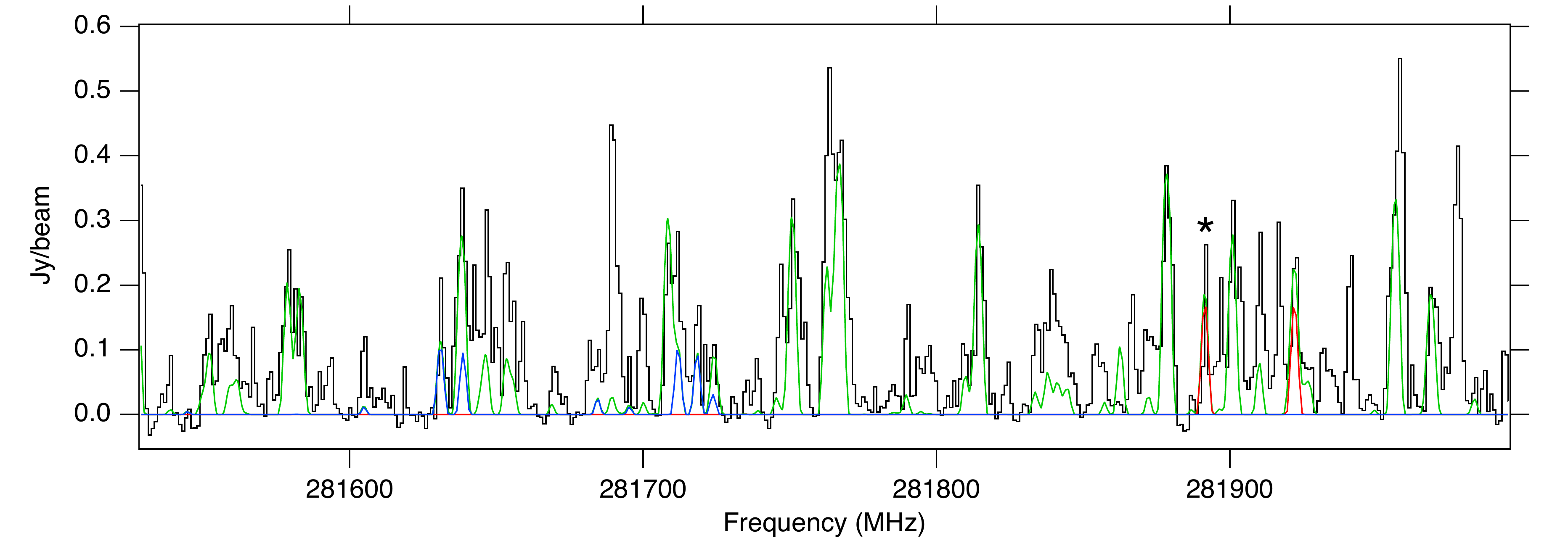}
    \includegraphics[width=0.9\textwidth]{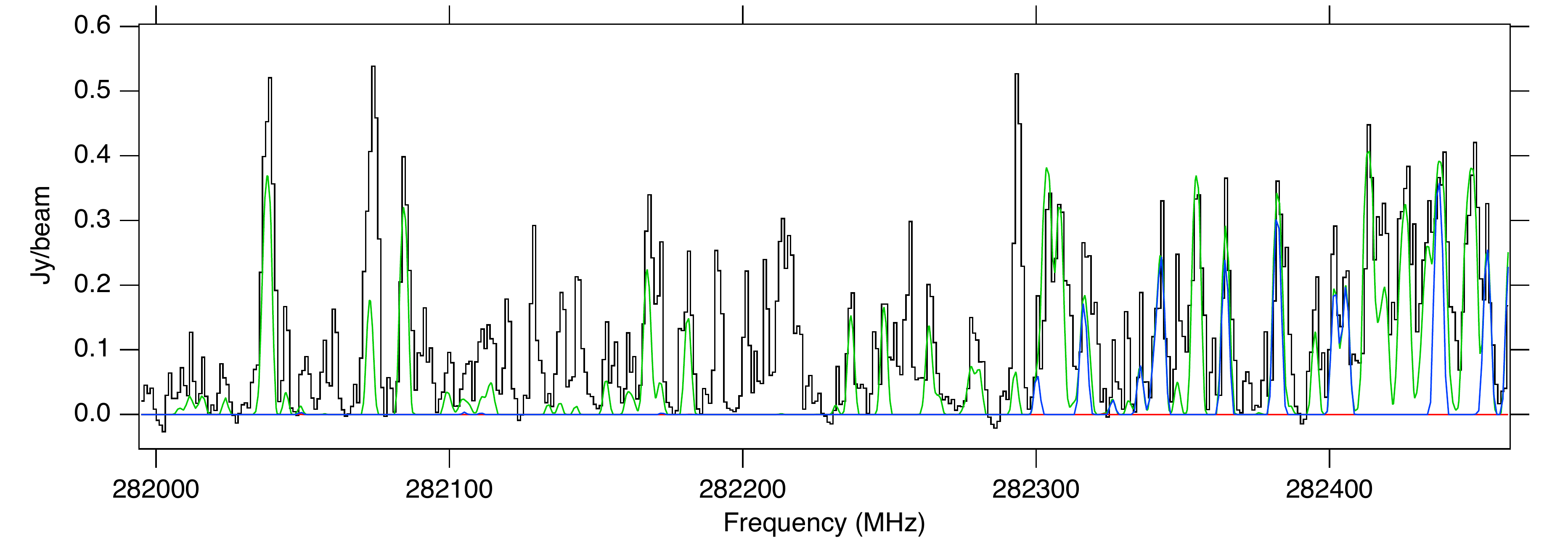}
    \includegraphics[width=0.9\textwidth]{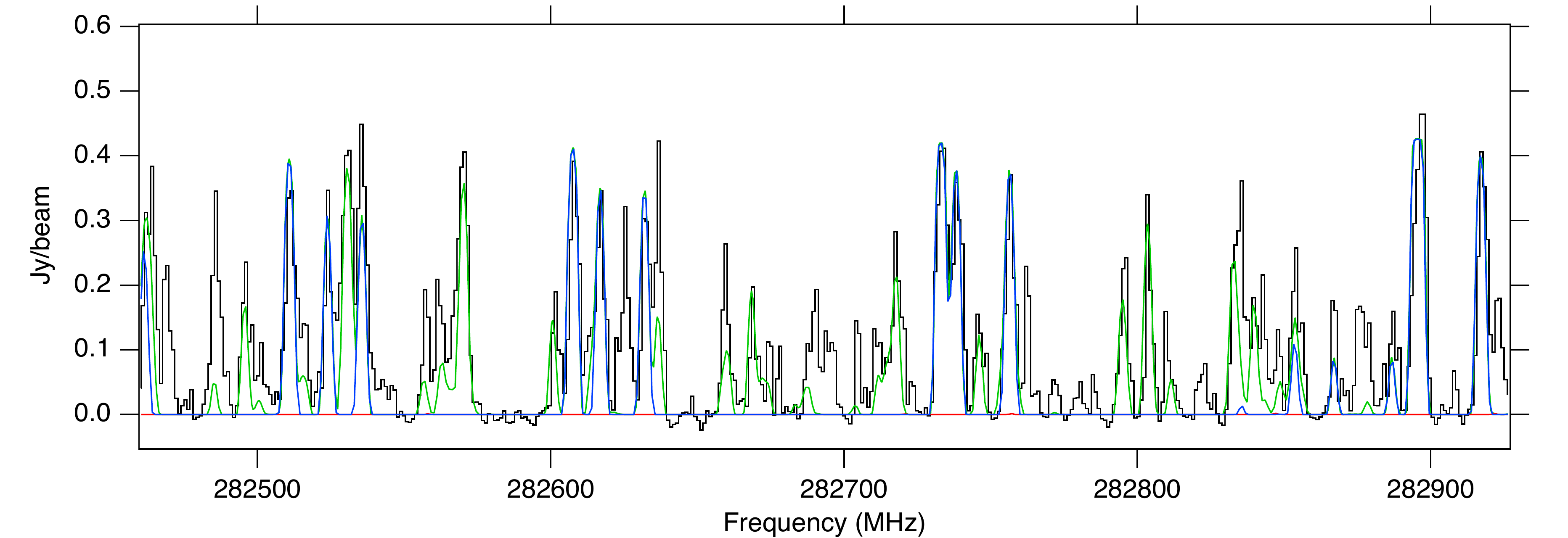}
    \caption{Spectra extracted toward NGC 6334I MM2-ii (black).  Overlaid in green is the full model of all assigned molecules in the spectrum (see text), and methyl formate and acetic acid are shown in color.  Transitions marked with an asterisk were identified as the least blended and optically thin, and were used for the column density analysis (see Table~\ref{freqs}).  Spectra were offset to a $v_{lsr}$~=~-9~km~s$^{-1}$.   }
    \label{mm2_2}
\end{figure}

\begin{figure}
    \centering
    \includegraphics[width=0.9\textwidth]{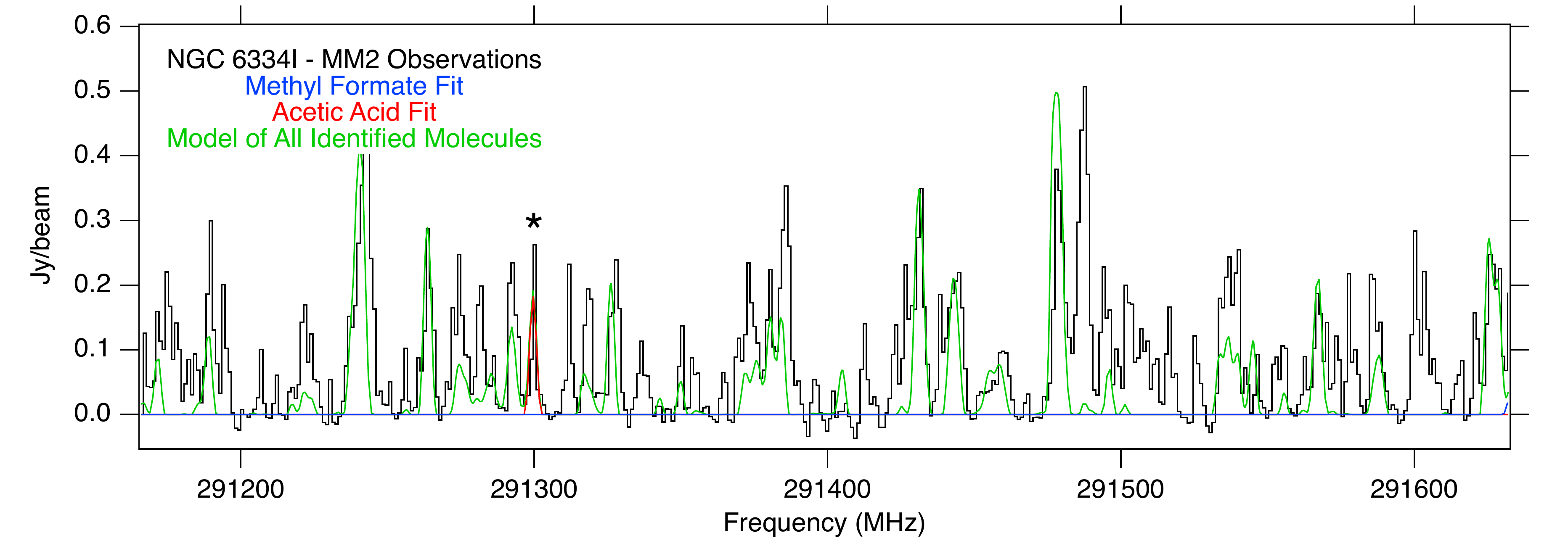}
    \includegraphics[width=0.9\textwidth]{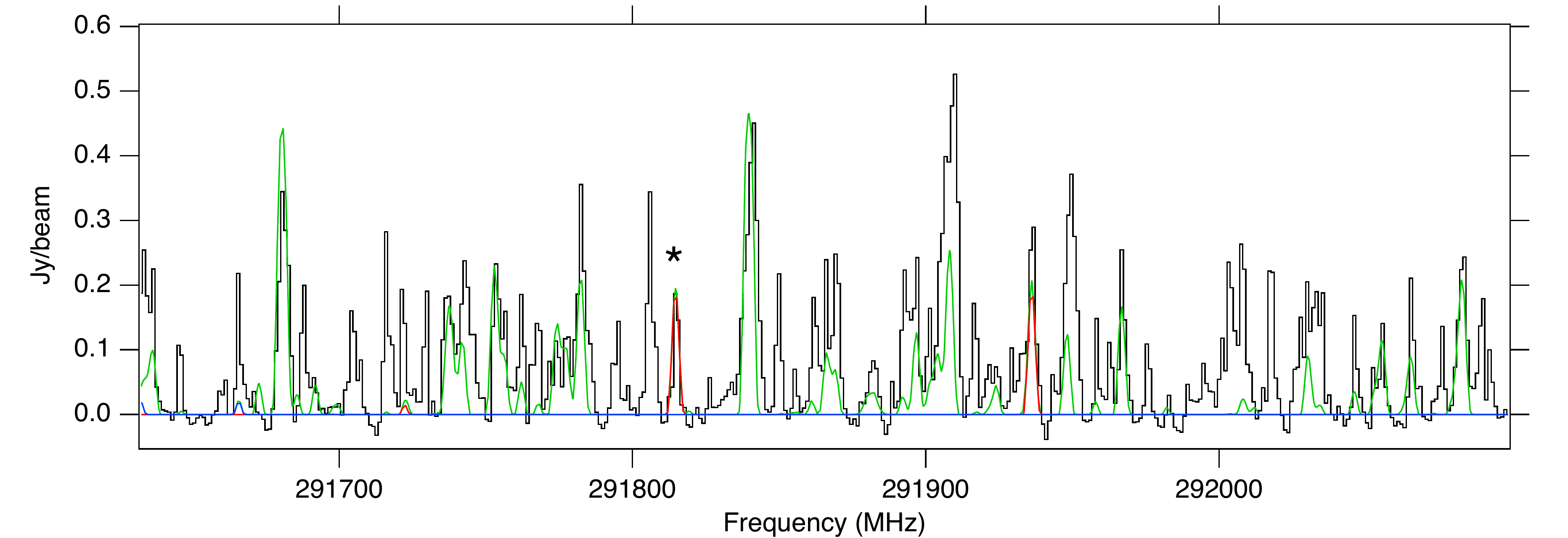}
    \includegraphics[width=0.9\textwidth]{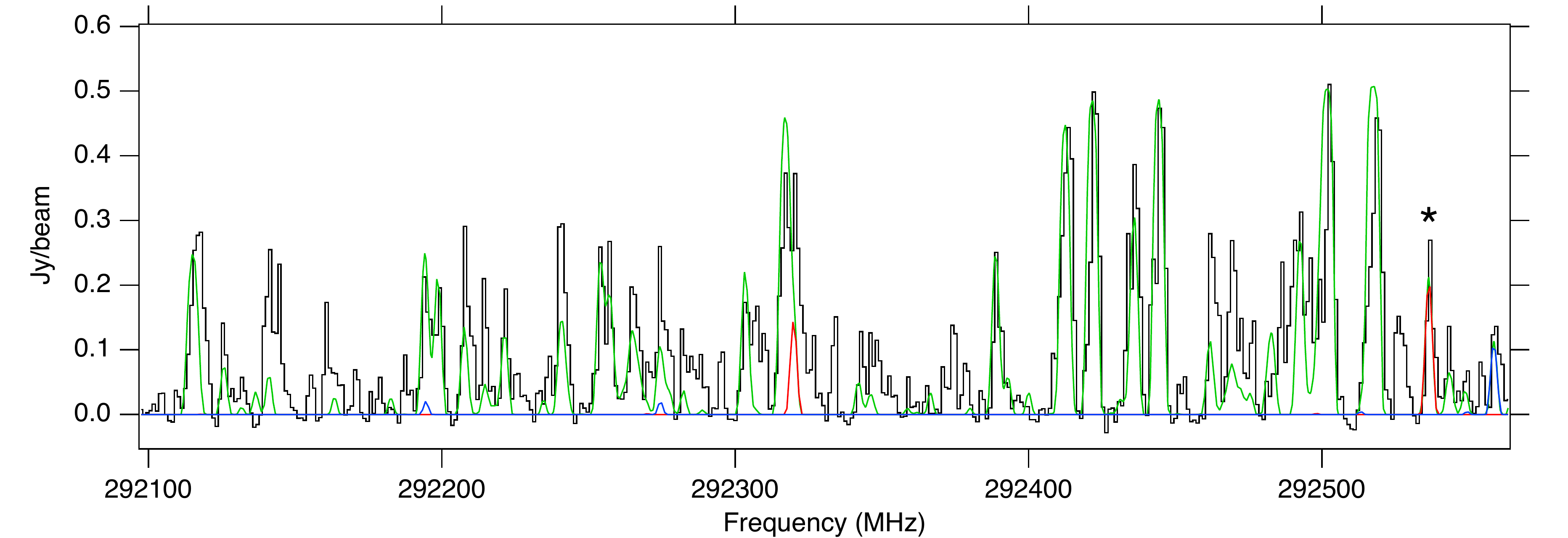}
    \includegraphics[width=0.9\textwidth]{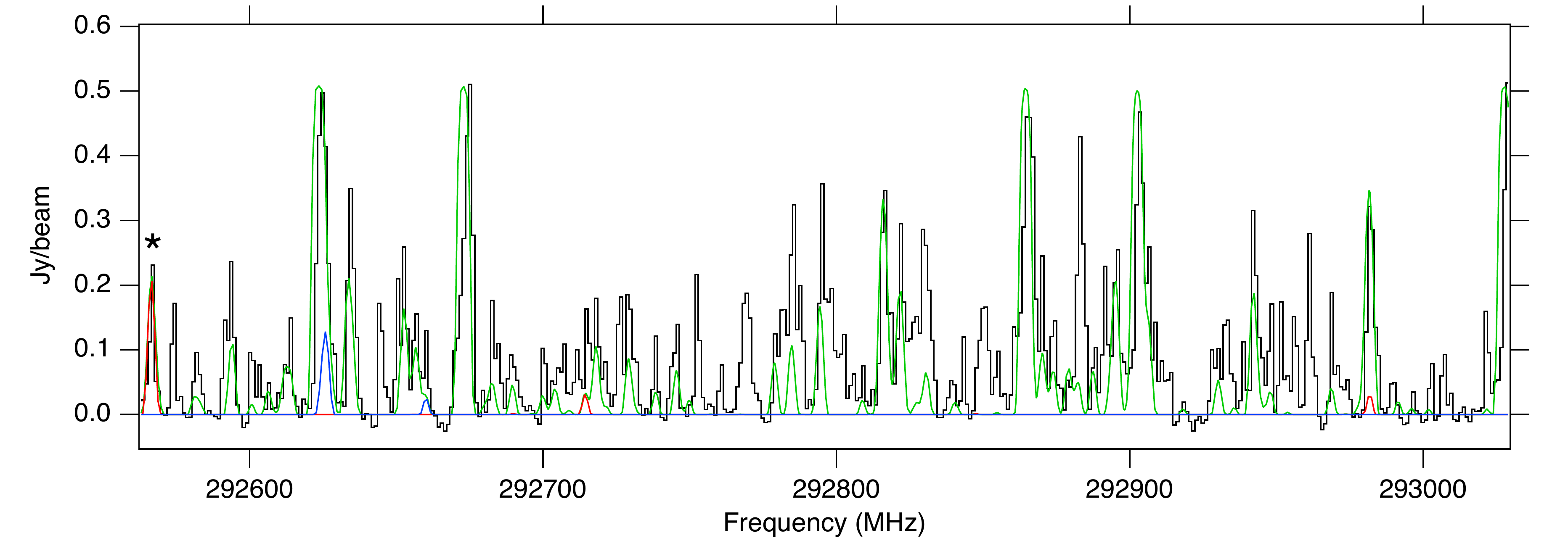}
    \caption{Spectra extracted toward NGC 6334I MM2-ii (black).  Overlaid in green is the full model of all assigned molecules in the spectrum (see text), and methyl formate and acetic acid are shown in color.  Transitions marked with an asterisk were identified as the least blended and optically thin, and were used for the column density analysis (see Table~\ref{freqs}).  Spectra were offset to a $v_{lsr}$~=~-9~km~s$^{-1}$.   }
    \label{mm2_3}
\end{figure}

\begin{figure}
    \centering
    \includegraphics[width=0.9\textwidth]{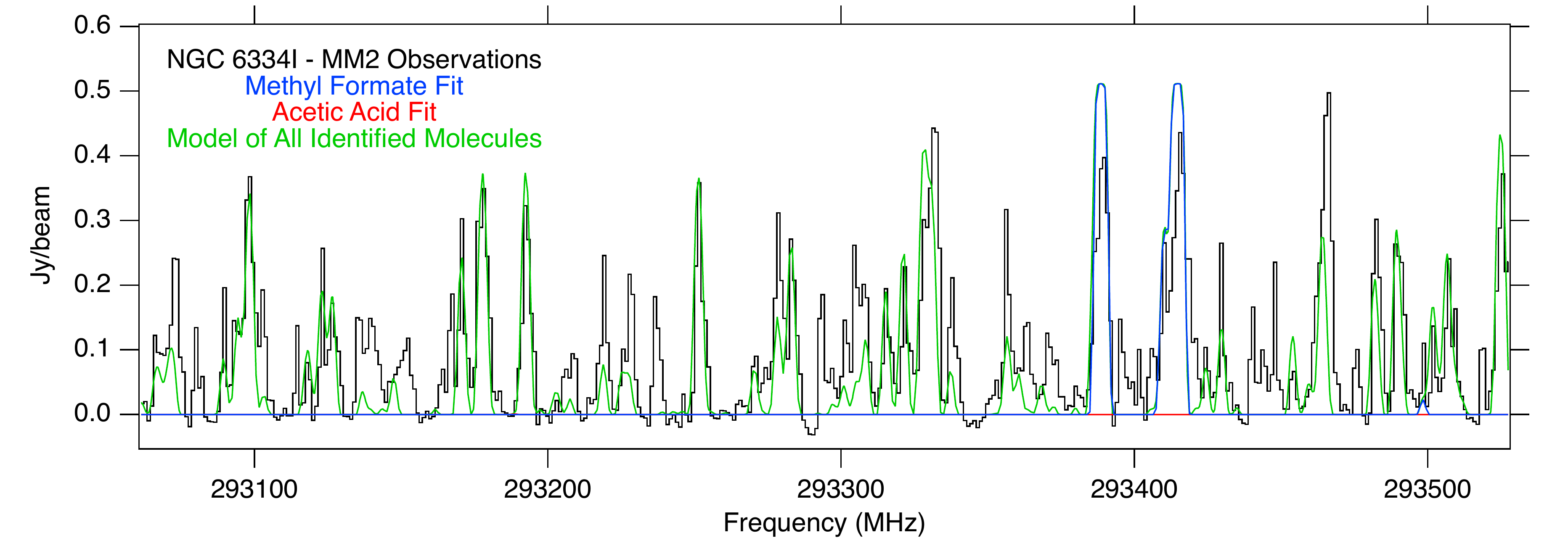}
    \includegraphics[width=0.9\textwidth]{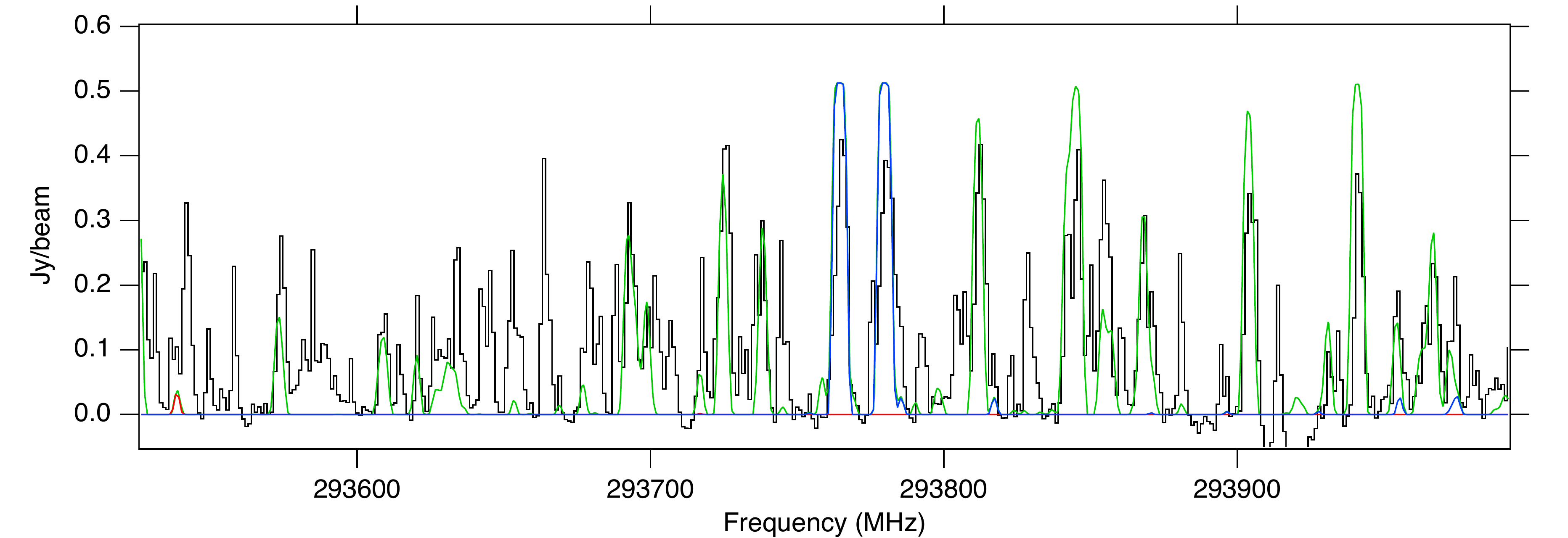}
    \includegraphics[width=0.9\textwidth]{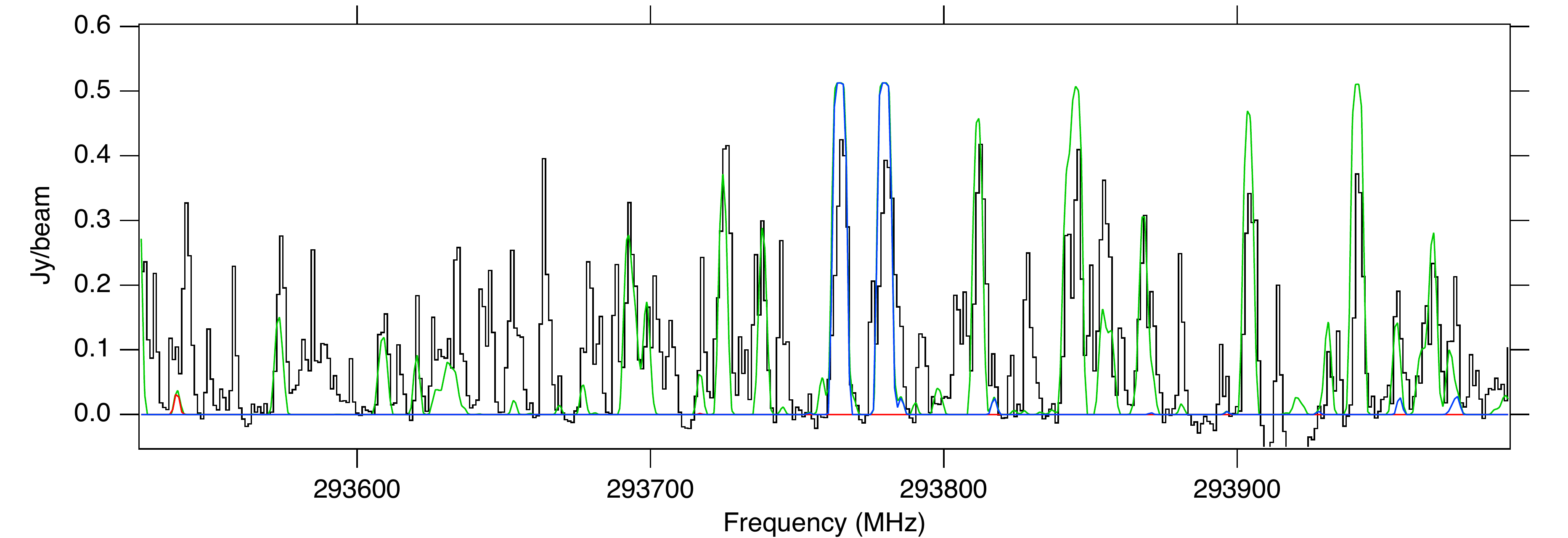}
    \includegraphics[width=0.9\textwidth]{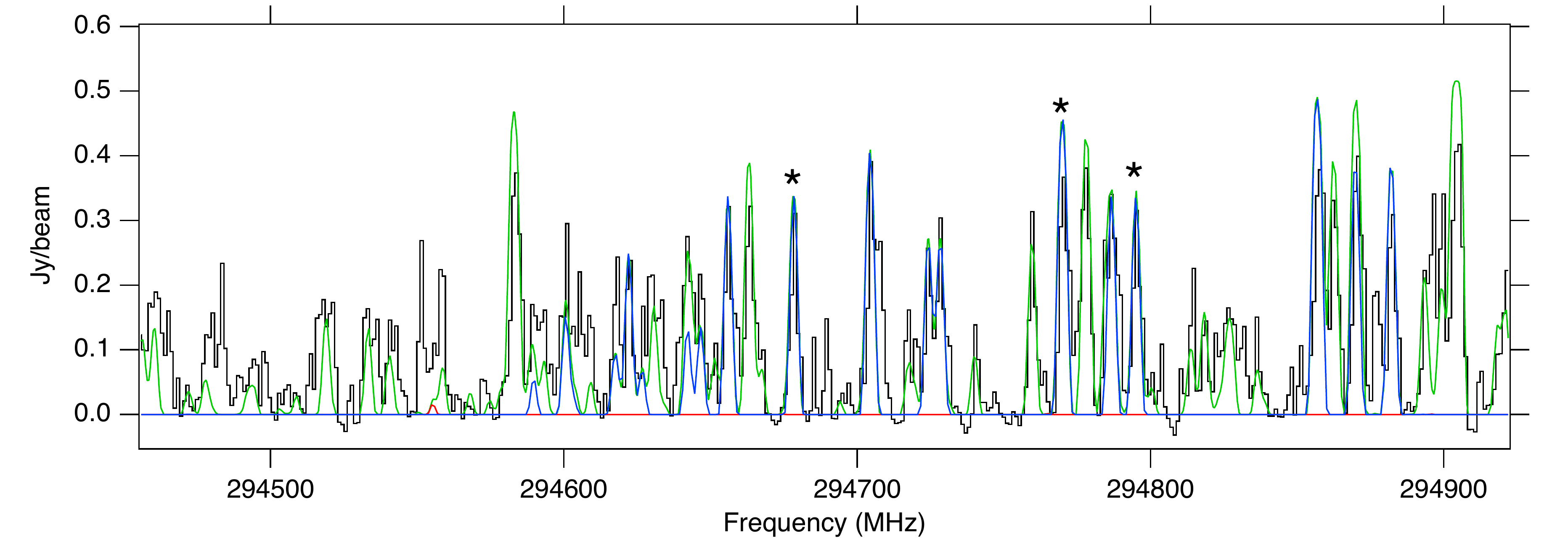}
    \caption{Spectra extracted toward NGC 6334I MM2-ii (black).  Overlaid in green is the full model of all assigned molecules in the spectrum (see text), and methyl formate and acetic acid are shown in color.  Transitions marked with an asterisk were identified as the least blended and optically thin, and were used for the column density analysis (see Table~\ref{freqs}).  Spectra were offset to a $v_{lsr}$~=~-9~km~s$^{-1}$.   }
    \label{mm2_4}
\end{figure}

\begin{figure}
    \centering
    \includegraphics[width=0.9\textwidth]{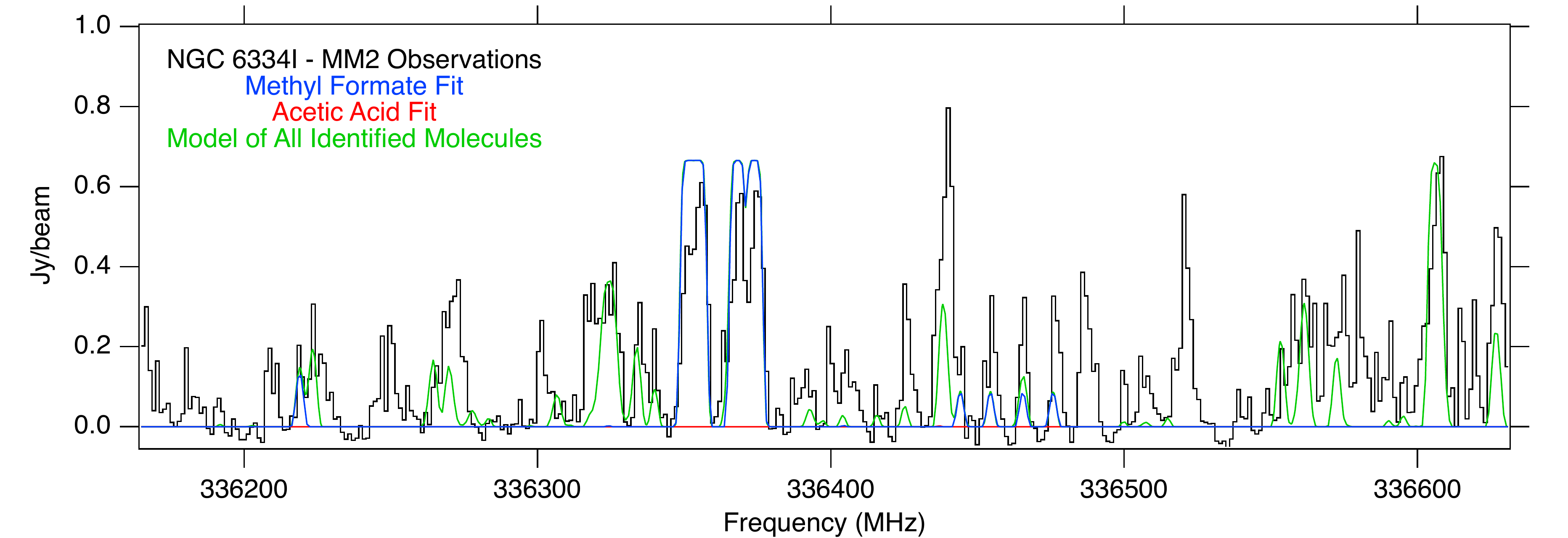}
    \includegraphics[width=0.9\textwidth]{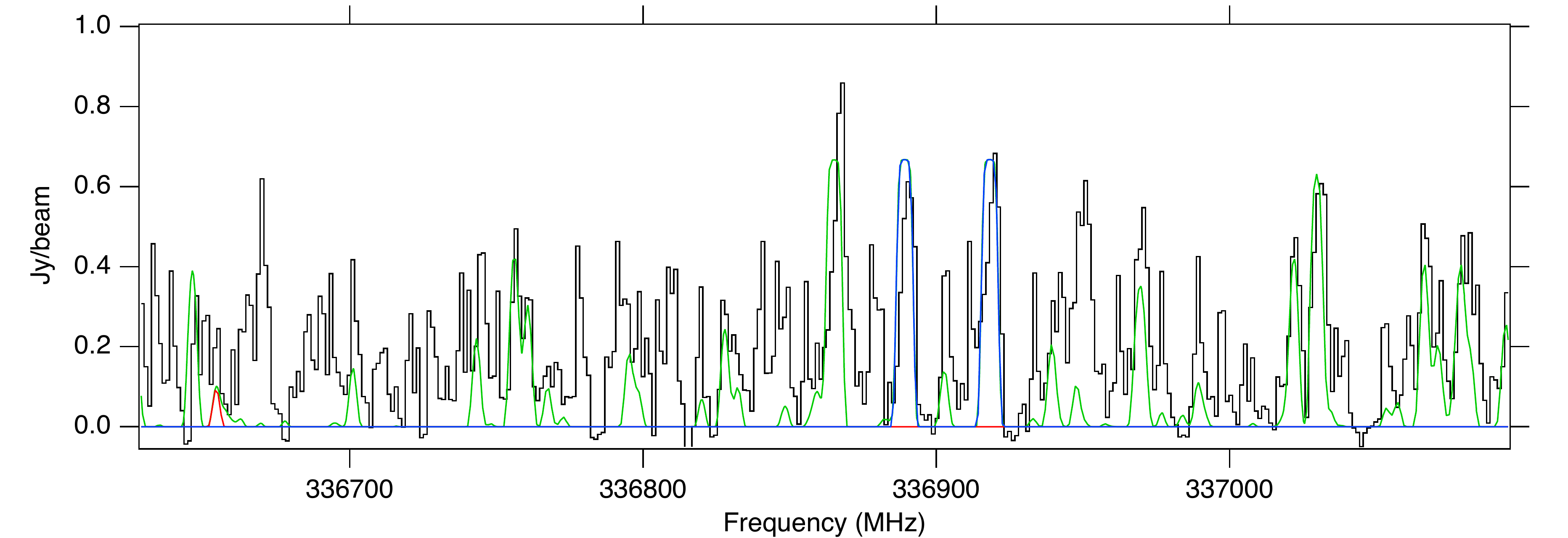}
    \includegraphics[width=0.9\textwidth]{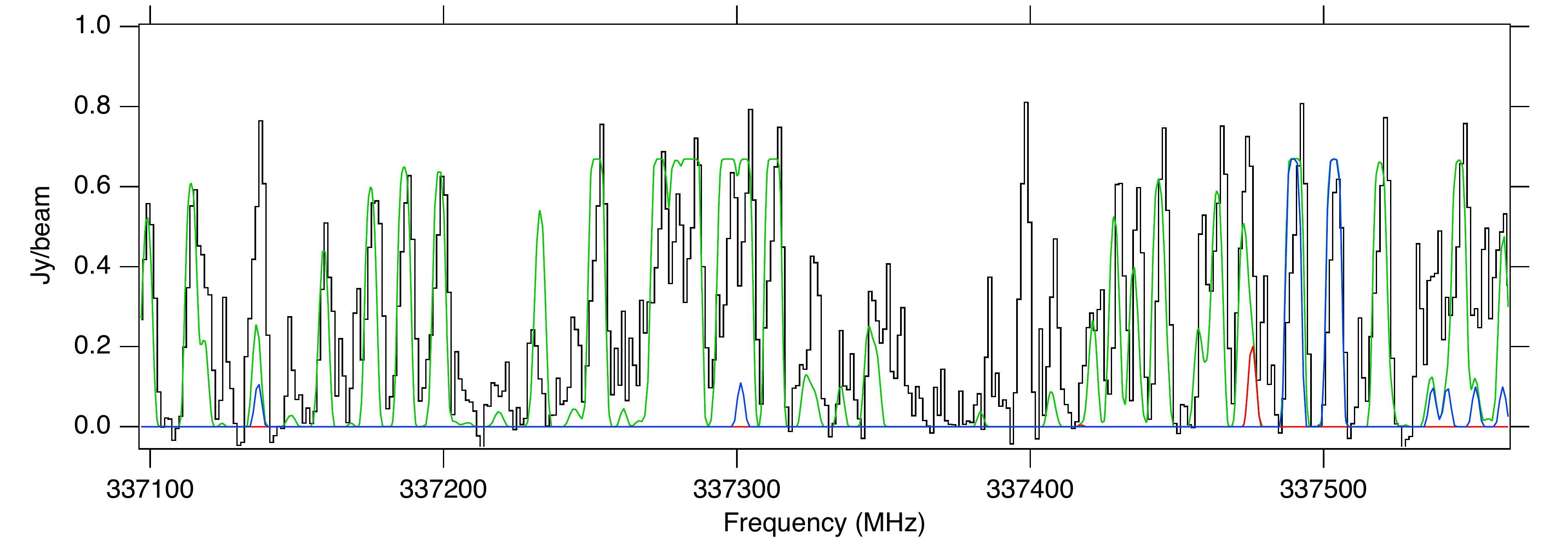}
    \includegraphics[width=0.9\textwidth]{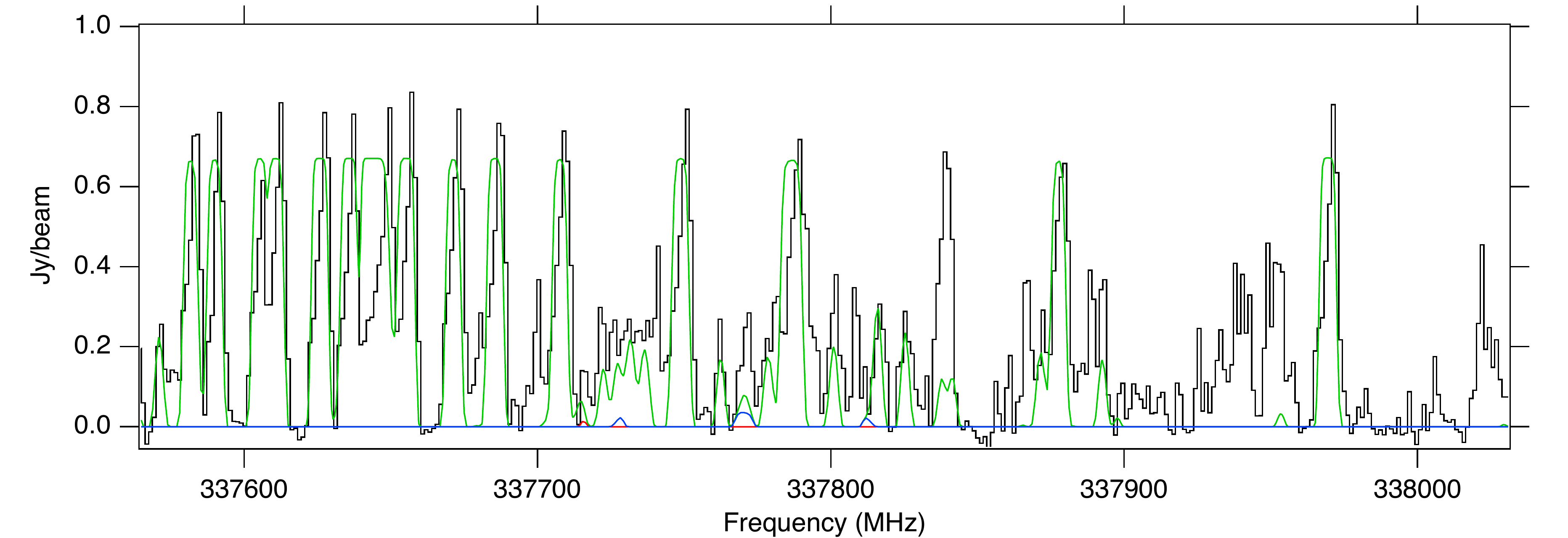}
    \caption{Spectra extracted toward NGC 6334I MM2-ii (black).  Overlaid in green is the full model of all assigned molecules in the spectrum (see text), and methyl formate and acetic acid are shown in color.  Transitions marked with an asterisk were identified as the least blended and optically thin, and were used for the column density analysis (see Table~\ref{freqs}).  Spectra were offset to a $v_{lsr}$~=~-9~km~s$^{-1}$.   }
    \label{mm2_5}
\end{figure}

\begin{figure}
    \centering
    \includegraphics[width=0.9\textwidth]{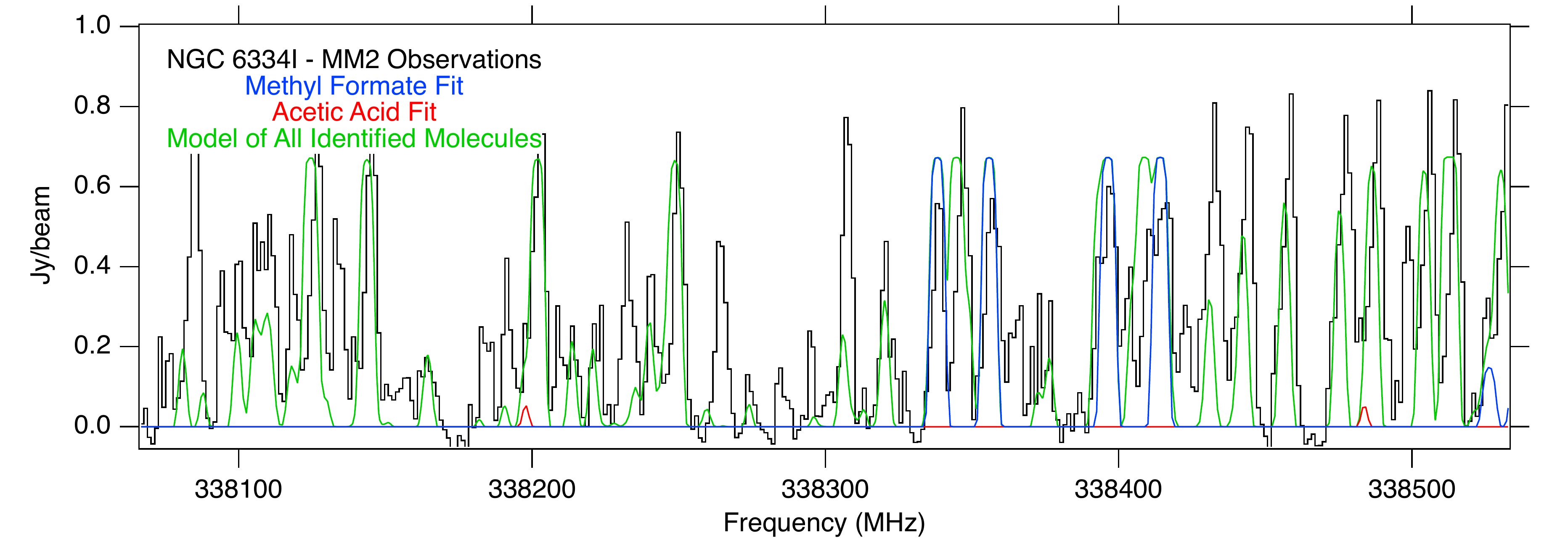}
    \includegraphics[width=0.9\textwidth]{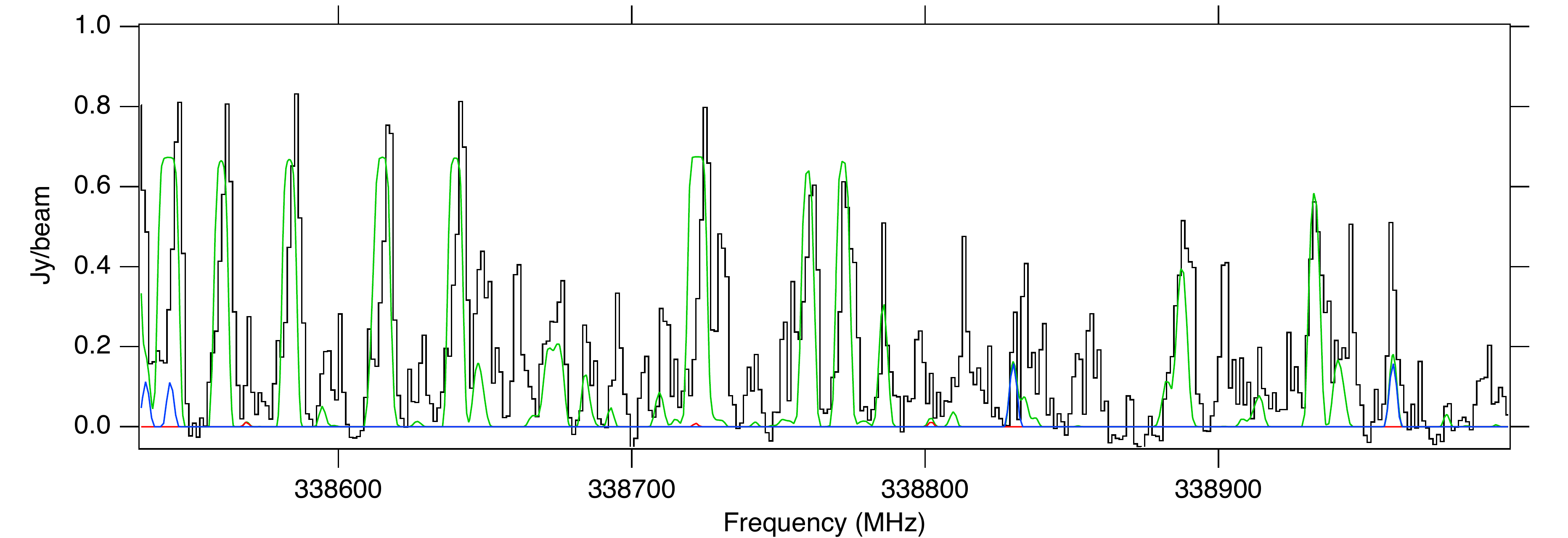}
    \includegraphics[width=0.9\textwidth]{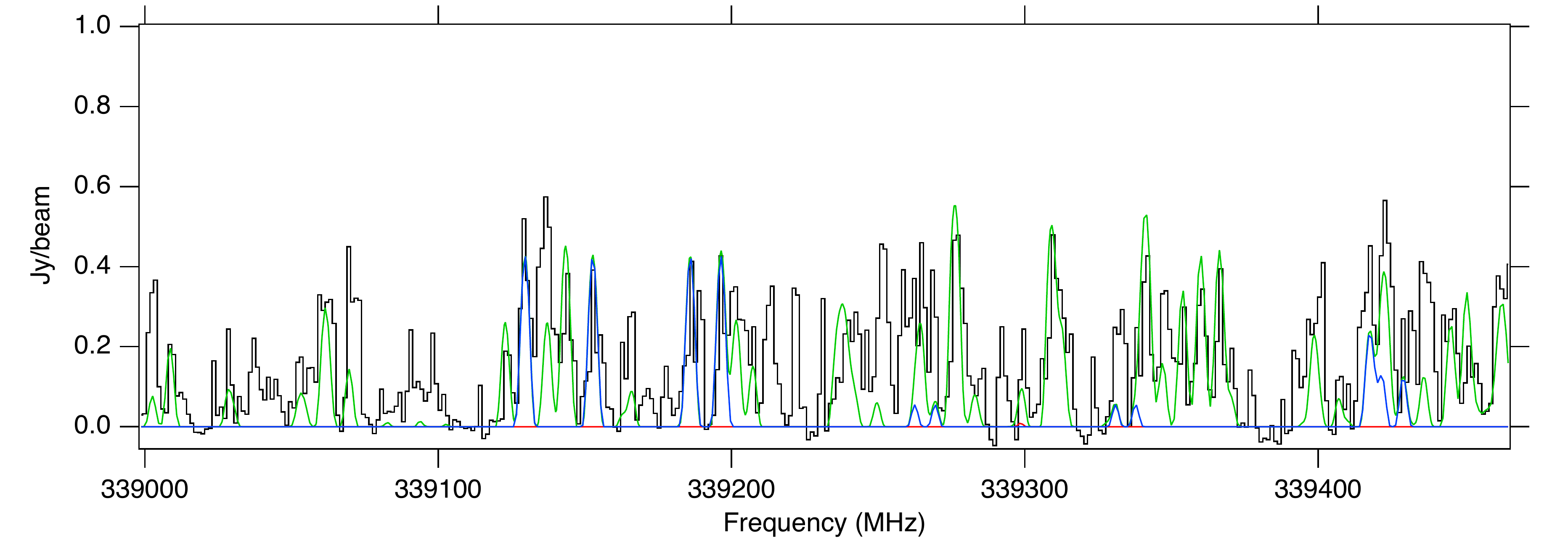}
    \includegraphics[width=0.9\textwidth]{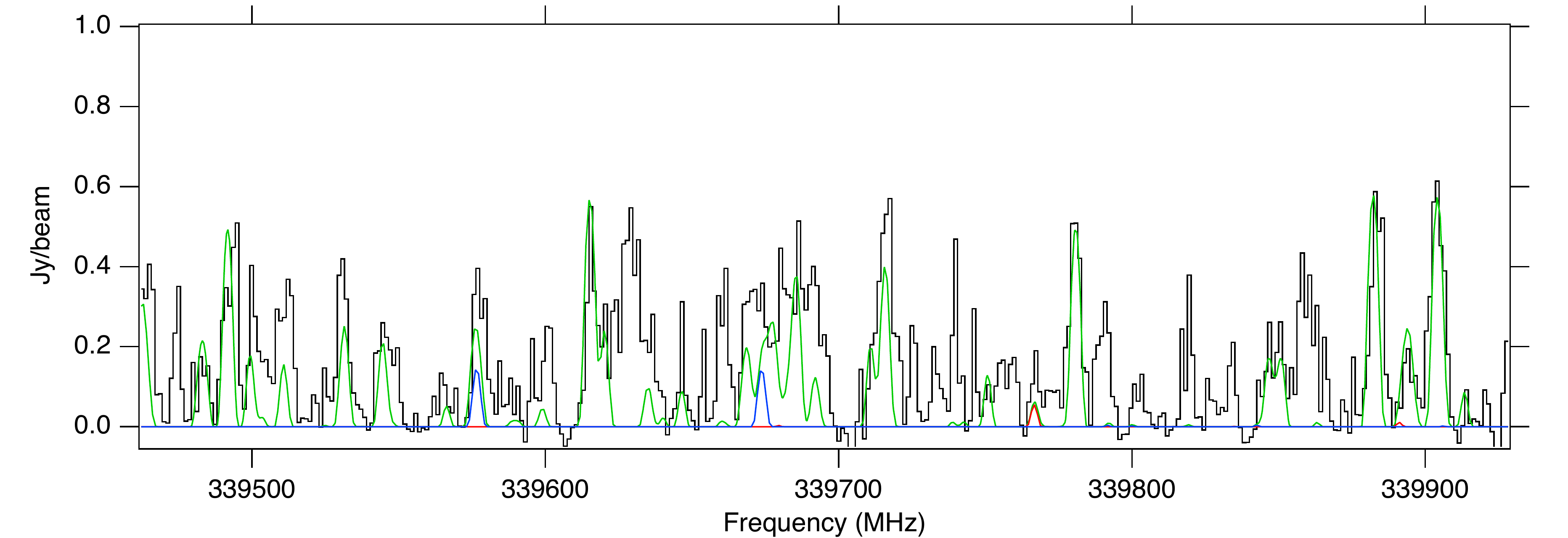}
    \caption{Spectra extracted toward NGC 6334I MM2-ii (black).  Overlaid in green is the full model of all assigned molecules in the spectrum (see text), and methyl formate and acetic acid are shown in color.  Transitions marked with an asterisk were identified as the least blended and optically thin, and were used for the column density analysis (see Table~\ref{freqs}).  Spectra were offset to a $v_{lsr}$~=~-9~km~s$^{-1}$.   }
    \label{mm2_6}
\end{figure}

\begin{figure}
    \centering
    \includegraphics[width=0.9\textwidth]{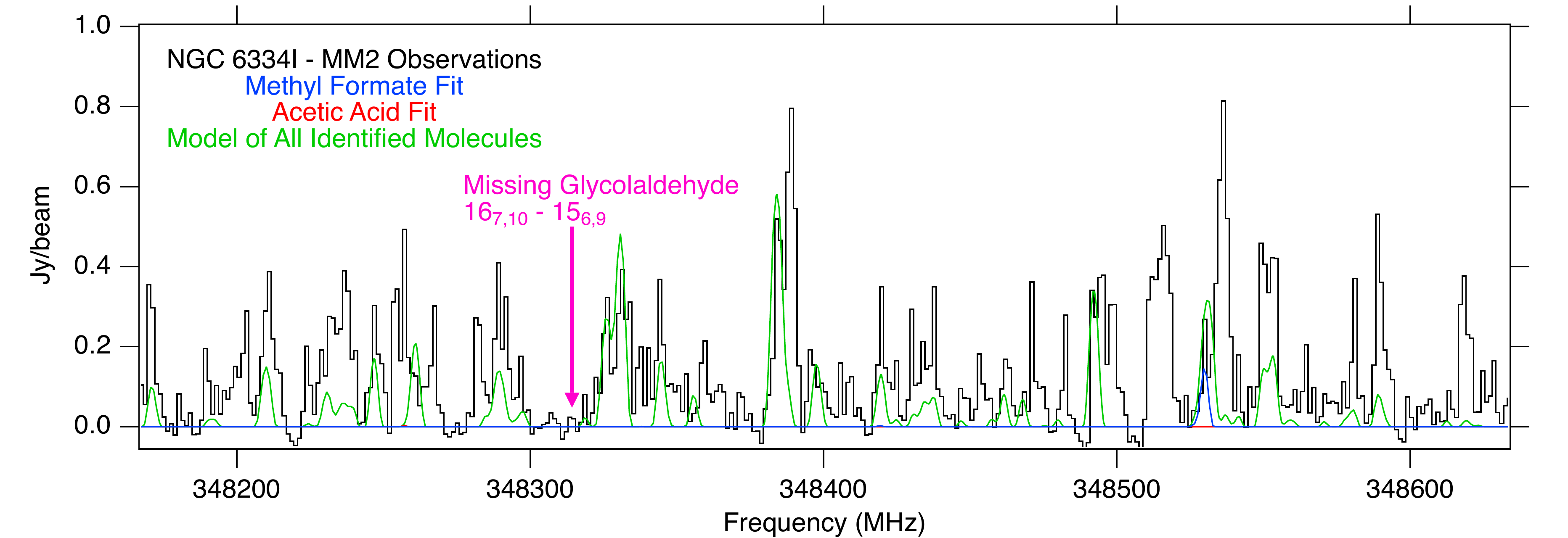}
    \includegraphics[width=0.9\textwidth]{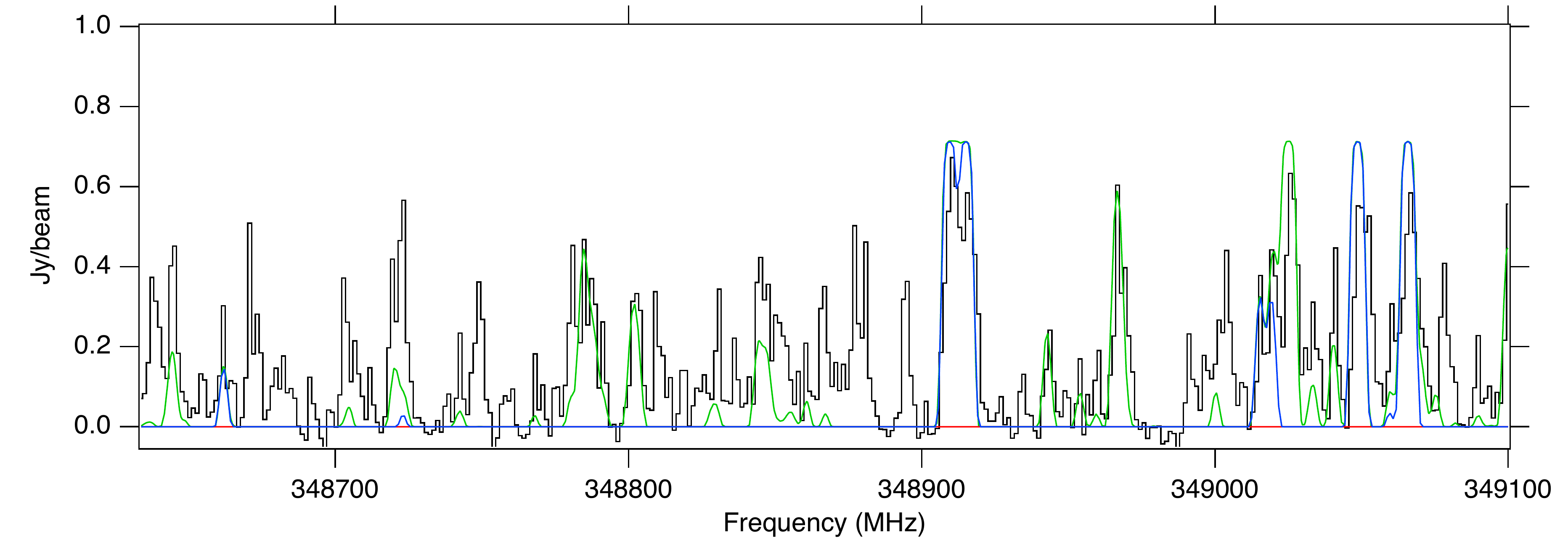}
    \includegraphics[width=0.9\textwidth]{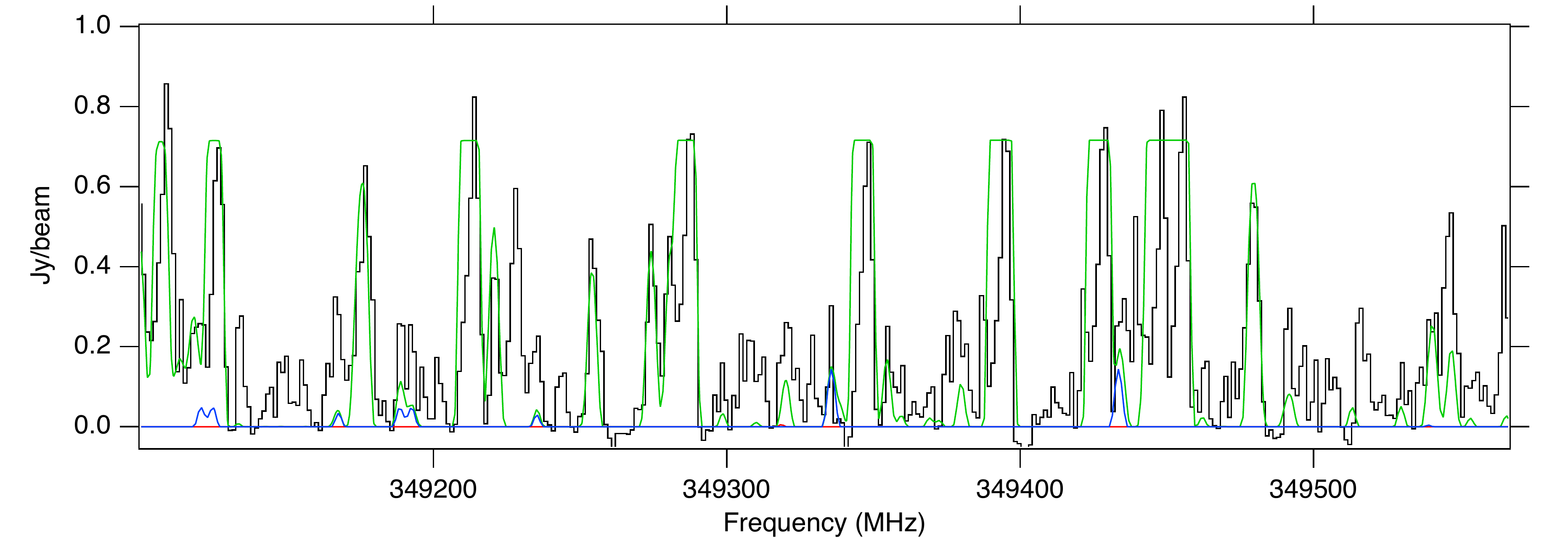}
    \includegraphics[width=0.9\textwidth]{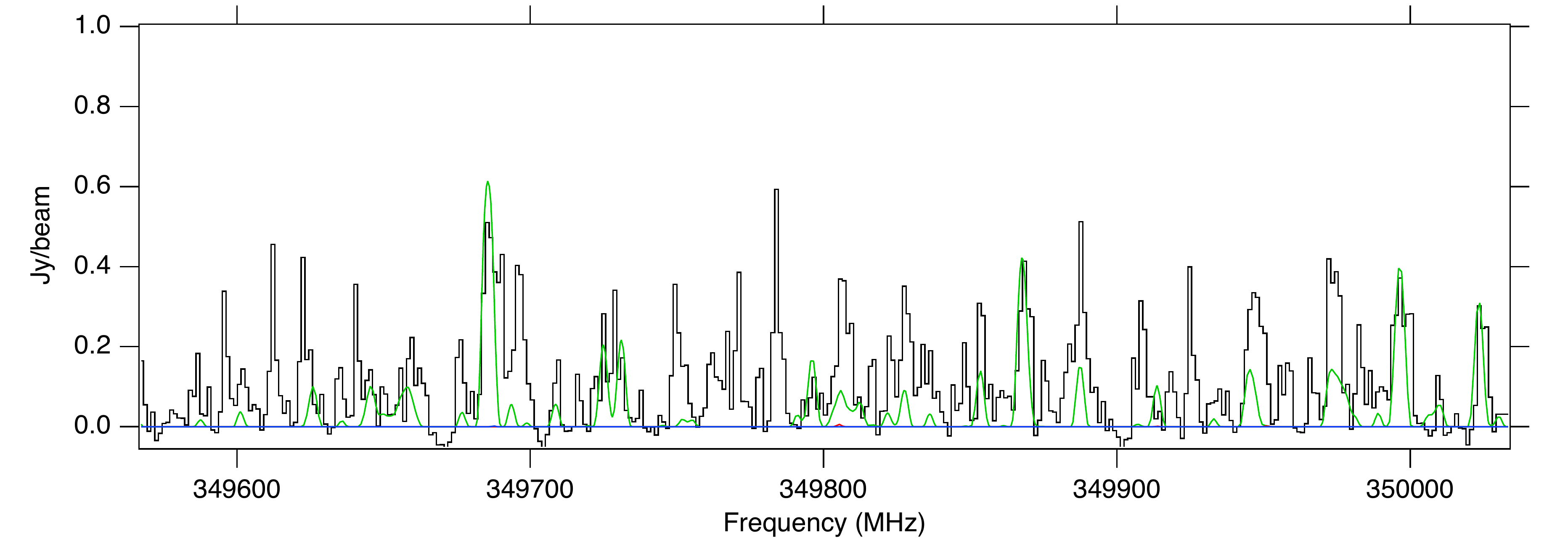}
    \caption{Spectra extracted toward NGC 6334I MM2-ii (black).  Overlaid in green is the full model of all assigned molecules in the spectrum (see text), and methyl formate and acetic acid are shown in color.  Transitions marked with an asterisk were identified as the least blended and optically thin, and were used for the column density analysis (see Table~\ref{freqs}).  The location of the missing glycolaldehyde transition used to determine the upper limit is marked with an arrow.  Spectra were offset to a $v_{lsr}$~=~-9~km~s$^{-1}$.   }
    \label{mm2_7}
\end{figure}

\begin{figure}
    \centering
    \includegraphics[width=0.9\textwidth]{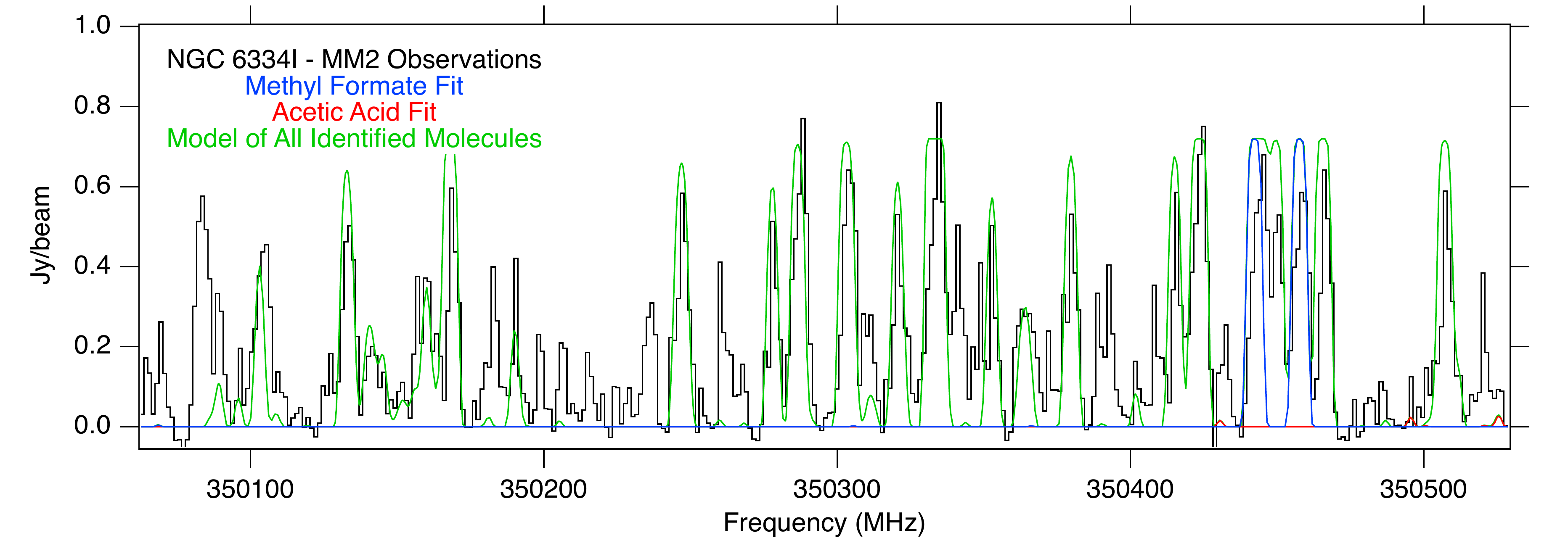}
    \includegraphics[width=0.9\textwidth]{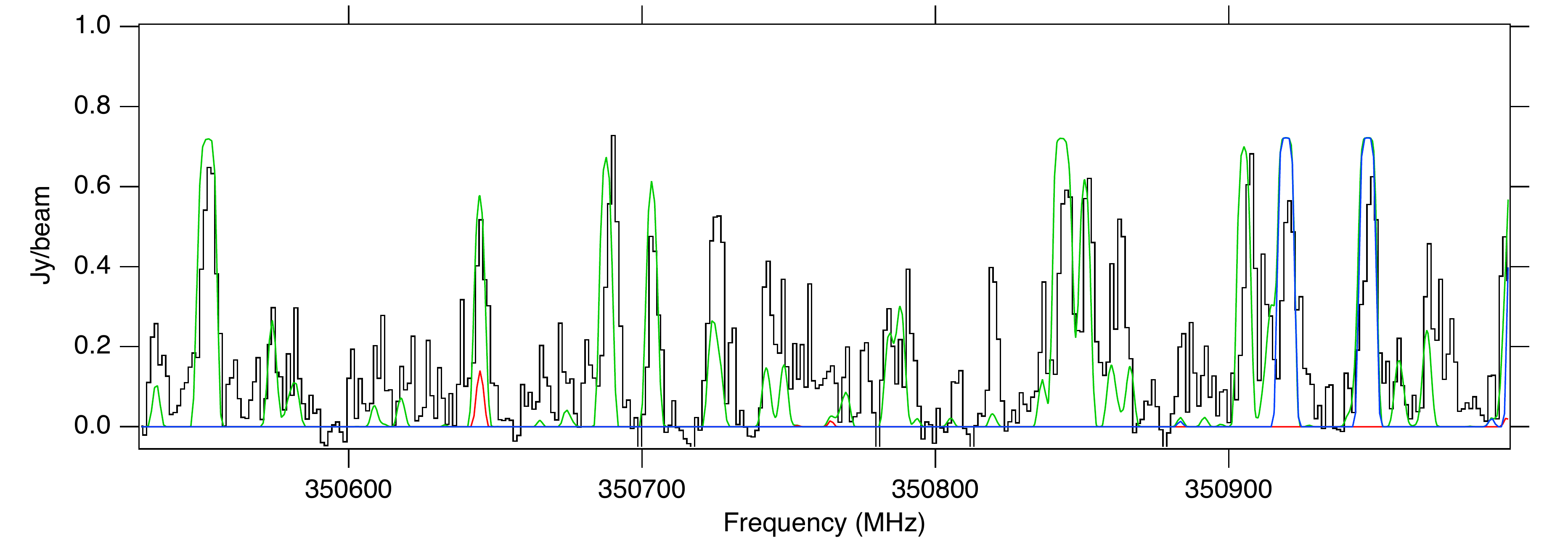}
    \includegraphics[width=0.9\textwidth]{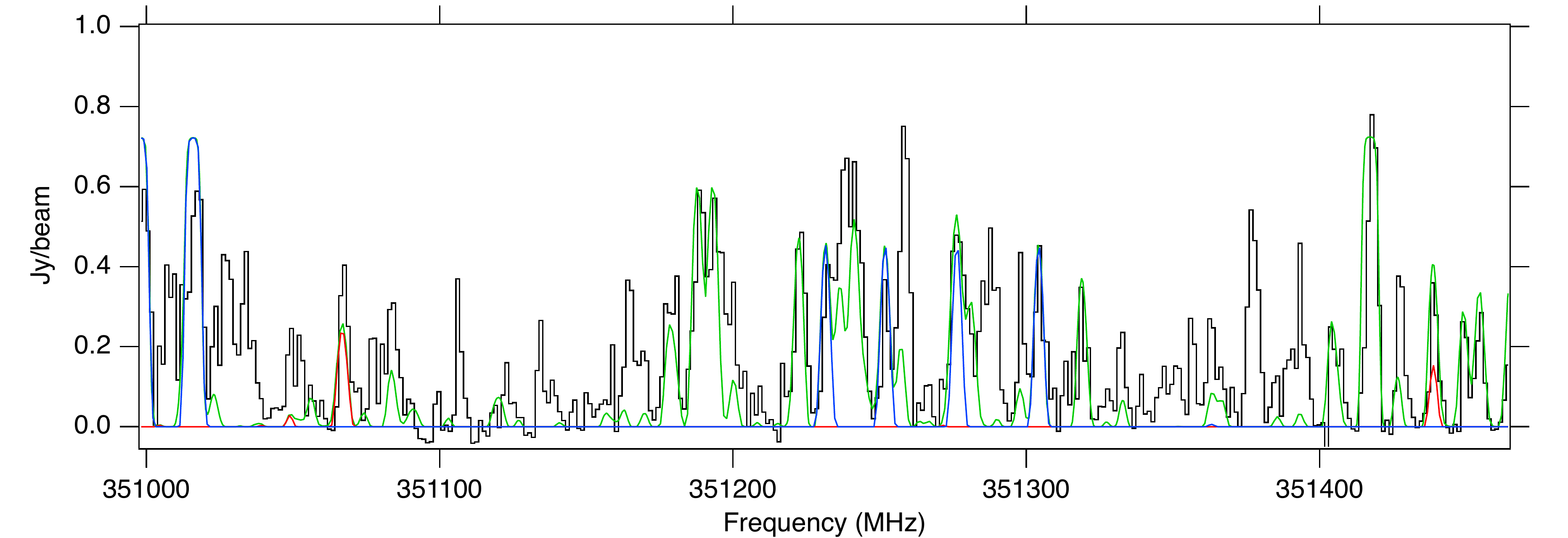}
    \includegraphics[width=0.9\textwidth]{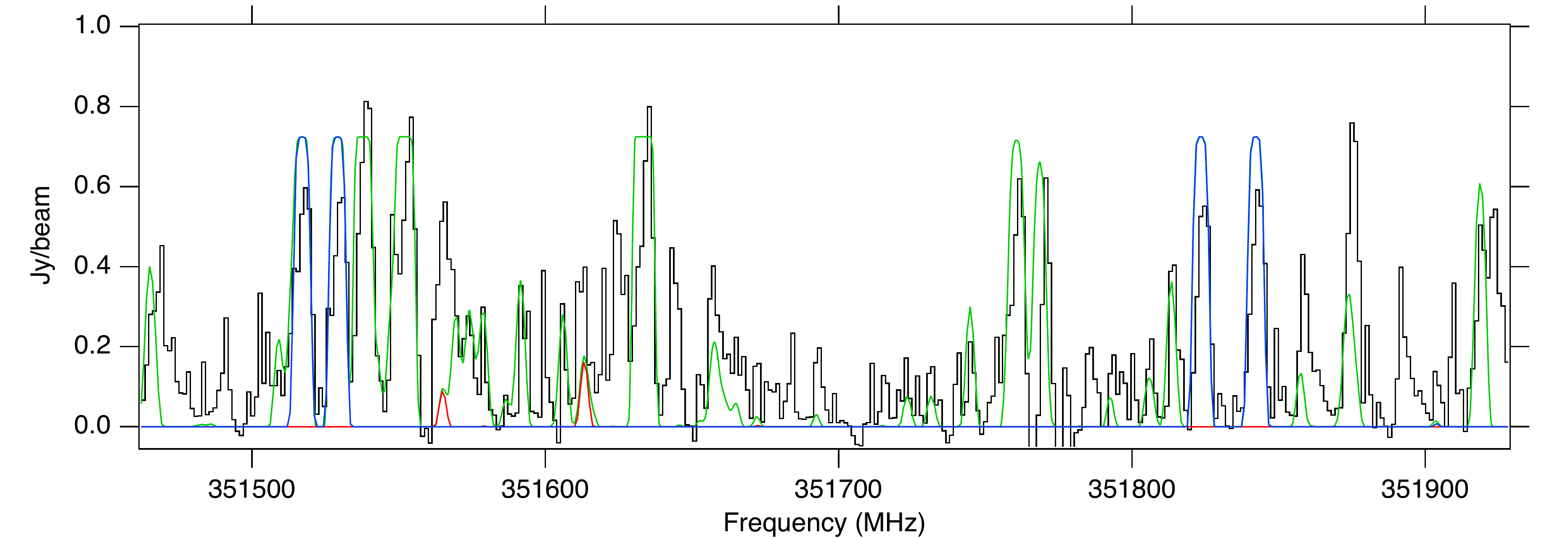}
    \caption{Spectra extracted toward NGC 6334I MM2-ii (black).  Overlaid in green is the full model of all assigned molecules in the spectrum (see text), and methyl formate and acetic acid are shown in color.  Transitions marked with an asterisk were identified as the least blended and optically thin, and were used for the column density analysis (see Table~\ref{freqs}).  Spectra were offset to a $v_{lsr}$~=~-9~km~s$^{-1}$.   }
    \label{mm2_8}
\end{figure}

The results are shown in the following pages. Figures~\ref{mm1_1} through \ref{mm1_8} display the observational spectra toward MM1-i, and Figures~\ref{mm2_1} through \ref{mm2_8} display those toward MM2-ii, with the total simulation as well as the individual simulations of methyl formate, glycolaldehyde (MM1-i only), and acetic acid overlaid.  The total simulation includes these species.  While the simulation accounts for opacity effects, there are several issues that cannot be properly accounted for without a full radiative transfer model that is beyond the scope of this analysis.  These are seen most prominently in the optically thick lines of \ce{CH3OH}.  First, for many of these lines there is substantial self-absorption that the model does not account for, resulting in `flat-topped' profiles in the model that do not match the observations for the highest optical-depth lines.  Second, these molecules are so abundant that their emission appears to also include non-trivial contributions from populations with $T_{ex}$ differing from 135~K by about $\pm$20~K.  As a result, some of these optically thick lines are over-predicted, and some under-predicted (depending on the species), by this comparatively simple model.

Thus, while the overall model is not perfect for the optically thick lines, it serves its purpose well: it is a zeroth-order analysis intended to help identify unblended lines.  Given the assumptions of uniform excitation and lineshapes for every species, and the by-eye analysis, we feel the fit is excellent, and completely fulfills its purpose.  These spectra demonstrate not only how line-rich the source is, but also how many lines are not easily assigned to known interstellar species.

The transitions of methyl formate, glycolaldehyde, and acetic acid used for the column density analysis are marked with asterisks.  As described in the Main Text, these were chosen as they appeared the least blended with other lines (including shoulders), and were not extremely optically thick.  

\clearpage

\clearpage

\clearpage

\clearpage

\clearpage

\clearpage

\clearpage

\clearpage

\clearpage

\clearpage

\clearpage

\clearpage

\clearpage

\clearpage

\clearpage

\end{document}